\documentclass[prd,aps,preprint,
tightenlines,
superscriptaddress,showpacs,nofootinbib,eqsecnum,amsfonts,amsmath,epsf]{revtex4}
\usepackage{bm,graphics,graphicx,epsfig,color}
\usepackage{subfigure}

\usepackage{amsfonts}
\usepackage{amsopn}
\usepackage{amssymb}
\usepackage{pxfonts}
\usepackage{rotating}
\usepackage{setspace}
\usepackage{amssymb}
\usepackage{lscape}
\usepackage{bm}
\usepackage[varg]{txfonts}
\usepackage{color}
\usepackage{pxfonts}

\newcommand{\ii}{\text{i}}
\newcommand{\ud}{\mathrm{d}}
\newcommand{\w}{\omega}

\allowdisplaybreaks

\begin{document}
\title{Third post-Newtonian angular momentum flux and the secular evolution of orbital elements for inspiralling compact binaries in quasi-elliptical orbits
}

\date{\today} \author{K G Arun} \email{arun@physics.wustl.edu} \affiliation{Raman
Research Institute, Bangalore 560 080, India}
\affiliation{$\mathcal{G}\mathbb{R}\varepsilon{\mathbb{C}}\mathcal{O}$,
Institut d'Astrophysique de Paris --- UMR 7095 du CNRS, \\ Universit\'e Pierre
\& Marie Curie, 98\textsuperscript{bis} boulevard Arago, 75014 Paris, France}
\affiliation{LAL, Universit{\'e} Paris-Sud, IN2P3/CNRS, Orsay, France}
\affiliation{McDonnell Centre for the Space Sciences, Department of Physics, Washington University, St Louis, Missouri 63130, USA. }
\author{Luc Blanchet} \email{blanchet@iap.fr} 
\affiliation{$\mathcal{G}\mathbb{R}\varepsilon{\mathbb{C}}\mathcal{O}$, Institut
d'Astrophysique de Paris --- UMR 7095 du CNRS, \\ Universit\'e Pierre \& Marie
Curie, 98\textsuperscript{bis} boulevard Arago, 75014 Paris, France}
\author{Bala R Iyer} \email{bri@rri.res.in} \affiliation{Raman Research
Institute, Bangalore 560 080, India} \author{Siddhartha Sinha} \email{p_siddhartha@rri.res.in} \affiliation{Raman Research Institute, Bangalore 560
080, India} \affiliation{Dept of Physics, Indian Institute of Science,
Bangalore, India}
\begin{abstract}
\end{abstract}
\preprint{gr-qc/yymmnnn}
\begin{abstract}
The angular momentum flux from an inspiralling binary system of compact objects moving in quasi-elliptical orbits is computed at the third post-Newtonian (3PN) order using the multipolar post-Minkowskian wave generation formalism. The 3PN angular momentum flux involves the instantaneous, tail, and tail-of-tails contributions as for the 3PN energy flux, and in addition a contribution due to non-linear memory. We average the angular momentum flux over the binary's orbit using the 3PN quasi-Keplerian representation of elliptical orbits. The averaged angular momentum flux provides the final input needed for gravitational wave phasing of binaries moving in quasi-elliptical orbits. We obtain the evolution of orbital elements under 3PN gravitational radiation reaction in the quasi-elliptic case. For small eccentricities, we give simpler limiting expressions relevant for phasing up to order $e^2$. This work is important for the construction of templates for quasi-eccentric binaries, and for the comparison of post-Newtonian results with the numerical relativity simulations of the plunge and merger of eccentric binaries.
\end{abstract}

\maketitle

%############################################################
\section{Introduction}\label{sec:intro}
%############################################################
The generation problem of gravitational waves (GWs) for inspiralling compact binaries has been completed at the third post-Newtonian (3PN) order both for the equation of motion of the binary and for its far-zone radiation field. The computations of the 3PN accurate equations of motion (EOM) and mass quadrupole moment were technically more involved than the corresponding 2PN cases due to the issues related to the ambiguities of self-field regularisation using Riesz or Hadamard regularisations~\cite{JS99,BFeom,DJSdim,BIJ02,BI04mult}. A deeper understanding of the cause of these ambiguities and the  use of the efficient dimensional regularisation scheme was crucial to the resolution of this problem~\cite{DJSdim,BDE04,BDEI04,BDEI05}. The 3.5PN phasing of inspiralling compact binaries (ICBs) moving in quasi-circular orbits is now complete and available for use in GW data analysis~\cite{BFIJ02,BDEI04,BDEI05}. This is timely since prototype binary GW sources for laser interferometer detectors are neutron star or black hole binaries close to their merger phase and consequently moving in quasi-circular orbits. However, astrophysical paradigms do exist that result in binaries with nonzero eccentricity in the sensitive bandwidth of both terrestrial and space-based GW detectors~\cite{KozaiOsc62,MillerHamilton02,Wen02,Benacquista01,GMHimbh04,GMHimbh06}.

More precisely, there currently exists a variety of astrophysical scenarios that can produce binaries which have residual ($\sim 0.01$), moderate ($\sim0.5$) or even very high ($\sim0.9$) eccentricities close to merger --- contrary to garden-variety ICB's mentioned earlier. For instance, the Kozai mechanism is one important scenario that produces eccentric binaries and involves the interaction between a pair of binaries in the dense cores of globular clusters~\cite{MillerHamilton02}. If the mutual inclination angle of the inner binary is strongly tilted 
with respect to the outer BH, then secular Kozai resonance~\cite{KozaiOsc62} can increase the eccentricity of the inner binary to large values. Numerical investigations~\cite{GMHimbh04,GMHimbh06} have shown that intermediate mass BH binaries can have eccentricity of order $0.9$ when they are visible in the LISA band. Even for the case of stellar mass BHs, the eccentricity may be about $0.1$ at 10Hz making them possible important sources for future ground-based GW detectors such as the Einstein Telescope (ET) which optimistically 
would attempt to achieve a seismic cut-off frequency around one Hertz. In the context of the ``final parsec problem'' for galaxy mergers, Ref.~\cite{ArmitageNatarajan05} pointed out that angular momentum loss to the circumbinary gas, which can provide a mechanism for overcoming this problem, can produce small but non zero eccentricity \textit{via} interaction of the binary with the gas disk. The resultant eccentricity can range from $0.01-0.1$ one week prior to merger depending on the binary's mass ratio and will have observable effects on the LISA signal. A more recent study of the scattering of stellar mass BHs in the galactic centres~\cite{OKL08} has found that more massive BHs dominate the scattering rate close to the central supermassive BH. These scatterings could give rise to bound binaries which will have a high eccentricity ($\sim 0.9$) when they enter the LIGO band. More importantly, due to higher harmonics present in the GW signals from eccentric binaries, these sources can be observed to larger distances and with larger masses ($\lesssim\,700\,M_\odot$) than for circular orbits, depending on the eccentricity~\cite{OKL08,YABW09}. This will have implications for sources in Advanced LIGO/VIRGO and ET detectors due to their very good proposed low frequency sensitivity~\cite{YABW09}. For a detailed discussion about various astrophysical mechanisms related to the eccentric orbit binaries see the introduction of Ref.~\cite{KG06} and appendix A of \cite{YABW09}.

The complete (ambiguity-free and fully determined) 3PN accurate EOM and mass quadrupole moment for compact binaries enable one to compute the 3PN energy and angular momentum fluxes for inspiralling compact binaries moving in general {\it non-circular} orbits. Recently, in two related papers~\cite{ABIQ07,ABIQ07tail}, we laid out the formalism and implemented the computation of the GW energy flux for non-circular orbits up to 3PN order. For non-circular orbits, to determine the orbital phasing, and the secular evolution of the orbital elements, the GW angular momentum flux needs to be known in addition to the energy flux. Of course, we also need the conserved center-of-mass energy and angular momentum of the orbit as deduced from the EOM.

In this paper, we compute the angular momentum flux of inspiralling compact binaries up to 3PN order generalising earlier work by Peters~\cite{Pe64} at Newtonian order, extended in Ref.~\cite{JunkS92} at 1PN order, 
in~\cite{RS97} at 1.5PN (tails)
and in~\cite{GI97} at 2PN. The 3PN contributions to energy and angular momentum fluxes come not only from {\it instantaneous} terms but also (non-linear) {\it hereditary} contributions~\cite{B98tail,B98quad}.
We shall find that for the angular momentum flux, the hereditary contributions comprise not only the tails, tails-of-tails and tail-square terms as for the energy flux but also formally an interesting {\it memory} contribution at 2.5PN order. One can then average the 3PN energy and angular momentum fluxes over an orbit thanks to the 3PN generalized quasi-Keplerian parametrization of the binary's orbital motion~\cite{MGS04}. Finally, we compute the secular evolution of orbital elements under 3PN gravitational radiation reaction (i.e. corresponding formally to 5.5PN terms in the EOM). This generalises the works of Peters and Mathews~\cite{PM63} at 2.5PN, Blanchet and Sch\"afer at 3.5PN~\cite{BS89} and 4PN~\cite{BS93,RS97} orders, and Gopakumar and Iyer~\cite{GI97} at 4.5PN. While~\cite{JunkS92,RS97} require the 1PN accurate
orbital description~\cite{DD85}, Ref.~\cite{GI97} crucially employs the generalised 2PN quasi-Keplerian parametrization of the binary's orbital motion in ADM coordinates as given in~\cite{DS88,SW93,Wex95}. In the present case the averaging of the instantaneous terms will require the full 3PN generalised quasi-Keplerian representation. However, the hereditary terms being relatively of higher PN orders, only require the 1PN parametrisation of the motion.

The secular evolution of orbital elements under gravitational radiation reaction provides the starting point for constructing templates for eccentric orbits. One of the first works in this direction was Ref.~\cite{MartPois99} which investigated the efficiency with which circular-orbit based templates would be able to detect an eccentric-orbit signal. To go beyond the secular evolution in the gravitational wave phasing one needs to include besides the averaged contribution the oscillatory terms in the evolution of orbital
elements. Damour, Gopakumar and Iyer~\cite{DGI04} discussed an analytic method for dealing with this issue at the leading radiation reaction order of 2.5PN, making possible the construction of high accuracy templates for the GW signals
from ICBs in quasi-elliptical orbits. This was extended to 3.5PN order in Ref.~\cite{KG06}, and the problem was revisited in a more elaborate way in ~\cite{TG07}, including
the computation of the noise-weighted overlaps for different astrophysical situations. Further investigations would be necessary to tackle the data analysis issues especially if the signals have moderate or high eccentricities. Including PN corrections to higher order in the evolution of orbital elements is a crucial step towards this.

With the recent advances in numerical relativity (NR) has emerged the possibility of comparing the NR waveforms to the PN results and exploring the regime of validity of the PN
approximation~\cite{BCP07NR,NRPNGoddard07,NRPNJena07,NRPNCaltech07}. These comparisons have been done for different source configurations such as nonprecessing spinning binaries~\cite{HHBG07}, precessing spins~\cite{CLNZ08}
and, very recently, eccentric binaries~\cite{HinderPNeccentric08}. It is obviously crucial to have a very accurate PN expression for the evolution of
the GW phase while comparing with the high accuracy NR simulations. For the circular orbit case we have 3.5PN accurate expressions in phasing~\cite{BDIWW95,B96,BFIJ02,BDEI04} and 3PN accurate ones in amplitude~\cite{BIWW96,ABIQ04,KBI07,K08,BFIS08}. Notably Ref.~\cite{HinderPNeccentric08} presented the first comparison between NR simulations of an eccentric binary black hole system with the corresponding current best PN results. The simulations relate to equal-mass, nonspinning
binaries but with an eccentricity $e\sim 0.1$ for about twenty GW cycles before merger, and comparison to a currently available 2PN eccentric binary. The work~\cite{HinderPNeccentric08} explores the parametrisation most suited for such a comparison and stands to improve with the 3PN model that our present paper
will now provide.

The organization of this paper is as follows. In Sec.~\ref{sec:structure} we start with the basic expression of the far-zone flux of angular momentum, employ expressions relating the radiative moments to the source moments and
decompose the angular momentum flux into its instantaneous and hereditary parts. Sec.~\ref{sec:AMF-instH} discusses the computation of the instantaneous terms in both standard harmonic coordinates and ADM coordinates. Using the 3PN
quasi-Keplerian representation, Sec.~\ref{sec:AMF-AvADM} computes the orbital average of the instantaneous part of the angular momentum flux in ADM coordinates. Sec.~\ref{sec:AMF-hered} deals with the computation of the hereditary contributions in the averaged angular momentum flux using a Fourier domain decomposition. The evolution of the orbital elements, in ADM coordinates, including both
instantaneous and hereditary terms, is presented in Sec.~\ref{sec:orbelem}. In the final Sec.~\ref{sec:small-ecc} we  provide simpler explicit expressions of various inputs up to first order in $e^2$ needed to deal with phasing in the small eccentricity limit $e\rightarrow 0$.
The paper concludes with three appendices.
Appendix \ref{appA} presents an analysis of  non-linear memory  leading to
a  DC term arising from  the dependence over the binary's (remote)
past history.
Appendix \ref{appB} includes tables of numerical values of the various 
``enhancement functions'' appearing in the paper to facilitate comparisons
with numerical relativity runs and use in data analysis applications.
The paper concludes with Appendix \ref{appC} where the important equations are also presented
in the modified harmonic coordinates for the convenience of the user.
%#################################################################
\section{The far-zone angular momentum flux}\label{sec:structure}
%#################################################################
In this section, we start from the angular momentum flux expressed in terms of the radiative multipole moments, use the relations connecting those radiative moments to the source moments, and rewrite the flux as a sum of the instantaneous terms which are functions of the retarded time, and hereditary terms which depend on the dynamics of the system in its entire past. The 3PN accurate angular momentum flux in the source's far-zone, denoted for convenience
\begin{equation}
{\cal G}_{i} \equiv \left( {\ud{\cal J}_{i}\over \ud t }\right)^\mathrm{GW}\,,
\label{eq:AMFnot}
\end{equation}
is expressed in terms of the mass and current type radiative multipole moments in radiative coordinates~\cite{Th80} as
\begin{eqnarray}
{\cal G}_{i} &=& {G\over c^5}\varepsilon_{iab} \biggl\{ {2\over 5} U_{aj}
U^{(1)}_{bj}\nonumber\\ &&+{1\over c^2} \left[ {1\over 63} U_{ajk}
U^{(1)}_{bjk} + {32\over 45} V_{aj} V^{(1)}_{bj}\right]+{1\over c^4} \left[
{1\over 2268} U_{ajkl} U^{(1)}_{bjkl} + {1\over 28} V_{ajk}
V^{(1)}_{bjk}\right] \nonumber\\ &&+{1\over c^6} \left[ {1\over 118800}
U_{ajklm} U^{(1)}_{bjklm} + {16\over 14175} V_{ajkl}
V^{(1)}_{bjkl}\right]+{\cal O}(8)\biggr\}\,.
\label{eq:AMFluxUV}
\end{eqnarray}
In the above $U_{L}$ and $V_{L}$ (with $L=i_1i_2\cdots i_l$ a multi-index composed of $l$ indices) are the symmetric-trace-free (STF) mass and current type radiative multipole moments respectively, and $U_{L}^{(p)}$ and
$V_{L}^{(p)}$ denote their $p^\text{th}$ time derivatives. The moments are functions of retarded time $T_R\equiv T-R/c$ in the radiative coordinates $(T,\mathbf{X})$, with $R=\vert\mathbf{X}\vert$ the distance of the source;
$\varepsilon_{iab}$ is the usual Levi-Civita symbol such that
$\varepsilon_{123}=+1$; the shorthand ${\cal O}(n)$ denotes a PN remainder of order of ${\cal O}(c^{-n})$.

%********************************************************
\subsection{ Radiative moments in terms of source moments}
\label{sec:UVIJ}
%*********************************************************
Using the MPM formalism~\cite{BD92,B98mult}, the radiative moments in Eq.~(\ref{eq:AMFluxUV}) can be computed in terms of the source moments to an accuracy sufficient for the computation of the angular momentum flux up to 3PN. One must compute the mass type radiative quadrupole $U_{ij}$ to 3PN
accuracy, mass octupole $U_{ijk}$ and current quadrupole $V_{ij}$ to 2PN, mass hexadecupole $U_{ijkm}$ and current octupole $V_{ijk}$ to 1PN, and finally $U_{ijkmn}$ and $V_{ijkm}$ to Newtonian order only. The relations connecting the radiative moments $U_{L}$ and $V_{L}$ to the corresponding source moments $I_L$ and $J_L$ (and also to the so-called gauge moments $W_L$, $X_L$, $Y_L$ and $Z_L$) are now given~\cite{BD92,B98tail,B98quad,B98mult}. For mass type moments we have (the brackets $<>$ denoting the STF projection)
\begin{subequations}\label{eq:u}
\begin{eqnarray}
U_{ij}(T_R) &=& I^{(2)}_{ij} (T_R) + {2GM\over c^3} \int_{-\infty}^{T_R} \ud V
\left[ \ln \left({T_R-V\over 2\tau_0}\right)+{11\over 12} \right] I^{(4)}_{ij}
(V) \nonumber \\ &+&\frac{G}{c^5}\left\{-\frac{2}{7}\int_{-\infty}^{T_R} \ud V
I^{(3)}_{a<i}(V)I^{(3)}_{j>a}(V) + {1
\over7}I^{(5)}_{a<i}I_{j>a} - {5 \over7} I^{(4)}_{a<i}I^{(1)}_{j>a} \right.\nonumber \\ &&\qquad~ \left.-{2
\over7} I^{(3)}_{a<i}I^{(2)}_{j>a} +{1
\over3}\varepsilon_{ab<i}I^{(4)}_{j>a}J_{b}
+4\left[W^{(2)}I_{ij}-W^{(1)}I_{ij}^{(1)}\right]^{(2)} \right\}\nonumber \\
&+&2\left(\frac{G M}{c^3}\right)^2\int_{-\infty}^{T_R}\ud V
I_{ij}^{(5)}\left(V\right)\left[\ln^2\left({T_R-V
\over2\tau_0}\right)+{57\over70}\ln\left({T_R-V\over2\tau_0}\right)
+{124627\over44100}\right] \nonumber\\ &+& {\cal O}(7)\,, \label{eq:uij}\\
U_{ijk} (T_R) &=& I^{(3)}_{ijk} (T_R) + {2GM\over c^3} \int_{-\infty}^{T_R}
\ud V\left[ \ln \left({T_R-V\over 2\tau_0}\right)+{97\over60} \right]
I^{(5)}_{ijk} (V) + {\cal O}(5)\,, \\ U_{ijkl} (T_R) &=& I^{(4)}_{ijkl} (T_R)
+ {\cal O}(3) \,, \\ U_{ijklm} (T_R) &=& I^{(5)}_{ijklm} (T_R) + {\cal O}(3)
\,.
\end{eqnarray}\end{subequations}
The $I_{L}$'s and $J_L$'s are the STF mass and current-type source moments; $M=I$ is the total mass monopole which is conserved; $W$ is the monopole corresponding to one type of the gauge moments, i.e. $W_L$. For the current-type moments we find 
\begin{subequations}\label{eq:v}
\begin{eqnarray}
V_{ij} (T_R) &=& J^{(2)}_{ij} (T_R) + {2GM\over c^3} \int_{-\infty}^{T_R} \ud V
\left[ \ln \left({T_R-V\over 2\tau_0}\right)+{7\over6} \right] J^{(4)}_{ij} (V) +
{\cal O}(5) \,,\label{eq:vij} \\ V_{ijk} (T_R) &=& J^{(3)}_{ijk} (T_R) + {\cal
O}(3) \,,\\ V_{ijkl} (T_R) &=& J^{(4)}_{ijkl} (T_R) + {\cal O}(3) \,.
\end{eqnarray}\end{subequations}

The radiative moments have two distinct contributions: ``instantaneous'', which is a snapshot function of the retarded instant $T_R$ only; and ``hereditary'', which depends on the dynamics of the system in its entire past $V\leq T_R$. The parameter $\tau_0$ appearing in the logarithms of Eqs.~\eqref{eq:u} and \eqref{eq:v} is a freely specifiable constant time scale, entering the relation between the retarded time $T_R=T-R/c$ in radiative coordinates and the corresponding time $t_\mathrm{H}-r_\mathrm{H}/c$ in harmonic coordinates (where $r_\mathrm{H}$ is the distance of the source in harmonic coordinates). Posing $r_0=c \tau_0$ we have
\begin{equation}
T_R = t_\mathrm{H}- \frac{r_\mathrm{H}}{c}
-\frac{2\,G\,M}{c^3}\ln\left(\frac{r_\mathrm{H}}{r_0}\right)\,.
\label{b}\end{equation}
We choose the constant $r_0$ scaling the logarithm to match with the choice made in the computation of tails-of-tails in~\cite{B98tail}.

\subsection{Structure of the 3PN angular momentum flux}

One can schematically split the total contribution to the angular momentum flux as the sum of the instantaneous and hereditary terms,
\begin{equation}
{\cal G}_i = {\cal G}_i^{\rm inst} + {\cal G}_i^{\rm hered}\,,
\end{equation}
where the instantantaneous terms are given by [not specifying the obvious $\mathcal{O}(n)$]
\begin{eqnarray}\label{eq:inst}
{\cal G}_i^{\rm inst} &=& {G\over c^5}\varepsilon_{iab} \Biggl\{ {2\over 5}
I^{(2)}_{aj} I^{(3)}_{bj}\nonumber\\ &+&{1\over c^2} \left[ {1\over 63}
I^{(3)}_{ajk} I^{(4)}_{bjk} + {32\over 45} J^{(2)}_{aj}
J^{(3)}_{bj}\right]+{1\over c^4} \left[ {1\over 2268} I^{(4)}_{ajkl}
I^{(5)}_{bjkl} + {1\over 28} J^{(3)}_{ajk} J^{(4)}_{bjk}\right] \nonumber\\
&+&\frac{2G}{5c^{5}}\left[4W^{(5)}I_{aj}^{(2)}I_{bj}+8W^{(4)}I_{aj}^{(2)}I_{bj}^{(1)}
-12W^{(2)}I_{aj}^{(2)}I_{bj}^{(3)}-4W^{(1)}I_{aj}^{(2)}I_{bj}^{(4)}
\right.\nonumber\\&&\quad+4W^{(4)}I_{aj}I_{bj}^{(3)}
+4W^{(3)}I_{aj}^{(1)}I_{bj}^{(3)} -4W^{(1)}I_{aj}^{(3)}I_{bj}^{(3)}
\nonumber\\ &&\quad+I_{bj}^{{(3)}} \bigg(-\frac{5}{7} I_{c<a}^{{(4)}}
I_{j>c}^{{(1)}}-\frac{2}{7} I_{c<a}^{{(3)}} I_{j>c}^{{(2)}}+\frac{1}{7}
I_{c<a}^{{(5)}} I_{j>c}+{1\over 3}\varepsilon_{cd<a}\,
I_{j>c}^{{(4)}}\,J_d\bigg) \nonumber \\&&\quad+
I_{aj}^{{(2)}}\bigg(-\frac{4}{7}
I_{c<b}^{{(5)}}I_{j>c}^{{(1)}}-I_{c<b}^{{(4)}}I_{j>c}^{{(2)}}
-\frac{4}{7}I_{c<b}^{{(3)}}I_{j>c}^{{(3)}}+\frac{1}{7} I_{c<b}^{{(6)}}I_{j>c}
+{1\over 3}\varepsilon_{cd<b}\,I_{j>c}^{{(5)}} J_d \bigg)\bigg]\nonumber\\
&+&{1\over c^6} \left[ {1\over 118800} I^{(5)}_{ajklm} I^{(6)}_{bjklm} +
{16\over 14175} J^{(4)}_{ajkl} J^{(5)}_{bjkl}\right] \Biggr\}\,.
\end{eqnarray}
Using \eqref{eq:u}--\eqref{eq:v} in Eq.~\eqref{eq:AMFluxUV} 
the hereditary part can be further decomposed as
\begin{equation}
{\cal G}_i^{\rm hered}={\cal G}_i^{\rm tail}+{\cal G}_i^{\rm tail(tail)}+{\cal
G}_i^{\rm (tail)^2}+{\cal G}_i^{\rm memory}\,,\label{eq:hered}
\end{equation}
where the quadratic tail integrals are given by\footnote{
Ref.~\cite{RS97} contains a typographical mistake at 1.5PN
in Eq.~(83). The first term in Eq.~(\ref{eq:tail}) below
containing ``$I^{(2)}(T_R)I^{(5)}(V)$" is missing. This has been
 independently pointed out recently in \cite{RBK08}.
However, the results in \cite{RS97} are correct  and do take
this term into account. 
Ref.~\cite{GI97} which  only quotes \cite{RS97} 
also contains this typo. }

\begin{eqnarray}
{\cal G}_i^{\rm tail}&=&\frac{G^2 M}{c^5}\,\varepsilon_{iab}\,\Biggl\{
\frac{4}{5c^3}I_{aj}^{{(2)}}(T_R)\int_{-\infty}^{T_R} \ud
V\left[{\ln}\left(\frac{T_R-V }{2 \tau_0}\right)+\frac{11}{12}\right]
I_{bj}^{{(5)}}(V)
\nonumber\\&&\qquad+\frac{4}{5c^3}I_{bj}^{{(3)}}(T_R)\int_{-\infty}^{T_R} \ud
V\left[{\ln}\left(\frac{T_R-V }{2 \tau_0}\right)+\frac{11}{12}\right]
I_{aj}^{{(4)}}(V) \nonumber\\&&\qquad+
\frac{64}{45c^5}J_{aj}^{{(2)}}(T_R)\int_{-\infty}^{T_R} \ud
V\left[{\ln}\left(\frac{T_R-V }{2 \tau_0}\right)+\frac{7}{6}\right]
J_{bj}^{{(5)}}(V)\nonumber\\&&\qquad+\frac{64}{45c^5}J_{bj}^{{(3)}}(T_R)
\int_{-\infty}^{T_R} \ud V\left[{\ln}\left(\frac{T_R-V }{2
\tau_0}\right)+\frac{7}{6}\right]
J_{aj}^{{(4)}}(V)\nonumber\\&&\qquad+\frac{2}{63c^5}I_{ajk}^{{(3)}}(T_R)
\int_{-\infty}^{T_R} \ud V\left[{\ln}\left(\frac{T_R-V }{2
\tau_0}\right)+\frac{97}{60}\right]
I_{bjk}^{{(6)}}(V)\nonumber\\&&\qquad+\frac{2}{63c^5}I_{bjk}^{{(4)}}(T_R)
\int_{-\infty}^{T_R} \ud V\left[{\ln}\left(\frac{T_R-V }{2
\tau_0}\right)+\frac{97}{60}\right] I_{ajk}^{{(5)}}(V) \Biggr\}\,,
\label{eq:tail}
\end{eqnarray}
and the cubic-order tail integrals are
\begin{subequations}\label{eq:tail-tail-sq}\begin{eqnarray}
{\cal G}_i^{\rm tail(tail)}&=&
\frac{4}{5}\frac{G^3
M^2}{c^{11}}\,\varepsilon_{iab}\,\Biggl\{I_{aj}^{{(2)}}(T_R)\int_{-\infty}^{T_R}
\ud V\left[\ln^2 \left(\frac{T_R-V}{2 \tau_0}\right) +\frac{57}{70} \ln
\left(\frac{T_R-V}{2 \tau_0}\right)+\frac{124627}{44100}\right]
I_{bj}^{{(6)}}(V)\nonumber\\&&\qquad+I_{bj}^{{(3)}}(T_R)\int_{-\infty}^{T_R}
\ud V\left[\ln^2 \left(\frac{T_R-V}{2 \tau_0}\right) +\frac{57}{70} \ln
\left(\frac{T_R-V}{2 \tau_0}\right)+\frac{124627}{44100}\right]
I_{aj}^{{(5)}}(V) \Biggr\}\,,\label{eq:tail-tail}\\\nonumber\\ {\cal G}_i^{\rm
(tail)^2}&=& \frac{8}{5}\frac{G^3
M^2}{c^{11}}\varepsilon_{iab}\left(\int_{-\infty}^{T_R} \ud V\left[\ln
\left(\frac{T_R-V}{2 \tau_0}\right)+\frac{11}{12}\right]
I_{aj}^{{(4)}}(V)\right)\left(\int_{-\infty}^{T_R} \ud V\left[\ln
\left(\frac{T_R-V}{2 \tau_0}\right)+\frac{11}{12}\right] I_{bj}^{{(5)}}(V)
\right)\,.\label{eq:tail-sq}\nonumber\\
\end{eqnarray}\end{subequations}
These tail and tail-of-tail integrals are similar to those occurring in the energy flux~\cite{ABIQ07tail}. However we have also the non-linear memory integral
\begin{eqnarray}\label{eq:memory}
{\cal G}_i^{\rm
memory}&=&\frac{4}{35}\frac{G^2}{c^{10}}\,\varepsilon_{iab}\,I_{aj}^{{(3)}}(T_R)
\int_{-\infty}^{T_R} \ud V I_{c<b}^{{(3)}}(V)~ I_{j>c}^{{(3)}}(V)\,.
\end{eqnarray}
Recall that the memory contributes to the radiative quadrupole moment $U_{ij}$ simply by an anti-derivative of source moments. It therefore becomes instantaneous when we consider the time derivative $U_{ij}^{(1)}$ of the radiative quadrupole moment. Hence we have incorporated in the instantaneous part of the angular momentum flux [Eq.~(\ref{eq:inst})] a term coming from the time derivative of the memory integral. The presence of the non-linear memory contribution~(\ref{eq:memory}) remaining in the angular momentum flux is to be noticed. This is in contrast with the case of the energy flux, where there is no memory contribution because the memory is time differentiated therein and therefore becomes instantaneous (see Ref.~\cite{ABIQ07tail}).

%%%%%%%%%%%%%%%%%%%%%%%%%%%%%%
\section{Instantaneous terms in the 3PN angular momentum flux}
\label{sec:AMF-instH}
%%%%%%%%%%%%%%%%%%%%%%%%%%%%%%%%%%%%%%%%%%%%%%%%%%%%%%%%%%%
The relevant source multipole moments needed for the computation of the angular momentum flux up to 3PN order are the same as for the energy flux. Hence we redirect the reader to Secs.~III and IV of Ref.~\cite{ABIQ07} for a detailed discussion of these moments as well as for the equation of motion up to 3PN order (see Ref.~\cite{BI03CM}) which is required when differentiating the source moments. Notice that the mass quadrupole moment $I_{ij}$ was available in~\cite{BI04mult} and was used in~\cite{ABIQ07tail}. Using all these source moments, and the EOM at 3PN order, it is possible to compute the different PN contributions to the instantaneous part~(\ref{eq:inst}) of the angular momentum flux.

The prominent application of the present computation will be discussed in Sec.~\ref{sec:orbelem} where the evolution of the orbital elements under gravitational radiation reaction will be investigated to 3PN order. This will be based on the solution of the motion in the form of the quasi-Keplerian representation of the orbit. The latter can be written up to 3PN order in ADM coordinates (and also in ``modified'' harmonic coordinates~\cite{ABIQ07}) but not in the standard harmonic coordinate system, due to the presence of gauge-dependent logarithms arising at 3PN order. We shall therefore present our results in the (standard) harmonic coordinate system and also in ADM coordinates, in which we give the evolution of the orbital elements (see Sec.~\ref{sec:orbelem}).

\subsection{Instantaneous flux in standard harmonic coordinates}
\label{sec:AMF-H}
The standard harmonic (SH) coordinate system is the one in which the original computations of the 3PN center-of-mass equations of motion~\cite{BI03CM} and 3PN source quadrupole moment~\cite{BI04mult} were given. It is known that the standard harmonic coordinate system develops some gauge-dependent logarithmic terms at 3PN order. This particular coordinate system is referred to as standard in order to distinguish it from the modified harmonic (MH) coordinate system, where the logarithms at 3PN order have been gauged away. Both the MH coordinates and the ADM coordinates are suitable for a 3PN quasi-Keplerian parametrization of the motion; they were reviewed and used in the computation of the energy flux in Ref.~\cite{ABIQ07}.

The angular momentum flux will be orthogonal to the orbital plane, and aligned with the orbital angular momentum. We introduce the unit direction along the orbital angular momentum,
\begin{equation}\label{Li}
\hat{L}_i \equiv \frac{\varepsilon_{ijk}\,x^j\,v^k}{r^2\dot{\varphi}}\,,
\end{equation}
where $x^j$ and $v^k=\ud x^k/\ud t$ are the relative separation and velocity of the two particles. Here  $\dot{\varphi}=\ud \varphi/\ud t$, where $\varphi$ denotes the orbital phase\footnote{Note that for non-circular orbits $\dot{\varphi}$ differs from the mean (``orbit-averaged'') angular frequency $\omega=K n$ we define below, namely $\dot{\varphi}=\omega+n \,\ud W/\ud\ell$, where $W(\ell)$ is periodic in $\ell=n(t-t_\text{P})$ with period $2\pi$, hence periodic in time with period $P$ (see e.g. Sec. II.A in~\cite{ABIQ07tail}). We can call $\dot{\varphi}$ the ``instantaneous'' angular frequency.}, and is linked to the orbital velocity by $v^2=\dot{r}^2+r^2\dot{\varphi}^2$ where $\dot{r}=\ud r/\ud t$. Posing
\begin{equation}\label{GiLi}
{\cal G}_i^{\rm inst} = \hat{L}_i\,{\cal G}_{\rm inst}\,,
\end{equation}
we look for the 3PN expansion 
\begin{equation}\label{fluxdef}
{\cal G}_{\rm inst} = {\cal G}_{\rm N}+{\cal G}_{\rm 1PN}+{\cal G}_{\rm 2PN}
+{\cal G}_{\rm 2.5PN}+{\cal G}_{\rm 3PN} + {\cal O}(7)\,,
\end{equation}
where as usual the Newtonian piece really corresponds to the dominant radiation reaction at 2.5PN order in the equations of motion. The results in SH coordinates then read\footnote{Mass parameters are the total mass $m=m_1+m_2$, the reduced mass $\mu=m_1m_2/m$ and the symmetric mass ratio $\nu=\mu/m$ which is such that $0<\nu\leq 1/4$.}
\begin{subequations}\label{fluxSH}
\begin{eqnarray}
{\cal G}_{\rm N} &=& \frac{ G^2 m^3 \nu ^2 }{ c^{5} r}\,\dot{\varphi}\,\Biggl\{ \frac{16}{5}
v^2 -\frac{24}{5}\dot{r}^2+\frac{16}{5}\frac{G\,m}{r} \Biggr\}\,, \\
%%%%%%%%%%%%%%%%%%%%%
{\cal G}_{\rm 1PN} &=& \frac{ G^2 m^3 \nu ^2 }{
c^{7} r} \,\dot{\varphi}\,\Biggl\{
v^4\left(\frac{614}{105}-\frac{1096}{105}\nu\right)
+v^2\,\dot{r}^2\left(-\frac{296}{35}+\frac{1108}{35}\nu\right) \nonumber\\ &&
+\frac{G\,m}{r}\,v^2\left(-\frac{464}{105}-\frac{152}{21}\nu\right)
+\dot{r}^4\left(\frac{38}{7}-\frac{144}{7}\nu\right)
\nonumber\\ &&
+\frac{G\,m}{r}\,\dot{r}^2\left(\frac{496}{35}+\frac{788}{105}\nu\right)
+\frac{G^2\,m^2}{r^2}\left(-\frac{596}{21}+\frac{8}{105}\nu\right) \Biggr\}\,,
\\
%%%%%%%%%%%%%%%%%%%%%
{\cal G}_{\rm 2PN} &=& \frac{ G^2 m^3 \nu ^2 }{ c^{9} r}\,\dot{\varphi}\,\Biggl\{
v^6\left(\frac{533}{63}-\frac{353}{9}\nu+\frac{614}{15}\nu^2\right)
+v^4\,\dot{r}^2\left(-\frac{2246}{105}+\frac{12653}{105}\nu-\frac{15637}{105}\nu^2\right)
\nonumber\\ &&
+\frac{G\,m}{r}\,v^4\left(\frac{11}{21}-\frac{491}{315}\nu+\frac{4022}{315}\nu^2\right)
+v^2\,\dot{r}^4\left(\frac{715}{21}-\frac{3361}{21}\nu+\frac{448}{3}\nu^2\right)
\nonumber\\ &&
+\frac{G\,m}{r}\,v^2\,\dot{r}^2\left(\frac{21853}{315}-\frac{7201}{105}\nu
+\frac{2551}{315}\nu^2\right) \nonumber\\ &&
+\frac{G^2\,m^2}{r^2}\,v^2\left(-\frac{21302}{315}+\frac{2262}{35}\nu-
\frac{6856}{315}\nu^2\right)
+\dot{r}^6\left(-\frac{52}{3}+\frac{652}{9}\nu-\frac{388}{9}\nu^2\right)
\nonumber\\ &&
+\frac{G\,m}{r}\,\dot{r}^4\left(-\frac{22312}{315}+\frac{5914}{45}\nu
-\frac{277}{9}\nu^2\right)
+\frac{G^2\,m^2}{r^2}\,\dot{r}^2\left(\frac{5624}{105}-\frac{7172}{45}\nu
+\frac{3058}{105}\nu^2\right) \nonumber\\ &&
+\frac{G^3\,m^3}{r^3}\left(\frac{340724}{2835}+\frac{15658}{315}\nu
+\frac{44}{45}\nu^2\right) \Biggr\}\,, \\
%%%%%%%%%%%%%%%%%%%%%
{\cal G}_{\rm 2.5PN} &=& \frac{ G^2 m^3 \nu ^2 }{5\, c^{10} r}
\,\dot{\varphi}\,\dot{r}\,\nu\,\Biggl\{-\frac{27744}{35}
\frac{G\,m}{r}\,v^4+\frac{19144}{7} \frac{G\,m}{r}\,v^2 \dot{r}^2
-\frac{116944}{105} \frac{G^2\, m^2}{r^2} v^2\nonumber\\ && +\frac{8976}{7}
\left(\frac{G m}{r}\right)^2 \dot{r}^2 - 1960
\frac{G\,m}{r}\,\dot{r}^4-\frac{22864}{105} \left(\frac{G
m}{r}\right)^3\Biggr\}\,,\label{eq:AMF2p5Har}\\
%%%%%%%%%%%%%%%%%%%%%
{\cal G}_{\rm 3PN} &=& \frac{G^2 m^3 \,\nu ^2}{c^{11}\,r}\,\dot{\varphi}\,\Biggl\{v^8
\left[\frac{145919}{13860}-\frac{110423  }{1260}\nu+\frac{1079083 
}{4620}\nu^2 -\frac{30229 }{165}\nu ^3\right]\nonumber\\ &&+\dot{r}^2 v^6
\left[-\frac{2473}{70}+\frac{763409  }{2310}\nu-\frac{2155249 }{2310}\nu ^2
+\frac{543171 }{770}\nu ^3\right]\nonumber\\ &&+\frac{G\,m}{r} v^6
\left[\frac{483097}{13860}-\frac{60913  }{1540}\nu+\frac{28711 }{4620}\nu ^2
+\frac{91 }{165}\nu ^3\right]\nonumber\\ &&+\dot{r}^4 v^4
\left[\frac{18695}{231}-\frac{632111  }{924}\nu+\frac{1552525 }{924}\nu ^2
-\frac{61970 }{77}\nu ^3\right]\nonumber\\ &&+\dot{r}^2
\frac{G\,m}{r} v^4 \left[\frac{205817}{13860}-\frac{74689  }{140}\nu
+\frac{6423539 }{13860}\nu ^2-\frac{9979 }{66}\nu ^3\right]\nonumber\\ &&+\frac{G^2\,m^2}{r^2} v^4
\left[\frac{5112059}{28875}-\frac{6848}{175}\ln\left({r\over
r_0}\right)\right.\nonumber\\ &&\left.+\left(\frac{3152431  }{6930}
-\frac{369   }{40}\pi ^2\right)\nu-\frac{407999 }{3465}\nu ^2+\frac{154421 }{1155}\nu ^3\right]\nonumber\\ &&+\dot{r}^6 v^2 \left(-\frac{451}{6}+\frac{59870
 }{99}\nu-\frac{14841 }{11}\nu ^2+\frac{32342 }{99}\nu ^3\right]\nonumber\\ &&+\dot{r}^4 \frac{G\,m}{r} v^2
\left[-\frac{100999}{1980}+\frac{3716239  }{2772}\nu-\frac{22275889 }{13860}\nu ^2+\frac{738973 }{3465}\nu ^3\right]\nonumber\\ &&+\dot{r}^2
\frac{G^2\,m^2}{r^2} v^2
\left[-\frac{14402404}{5775}+\frac{1712}{5}\ln\left({r\over r_0}\right)
\right.\nonumber\\ &&\left.+\left(-\frac{1394339  }{495}+\frac{369  }{4}\pi ^2\right)\nu+\frac{317813 }{495}\nu ^2-\frac{869048 }{3465}\nu ^3\right]\nonumber\\ &&+\frac{G^3\,m^3}{r^3} v^2
\left[\frac{1229915081}{1559250}-\frac{37664}{525}\ln\left({r\over
r_0}\right)\right.\nonumber\\ 
&&\left.-\left(\frac{8689013  }{28350}+\frac{41  }{10}\pi ^2+\frac{352 }{15}\ln\left({r\over r_0'}\right)\right)\nu+\frac{184003 }{990}\nu ^2-\frac{155339 }{3465}\nu ^3\right]\nonumber\\ &&+\dot{r}^8 \left[\frac{93}{4}-\frac{4035  }{22}\nu+\frac{4294 }{11}\nu ^2-\frac{410 }{11}\nu ^3\right]\nonumber\\
&&+\dot{r}^6 \frac{G\,m}{r} \left[\frac{1932907}{69300}-\frac{106499 
}{140}\nu+\frac{13641581 }{13860}\nu ^2-\frac{50525 }{1386}\nu ^3\right]\nonumber\\ &&+\dot{r}^4 \frac{G^2\,m^2}{r^2}
\left[\frac{46903957}{17325}-\frac{1712}{5}\ln\left({r\over
r_0}\right) \right.\nonumber\\ &&\left.
+\left(\frac{3002737  }{1155}-\frac{861  }{8}\pi ^2\right)\nu
-\frac{28913 }{330}\nu ^2
+\frac{85543 }{3465}\nu ^3\right] \nonumber\\ &&+\dot{r}^2 \frac{G^3\,m^3}{r^3}
\left[-\frac{152347309}{103950}+\frac{5136}{35}\ln\left({r\over
r_0}\right)\right.\nonumber\\ 
&&\left.+\left(\frac{102197341  }{103950}+\frac{123  }{20}\pi ^2-\frac{176  }{5}\ln\left({r\over r_0'}\right)\right)\nu-\frac{3265967 }{6930}\nu ^2+\frac{267188 }{3465}\nu ^3\right]\nonumber\\ &&+\frac{G^4\,m^4}{r^4}
\left[-\frac{47779894}{111375}+\frac{3424}{525} \ln\left({r\over
r_0}\right) \right.\nonumber\\ 
&&\left.+\left(-\frac{49735624  }{51975}+\frac{779   }{40}\pi ^2
+\frac{704  }{15}\ln\left({r\over r_0'}\right) \right)\nu
-\frac{818 }{63}\nu ^2+\frac{1780 }{693}\nu ^3\right]\Biggr\}\,.
\end{eqnarray}\end{subequations}
We reproduce the terms computed earlier~\cite{Pe64,JunkS92,GI97} in the angular momentum flux up to 2PN order. As can be seen from above, the 3PN terms contain two kinds of logarithms, $\ln(r/r_0)$ and $\ln(r/r_0')$. The logarithms $\ln(r/r_0')$ are specific to the standard harmonic coordinate system. We discuss later in more detail these different kinds of logarithmic terms.

\subsection{Instantaneous flux in ADM coordinates}
\label{sec:AMF-ADM}
Since many related numerical relativity studies are in ADM-type coordinates, we shall present the applications in later sections of this paper in ADM coordinates. Going from SH to ADM coordinates removes the gauge-dependent logarithms $\ln(r/r_0')$ which are not very convenient to handle in numerical calculations. We shall thus re-express the instantaneous flux \eqref{fluxdef}--\eqref{fluxSH} in terms of the variables in ADM coordinates. This requires the use of the so-called contact transformation of these variables ($x^i$, $v^i$ and ${\dot r}$), linking the standard harmonic coordinates to the ADM ones. We refer to~\cite{ABIQ07} (see Sec.~VI.B) for the details of that transformation. The contact transformation equation will depend on some $\ln(r/r_0')$'s and that dependence will be such that the flux in ADM coordinates becomes independent of those logarithms $\ln(r/r_0')$, revealing the gauge nature of the constant $r_0'$. However, as we shall comment below, the other logarithms of type $\ln(r/r_0)$ will remain in the instantaneous part of the flux but will be cancelled by related contributions coming from the tail integrals (and more precisely the tails-of-tails); indeed, see the constant $\tau_0=r_0/c$ in Eqs.~\eqref{eq:tail}--\eqref{eq:tail-tail-sq}.

The angular momentum flux in ADM coordinates admits the same type of PN expansion as~\eqref{fluxdef}. Since the transformation between the SH and ADM coordinate systems starts at the 2PN order, only the 2PN and 3PN terms in the flux get modified and we now list these two terms, labelled by ``ADM'' to remember that all variables therein correspond to ADM coordinates. Let us recall that the transformation formulas between the standard harmonic and ADM coordinates which were given in~\cite{ABIQ07} concerned only the conservative part of the dynamics. We thus give here only the 2PN and 3PN terms in the angular-momentum
flux following these transformations,
\begin{subequations}\label{fluxADM}
\begin{eqnarray}
{\cal G}_{\rm 2PN}^{\rm ADM} &=& \frac{G^2 m^3 \,\,\nu
^2}{c^{9}\,r}\,\dot{\varphi}\,\Biggl\{v^6 \left[\frac{533}{63}
-\frac{353  }{9}\nu+\frac{614 }{15}\nu ^2\right]\nonumber\\ &&+\,\dot{r}^2 v^4
\left[-\frac{2246}{105}+\frac{12653  }{105}\nu-\frac{15637 }{105}\nu ^2\right]\nonumber\\ 
&&+\left(\frac{G\,m}{r}\right) v^4
\left[\frac{11}{21}-\frac{1333  }{63}\nu+\frac{4022 }{315}\nu ^2\right]\nonumber\\ &&+\,\dot{r}^4 v^2 \left[\frac{715}{21}-\frac{3361
 }{21}\nu+\frac{448 }{3}\nu ^2\right]\nonumber\\
&&+\left(\frac{G\,m}{r}\right) \,\dot{r}^2 v^2
\left[\frac{21853}{315}+\frac{2942  }{105}\nu+\frac{2551
}{315}\nu^2\right]\nonumber\\ &&+\left(\frac{G\,m}{r}\right)^2 v^2
\left[-\frac{4210}{63}+\frac{2962  }{35}\nu-\frac{6856
}{315}\nu^2\right]\nonumber\\ &&+\,\dot{r}^6 \left[-\frac{52}{3}+\frac{652
 }{9}\nu-\frac{388 }{9}\nu ^2\right]\nonumber\\
&&+\left(\frac{G\,m}{r}\right) \,\dot{r}^4 \left[-\frac{22312}{315}+\frac{1999
 }{45}\nu-\frac{277 }{9}\nu ^2\right]\nonumber\\
&&+\left(\frac{G\,m}{r}\right)^2 \,\dot{r}^2 \left[\frac{5666}{105}-\frac{6938
 }{45}\nu+\frac{3058 }{105}\nu ^2\right]\nonumber\\
&&+\left(\frac{G\,m}{r}\right)^3 \left[\frac{336188}{2835}+\frac{11878  }{315}\nu+\frac{44 }{45}\nu ^2\right]\Biggr\}\,,\\
%%%%%%%%%%%%%%%%%%%%%%%%%%%%%
{\cal G}_{\rm 3PN}^{\rm ADM} &=&\frac{G^2 m^3 \,\,\nu ^2}{c^{11}\,r}\,\dot{\varphi}\,
\Biggl\{v^8 \left[\frac{145919}{13860}-\frac{110423  }{1260}\nu+\frac{1079083 }{4620}\nu^2-\frac{30229 }{165}\nu ^3\right]\nonumber\\
&&+\,\dot{r}^2 v^6 \left[-\frac{2473}{70}+\frac{763409  }{2310}\nu-\frac{2155249 }{2310}\nu ^2
+\frac{543171 }{770}\nu ^3\right]\nonumber\\ &&+\left(\frac{G\,m}{r}\right) v^6
\left[\frac{483097}{13860}-\frac{17429  }{154}\nu+\frac{693331 }{4620}\nu ^2+\frac{91 }{165}\nu ^3\right]\nonumber\\ &&+\,\dot{r}^4 v^4
\left[\frac{18695}{231}-\frac{632111  }{924}\nu+\frac{1552525 }{924}\nu ^2-\frac{61970 }{77}\nu ^3\right]\nonumber\\
&&+\left(\frac{G\,m}{r}\right) \,\dot{r}^2 v^4
\left[\frac{205817}{13860}-\frac{22549  }{105}\nu-\frac{6112567 }{13860}\nu ^2-\frac{9979 }{66}\nu ^3\right]\nonumber\\
&&+\left(\frac{G\,m}{r}\right)^2 v^4
\left[\frac{10477393}{57750} -\frac{6848}{175} \ln\left(\frac{r}{r_0}\right)
\right.\nonumber\\ &&\left.
+\left(\frac{1801028  }{3465}-\frac{369   }{40}\pi ^2\right)\nu-\frac{994673 }{3465}\nu ^2
+\frac{154421 }{1155}\nu ^3 \right]\nonumber\\
&&+\,\dot{r}^6 v^2 \left[-\frac{451}{6}+\frac{59870  }{99}\nu-\frac{14841 }{11}\nu ^2+\frac{32342 }{99}\nu ^3\right]\nonumber\\
&&+\left(\frac{G\,m}{r}\right) \,\dot{r}^4 v^2
\left[-\frac{100999}{1980}+\frac{6446971  }{6930}\nu-\frac{512285 }{2772}\nu ^2+\frac{738973 }{3465}\nu ^3\right]\nonumber\\
&&+\left(\frac{G\,m}{r}\right)^2 \,\dot{r}^2 v^2
\left[-\frac{14457734}{5775} +\frac{1712}{5} \ln\left({r\over r_0}\right)
\right.\nonumber\\ &&\left.
+\left(-\frac{18425707  }{6930}+\frac{369   }{4}\pi ^2\right)\nu+\frac{3826709 }{3465}\nu ^2 -\frac{869048 }{3465}\nu ^3
\right]\nonumber\\
&&+\left(\frac{G\,m}{r}\right)^3 v^2
\left[\frac{1229915081}{1559250} -\frac{37664}{525} \ln\left(\frac{r}{r_0}\right)
\right.\nonumber\\ &&\left.
-\left(\frac{827081  }{2835}+\frac{31  }{5}\pi ^2\right)\nu+\frac{886279 }{6930}\nu ^2
-\frac{155339 }{3465}\nu ^3
\right]\nonumber\\ &&+\,\dot{r}^8
\left[\frac{93}{4}-\frac{4035  }{22}\nu+\frac{4294 }{11}\nu ^2-\frac{410 }{11}\nu ^3\right]\nonumber\\ &&+\left(\frac{G\,m}{r}\right) \,\dot{r}^6
\left[\frac{1932907}{69300}-\frac{40997  }{70}\nu+\frac{955543 }{2772}\nu ^2-\frac{50525 }{1386}\nu ^3\right]\nonumber\\
&&+\left(\frac{G\,m}{r}\right)^2 \,\dot{r}^4
\left[\frac{93865829}{34650} -\frac{1712}{5} \ln\left(\frac{r}{r_0}\right)
\right.\nonumber\\ 
&&\left.
+\left(\frac{924466  }{385}-\frac{861   }{8}\pi ^2\right)\nu-\frac{210811 }{462}\nu ^2
+\frac{85543 }{3465}\nu ^3
\right]\nonumber\\
&&+\left(\frac{G\,m}{r}\right)^3 \,\dot{r}^2
\left[-\frac{153361069}{103950}
+\frac{5136}{35} \ln\left({r\over r_0}\right) 
\right.\nonumber\\ &&\left.
+\left(\frac{7294789  }{10395}+3 \pi ^2\right) \nu 
-\frac{2376287 }{6930}\nu^2+\frac{267188 }{3465}\nu^3 
\right]\nonumber\\
&&+\left(\frac{G\,m}{r}\right)^4 \left[-\frac{317864383}{779625}
+\frac{3424}{525} \ln\left(\frac{r}{r_0}\right)
\right.\nonumber\\ &&\left.
+\left(-\frac{7120061  }{10395}+\frac{947   }{40}\pi^2\right)\nu-\frac{3748 }{315}\nu^2
+\frac{1780 }{693}\nu^3
\right]\Biggr\}\,.
\end{eqnarray}\end{subequations}

\subsection{2.5PN terms in the fluxes in ADM coordinates}
\label{sec:IW}
In the previous section, we have taken into account the contact transformation involving ``conservative'' orders up to 3PN required to go from the standard harmonic coordinates to the ADM coordinates. However, there still remains the possible change of gauge in the radiation reaction (dissipative) terms at order 2.5PN. We discuss these terms here since the corresponding transformation law was not given in Ref.~\cite{ABIQ07}.\footnote{Actually the dissipative terms at 2.5PN order in the fluxes are not very important for the present purpose because they will average to zero and thus will not contribute to the balance equations. The present discussion is for completeness and future use.} We recall that in the standard harmonic (SH) coordinate system the lowest-order dissipative part of the equations of motion, i.e. the 2.5PN acceleration term, is given by (with boldface letters indicating ordinary three-dimensional vectors)
\begin{equation}\label{IWshar}
\mathbf{a}_{\rm 2.5PN}^{\rm SH} =\frac{8}{5}\frac{G^2 m^2\nu}{c^5 r^3}\left(\left[3\,v^2+\frac{17}{3}
\frac{G m}{r}\right] \,\dot{r}\,\mathbf{n} + \left[
-\,v^2 - 3\frac{G m}{r}\right]\,\mathbf{v} \right)\,.
\end{equation}
However, one may choose to work in alternative radiation gauges and a convenient characterisation at 2.5PN has been investigated  earlier in~\cite{IW93,IW95} (see also~\cite{GII97}). Following this work the most general form of the relative acceleration is specified by the two-parameter family written as,
\begin{subequations}\label{IW}\begin{eqnarray}
\mathbf{a}_{\rm 2.5PN}^{\rm gen} &=& \frac{8}{5}\frac{G^2 m^2\nu}{c^5 r^3}\Bigl(A_{\rm 2.5PN}\,\dot{r}\,\mathbf{n}+B_{\rm 2.5PN}\,\mathbf{v}\Bigr)\,,\\
A_{\rm 2.5PN} &\equiv& 3(1+\beta)v^2+\frac{1}{3}\left(23+6\alpha-9\beta\right)\frac{G m}{r}-5\beta\dot{r}^2\,,\\
B_{\rm 2.5PN} &\equiv& -(2+\alpha)v^2-(2-\alpha)\frac{G m}{r}+3(1+\alpha)\dot{r}^2 \,.
\end{eqnarray}\end{subequations}
The general 2.5PN gauge is parametrized by the two numerical constants $\alpha$ and $\beta$. The SH gauge in which the acceleration is given by \eqref{IWshar} corresponds to setting $\alpha=-1$ and $\beta=0$; the ADM gauge is obtained by using $\alpha=5/3$ and $\beta=3$, in which case the 2.5PN acceleration becomes~\cite{schafer82}
\begin{equation}\label{IWadm}
\mathbf{a}_{\rm 2.5PN}^{\rm ADM} =\frac{8}{5}\frac{G^2 m^2\nu}{c^5 r^3}\left(\left[12\,v^2+2\frac{G m}{r}-15\dot{r}^2\right] \,\dot{r}\,\mathbf{n} + \left[
-\frac{11}{3}\,v^2-\frac{1}{3}
\frac{G m}{r} + 8\dot{r}^2\right]\,\mathbf{v} \right)\,.
\end{equation}

To obtain the energy and angular momentum fluxes in a general 2.5PN gauge parametrised by $\alpha$ and $\beta$ we apply upon the SH particle's worldlines the $(\alpha, \beta)$-dependent shift~\cite{IW93,IW95}
\begin{equation}\label{IWshift}
\bm{\epsilon} =\frac{8}{15}\frac{G^2 m^2\nu}{c^5 r}\Bigl(-\beta\,\dot{r}\,\mathbf{n} + \left(3+3\alpha-2\beta\right)\,\mathbf{v} \Bigr)\,,
\end{equation}
such that the relative position of the bodies in a general gauge is given by $\mathbf{x}^{\rm gen}(t)=\mathbf{x}^{\rm SH}(t)+\bm{\epsilon}(t)$. The relative velocity and acceleration of the bodies are then $\mathbf{v}^{\rm gen}(t)=\mathbf{v}^{\rm SH}(t)+\ud\bm{\epsilon}/\ud t$ and $\mathbf{a}^{\rm gen}(t)=\mathbf{a}^{\rm SH}(t)+\ud^2\bm{\epsilon}/\ud t^2$. However, the effect of this shift on the acceleration is more subtle. Indeed, for the acceleration, one must remember that it is a functional of the position and velocity coming from the equation of motion and consequently the acceleration must consistently be expressed in terms of the position and velocity in the corresponding frame. Thus $\mathbf{a}^{\rm gen}(t)=\mathbf{a}^{\rm SH}(t)+\ud^2\bm{\epsilon}/\ud t^2$ is correct, but here our too condensed notation for the acceleration means in fact that $\mathbf{a}^{\rm gen}(t)\equiv\mathbf{a}^{\rm gen}[\mathbf{x}^{\rm gen}(t),\mathbf{v}^{\rm gen}(t)]$ in the general frame and $\mathbf{a}^{\rm SH}(t)\equiv\mathbf{a}^{\rm SH}[\mathbf{x}^{\rm SH}(t),\mathbf{v}^{\rm SH}(t)]$ in the SH frame. Re-expressing the latter relation in terms of ``dummy'' variables $\mathbf{x}(t)$ and $\mathbf{v}(t)$ we then get the \textit{functional} relation $\mathbf{a}^{\rm gen}[\mathbf{x}(t),\mathbf{v}(t)]=\mathbf{a}^{\rm SH}[\mathbf{x}(t),\mathbf{v}(t)]+\delta_\epsilon\mathbf{a}(t)$ where the change in the acceleration is given by
\begin{equation}\label{deltaaIW}
\delta_{\epsilon}\mathbf{a} = \frac{\ud^2\bm{\epsilon}}{\ud t^2} -
\epsilon^i\,\frac{\partial\mathbf{a}_{\rm N}}{\partial x^i}\,.
\end{equation}
Here, neglecting higher-order terms, $\mathbf{a}_{\rm N}$ denotes the ordinary Newtonian acceleration which depends on $\mathbf{x}$ but not on $\mathbf{v}$. 

Consider now the effect of the shift on the components of the quadrupole moment (the generalization to higher moments is trivial). Under this shift the mass quadrupole moment $I_{ij}^{\rm SH}$ in SH coordinates will be changed into $I_{ij}^{\rm gen}$ such that $I_{ij}^{\rm gen}[\mathbf{x}^{\rm gen},\mathbf{v}^{\rm gen}]=I_{ij}^{\rm SH}[\mathbf{x}^{\rm SH},\mathbf{v}^{\rm SH}]$. Hence, introducing dummy variables $\mathbf{x}$, $\mathbf{v}$ we readily obtain $I_{ij}^{\rm gen}[\mathbf{x},\mathbf{v}]=I_{ij}^{\rm SH}[\mathbf{x},\mathbf{v}]+\delta_{\epsilon}I_{ij}$ where the change can be expressed in terms of the Newtonian quadrupole moment $I^{\rm (N)}_{ij}$ as $\delta_{\epsilon}I_{ij}=-\epsilon^k\partial I^{\rm (N)}_{ij}/\partial x^k$ (neglecting higher-order PN terms). Using $I^{\rm (N)}_{ij} = m\,\nu\,x^{<i}x^{j>}$ (which depends only on $\mathbf{x}$) we get,\footnote{Some signs should be corrected in Ref.~\cite{ABIQ07}: namely, Eq.~(6.4) should have a minus sign, Eqs.~(6.5) and (6.7) have plus signs, and the second term of Eq.~(6.6) should have a plus sign. The results in Ref.~\cite{ABIQ07}, notably Eq.~(6.8), are unchanged.}
\begin{equation}\label{deltaIij}
\delta_{\epsilon}I_{ij} = - 2 m\,\nu\,x^{<i}\epsilon^{j>}\,.
\end{equation}
Next, we compute the successive time derivatives of this quadrupole moment, being careful that when reducing the result by means of the equation of motion the acceleration is modified by the amount \eqref{deltaaIW}. This order reduction starts from the second time derivative. We readily obtain $\ud^2 I_{ij}^{\rm gen}[\mathbf{x},\mathbf{v}]/\ud t^2=(a_{\rm gen}^k[\mathbf{x},\mathbf{v}]\partial I_{ij}^{\rm SH}[\mathbf{x},\mathbf{v}]/\partial x^k+\cdots)+\ud^2\delta_{\epsilon}I_{ij}/\ud t^2$, where we have explicitly replaced the acceleration by the general-frame one (since we are evaluating in the left-hand-side the time derivative of the general-frame moment), and where the dots indicate other terms which do not need any replacement of accelerations. Hence, using the modification of the acceleration as given by \eqref{deltaaIW}, we can connect the right-hand-side to what we would get for the time derivatives of the SH moment using the SH acceleration. This yields $\ddot{I}_{ij}^{\rm gen}[\mathbf{x},\mathbf{v}]=\ddot{I}_{ij}^{\rm SH}[\mathbf{x},\mathbf{v}] + \delta_\epsilon\ddot{I}_{ij}$ where the total change is now given by $\delta_\epsilon\ddot{I}_{ij} = \delta_\epsilon a^k\partial I_{ij}^{\rm N}/\partial x^k+\ud^2\delta_{\epsilon}I_{ij}/\ud t^2$ (the dots mean the time derivatives). The results, which proceed in the same way for the third time-derivative, read
\begin{subequations}\label{modifs}\begin{align}
\delta_\epsilon\ddot{I}_{ij} &= 2 m\,\nu\,x^{<i}\,\delta_\epsilon
a^{j>} + \frac{\ud^2}{\ud
t^2}\left(\delta_\epsilon I_{ij}\right) \,,\\
\delta_\epsilon\!\dddot{I}_{ij} &= 2 m\,\nu\left[3 v^{<i}\,\delta_\epsilon
a^{j>} + x^{<i}\,\frac{\ud\delta_\epsilon a^{j>}}{\ud t}\right] + \frac{\ud^3}{\ud
t^3}\left(\delta_\epsilon I_{ij}\right)\,.
\end{align}\end{subequations}
Beware that the total change in the time derivative of moments $\delta_\epsilon\ddot{I}_{ij}$ is different from the time derivative of the change of the moments $\ud^2\delta_\epsilon I_{ij}/\ud t^2$.

Finally we obtain the shifts of the energy and angular momentum fluxes. Since they are given in first approximation by the quadrupole formulas, for instance $\mathcal{F}^{\rm gen}[\mathbf{x},\mathbf{v}] = \frac{G}{5c^5}\dddot{I}_{ij}^{\rm gen}[\mathbf{x},\mathbf{v}]\,\dddot{I}_{ij}^{\rm gen}[\mathbf{x},\mathbf{v}]$, we can immediately use the previous link $\dddot{I}_{ij}^{\rm gen}[\mathbf{x},\mathbf{v}]=\dddot{I}_{ij}^{\rm SH}[\mathbf{x},\mathbf{v}] + \delta_\epsilon\!\dddot{I}_{ij}$ to conclude that the associated changes in the fluxes, say ${\cal F}^{\rm gen}[\mathbf{x},\mathbf{v}]={\cal F}^{\rm SH}[\mathbf{x},\mathbf{v}] + \delta_\epsilon\mathcal{F}$, are
\begin{subequations}\begin{align}
\delta_\epsilon\mathcal{F} &= \frac{2G}{5c^5}\,\dddot{I}_{ij}^\mathrm{(N)}\,\delta_\epsilon\!\dddot{I}_{ij}\,,\\
\delta_\epsilon\mathcal{G}_i &= \frac{2G}{5c^5}\,\varepsilon_{iab}\left[\ddot{I}_{ac}^\mathrm{(N)}\,\delta_\epsilon\!\dddot{I}_{bc}-\dddot{I}_{ac}^\mathrm{(N)}\,\delta_\epsilon\ddot{I}_{bc}\right]\,.
\end{align}\end{subequations}
Evidently the factors of the gauge modified multipole moments are Newtonian at this order and straightforwardly computed. Explicit results for the 2.5PN terms in the energy and angular momentum fluxes in the $(\alpha, \beta)$-dependent gauge are then\footnote{
The results  can  also be derived directly
in the general radiation gauge by explicit use of the mass quadrupole
expression in the general radiation gauge Eq.~(\ref{deltaIij}) and the
2.5PN equation of
motion terms in the general radiation gauge Eq.~(\ref{IW}).
Alternatively, it can  also be obtained by a direct transformation of the flux expressions in SH gauge to general gauge using the shift \eqref{IWshift} and the equations (6.1)--(6.3) from~\cite{ABIQ07} for the changes in $r$, $\dot{r}$ and $v^2$, together with a similar relation for $\dot{\varphi}$. We have verified that these various methods lead to the same results.}
\begin{subequations}\begin{eqnarray}
{\cal F}_{\rm 2.5PN}^{\rm gen} &=& \frac{32G^4m^5\nu^3\dot{r}}{5c^{10}r^5}\left\{-\frac{12349}{210}\,v^4 +\left[\frac{G m}{r}\left(-\frac{5869}{630}+\frac{188}{15}\alpha -\frac{28}{3} \beta\right)+\frac{4524}{35}\,\dot{r}^2\right]\,v^2\right.\nonumber\\
&&\left.-\frac{985}{14}\,\dot{r}^4+\frac{G m}{r}\left(\frac{6589}{630}-\frac{176}{15}\alpha +\frac{128}{15}\beta
\right)\,\dot{r}^2+\frac{G^2m^2}{r^2}\,\left(\frac{83}{315}+\frac{4}{15}\alpha-\frac{4}{15}\beta\right)\right\}\,,\\
%%%%%%%%%%%%%%%%%%%%%%%%%%%%%%%%%%%%%%%%%%%%%%%%%%%%%%%%%%%%%
\cal{G}_{\rm 2.5PN}^{\rm gen} &=& \frac{G^3m^4\nu^3\dot{\varphi}\,\dot{r}}{5c^{10}r^2}\left\{
-\frac{27744}{35}\,v^4+\left[\frac{G m}{r}\left(-\frac{92752}{105}+\frac{1152}{5}\alpha -\frac{768}{5}\beta\right)+\frac{19144}{7}\,\dot{r}^2\right]v^2\right.\nonumber\\
&&\left.-1960
  \,\dot{r}^4+\frac{G m}{r}\left(\frac{34128}{35}-\frac{1536}{5} \alpha 
  +\frac{1088}{5} \beta \right)\dot{r}^2 \right.\nonumber\\
&&+\left.\frac{G^2m^2}{r^2} \left(-\frac{12112}{105}+\frac{512}{5} \alpha -\frac{384}{5} \beta\right)\right\}\,.
\end{eqnarray}\end{subequations}
These expressions could be of use in the discussion of non-quasi-circular effects required to match high accuracy PN waveforms to those of numerical relativity in the effective-one-body formalism~\cite{Damour:2009b,Buonanno:2009qa}. By setting $\alpha=-1$ and $\beta=0$ one recovers the SH coordinates results of Eq.~(5.2d) of Ref.~\cite{ABIQ07} and \eqref{eq:AMF2p5Har} of this paper. On the other hand, for $\alpha=5/3$ and $\beta=3$ we get the ADM results given by
\begin{subequations}\begin{eqnarray}
{\cal F}_{\rm 2.5PN}^{\rm ADM} &=& \frac{32G^4m^5\nu^3\dot{r}}{5c^{10}r^5}\left\{-\frac{12349}{210}\,v^4 +\left[-\frac{10349}{630}\frac{G m}{r}+\frac{4524}{35}\,\dot{r}^2\right]\,v^2\right.\nonumber\\
&&\left.-\frac{985}{14}\,\dot{r}^4+\frac{10397}{630}\frac{G m}{r}
\,\dot{r}^2-\frac{29}{315}\frac{G^2m^2}{r^2}\right\}\,,\\
%%%%%%%%%%%%%%%%%%%%%%%%%%%%%%%%%%%%%%%%%%%%%%%%%%%%%%%%%%%%%
\cal{G}_{\rm 2.5PN}^{\rm ADM} &=& \frac{G^3m^4\nu^3\dot{\varphi}\,\dot{r}}{5c^{10}r^2}\left\{
-\frac{27744}{35}\,v^4+\left[-\frac{100816}{105}\frac{G m}{r}+\frac{19144}{7}\,\dot{r}^2\right]v^2\right.\nonumber\\
&&\left.-1960
  \,\dot{r}^4+\frac{39056}{35}\frac{G m}{r}\,\dot{r}^2 -\frac{6128}{35}\frac{G^2m^2}{r^2}\right\}\,.
\end{eqnarray}\end{subequations}
From now on our calculations will be in ADM coordinates, so we henceforth suppress all indications regarding the ADM coordinate system.

%##############################################################
\section{Orbital average of the instantaneous angular momentum flux in ADM coordinates}
\label{sec:AMF-AvADM}
%###############################################################
At this juncture let us remind where we are eventually headed. We are interested in the phasing of the binaries moving in quasi-eccentric orbits and in the first instance, as for quasi-circular orbits, we work in the adiabatic approximation. In this limit the radiation time scale would be much longer than the orbital time scale and consequently we would require an averaged description of the radiation reaction over an orbital period; so we average the flux over one orbit. To deal with 3PN radiation reaction (corresponding
formally to 5.5PN terms in the equation of motion) we require the description of the motion to be 3PN accurate. Such a description is available from the work of Memmesheimer, Gopakumar and Sch\"afer~\cite{MGS04} based on the known 3PN equations of motion~\cite{JS99,BFeom,DJSdim,BDE04}, and we shall use this generalized 3PN quasi-Keplerian (QK) representation of the motion in ADM coordinates to average our angular momentum flux. The details of the expressions we use are available in Ref.~\cite{MGS04} and we have recast these into forms better suited for the present context in Ref.~\cite{ABIQ07} (see Sec. VII there).

Since the quasi-Keplerian orbit is planar, to quantify the evolution of the orbital elements under radiation reaction we only need to average the magnitude of the angular momentum flux over an orbit. The computation thus becomes a generalisation of our earlier computation of the average of the energy flux in~\cite{ABIQ07} and requires similar intermediates. Using the QK representation of the orbit in ADM coordinates and the instantaneous angular momentum flux in ADM coordinates obtained in Sec.~\ref{sec:AMF-instH}, one transforms the expression for the norm of the angular momentum flux ${\cal
G}_{\rm inst}\,(r,\dot {r},\dot{\varphi})$ into another expression ${\cal G}_{\rm inst}\,(E,h,e_t,u)$ depending on the 3PN conserved orbital energy $E$, on the norm of the 3PN conserved angular momentum $J=\vert J^i\vert$ as rescaled by $G \,m$
(thus $h\equiv J/G \,m$), on that particular choice of eccentricity $e_t$ which parametrizes the (3PN-generalized) Kepler equation, and on the eccentric anomaly $u$; see Eqs.~(7.1)--(7.4) in~\cite{ABIQ07}. Like for the case of the
energy flux we find that the general structure of the angular momentum flux in terms of these variables reads up to 3PN order,
\begin{equation}
\label{structGinst}
{\cal G}_{\rm inst} = \frac{\ud u}{\ud\ell}\sum_{N=2}^{10}\left[
\frac{\alpha_N(E,h)}{(1- e_t\cos u)^{N}}+{\beta_N(E,h)\,\frac{\sin
u}{(1-e_t\,\cos u)^N}}+{\gamma_N(E,h)\,\frac{\ln(1-e_t\cos u)}{(1-e_t\cos
u)^N}}\right]\,.
\end{equation}
For later convenience we have factored out $\ud u/\ud\ell$, where $\ell=n(t-t_\text{P})$ denotes the mean anomaly, with $t_\text{P}$ the instant of passage to the periastron, and $n=2\pi/P$ is the mean motion, with $P$ the orbital period. The coefficients $\alpha_N$, $\beta_N$ and $\gamma_N$ are explicit functions of the invariants $E$ and $h$,\footnote{Notice that the structure of Eq.~\eqref{structGinst} requires $\alpha_N$ and $\beta_N$ to be functions of $E$ and $h$. The angular momentum flux could be averaged without reducing it to this form, though in alternative forms the average of the equivalent of the $\gamma_N$ term may require another standard integral different from that given in Eq.~\eqref{int2}.} and are straightforwardly deduced from the QK parametrization in the form of a PN series --- see Eq.~(4.12) of~\cite{GI97} for instance --- but too long to be listed here. Rewriting the angular momentum flux using the generalized QK representation, the flux can be averaged over an orbit to order 3PN extending the results of~\cite{GI97} at 2PN. The orbital average is (setting $t_\text{P}=0$ for definiteness)
\begin{equation}
\langle{\cal G}_{\rm inst}\rangle ={1 \over P}\int^{P}_{0}{\cal G}_{\rm
inst}(t )\,\ud t\, ={ 1 \over 2\,\pi}\int ^{2\,\pi}_{0}{\cal G}_{\rm inst}(u)
{\ud\ell\over \ud u}\ud u \,.
\label{eq:avg-integral}
\end{equation}
To compute it we use some integration formulas. First it is clear that the second term in \eqref{structGinst} will not contribute,
\begin{equation}\label{int0}
\frac{1}{2\pi}\int_0^{2\pi}\frac{\sin u\,\ud u}{(1-e \cos u)^{N}} =0\,.
\end{equation}
Note that this term corresponds to the 2.5PN contribution in the angular momentum flux which is therefore seen not to contribute to the average. To get the first term we have the useful formula
\begin{equation}\label{int1}
\frac{1}{2\pi}\int_0^{2\pi}\frac{\ud u}{(1-e \cos u)^{N}} =
\frac{(-)^{N-1}}{(N-1)!}
\left(\frac{\ud^{N-1}}{\ud y^{N-1}}\left[\frac{1}{\sqrt{y^2-e^2}}
\right]\right)_{y=1} \,,
\end{equation}
where the right-hand-side (RHS) is to be evaluated at the value $y=1$ after the $N-1$ differentiations. We have also an alternative formulation in terms of the Legendre polynomial $P_{N-1}$,
\begin{equation}\label{int11}
\frac{1}{2\pi}\int_0^{2\pi}\frac{\ud u}{(1-e \cos u)^{N}} =
\frac{1}{(1-e^2)^{N/2}} P_{N-1} \left(\frac{1}{\sqrt{1-e^2}}\right)\,.
\end{equation}
Furthermore, we dispose of the more complicated formula~\cite{ABIQ07}
\begin{equation}\label{int2}
\frac{1}{2\pi}\int_0^{2\pi}\frac{\ln(1-e \cos u)}{(1-e \cos
u)^{N}}\,\ud u = \frac{(-)^{N-1}}{(N-1)!}
\left(\frac{\ud^{N-1}Y(y,e)}{\ud y^{N-1}}\right)_{y=1}\,,
\end{equation}
where
\begin{equation}\label{Yy}Y(y,e)=\frac{1}{\sqrt{y^2-e^2}}\left\{\ln
\left[\frac{\sqrt{1-e^2}+1}{2}\right]+2\ln\left[1 +
\frac{\sqrt{1-e^2}-1}{y +\sqrt{y^2-e^2}}\right]\right\}\,.
\end{equation}
These formulas, and notably the third one~\eqref{int2}--\eqref{Yy}, permit one to display the result in a completely closed form, and give 
\begin{equation}
\label{moyenneres}
\langle{\cal G}_{\rm inst}\rangle =
\sum_{N=2}^{10}\frac{(-)^{N-1}}{(N-1)!}\left(\frac{\ud^{N-1}}{\ud y^{N-1}}
\left[\frac{\alpha_N(E,h)}{\sqrt{y^2-e^2}} +
\gamma_N(E,h)\,Y(y,e)\right]\right)_{y=1}\,.
\end{equation}

The results can be expressed in terms of different choices of variables as for the energy flux in~\cite{ABIQ07}. For reasons discussed there, we choose the pair consisting of $(e_t,x)$ with
\begin{equation}\label{x}
x\equiv\left(\frac{Gm\omega}{c^3}\right)^{2/3}\,,
\end{equation}
where the mean orbital frequency is $\omega=\langle\dot{\varphi}\rangle=K\,n$, equal to the mean
motion $n$ times the periastron precession $K=1+k$ (with $k$ denoting the relativistic precession). The expression for the orbital averaged angular momentum flux can be finally written (in ADM coordinates) as
\begin{equation}
\langle{\cal G}_{\rm inst}\rangle = \frac{4}{5}\, c^2\,m\,\nu ^2\,x^{7/2}\biggl(
{\cal H}_{\rm N} +x\,{\cal H}_{\rm 1PN} +x^2\,{\cal H}_{\rm
2PN} +x^3\,{\cal H}_{\rm 3PN} \biggr)\,,\label{eq:AvAMFx}
\end{equation}
where the individual PN terms read as
\begin{subequations}\label{AMFxPN}
\begin{eqnarray}
{\cal H}_{\rm N}&=&\frac{8+7
e_t^2}{\left(1-e_t^2\right)^{2}}\label{eq:AvAMFx0}\,,\label{AMFxN}\\ {\cal
H}_{\rm 1PN}&=&\,\frac{1}{(1-e_t^2)^{3}}\,\Biggl\{-\frac{1247}{42}
-\frac{70}{3}\nu+\left(\frac{3019}{42}-\frac{335 }{3}\nu\right)
e_t^2+\left(\frac{8399}{336}-\frac{275 }{12}\nu\right)
e_t^4\Biggr\}\label{AMFx1PN}\\ {\cal H}_{\rm
2PN}&=&\frac{1}{(1-e_t^2)^{4}}\,\Biggl\{-\frac{135431}{1134}+\frac{11287
}{63}\nu+\frac{260}{9}\nu^2+\left(-\frac{607129}{756}-\frac{6925
}{84}\nu+\frac{1546}{3}\nu^2\right) e_t^2\nonumber\\
&&+\left(\frac{28759}{432}-\frac{116377 }{168}\nu+569 \nu ^2\right)
e_t^4+\left(\frac{30505}{2016}-\frac{2201}{56}\nu+\frac{1519}{36}\nu^2\right)
e_t^6\nonumber\\ &&+\sqrt{1-e_t^2}\left[80-32\nu+(335-134 \nu ) e_t^2+(35-14
\nu ) e_t^4\right]\Biggr\}\label{AMFx2PN}\\
%%%%%%%%%%%
{\cal H}_{\rm 3PN}&=&\frac{1}{\left(1-e_t^2\right)^{5}}\,\Biggl\{
\frac{2017023341}{1247400}+\frac{4340155 }{6804}\nu -\frac{167483 }{378}\nu
^2-\frac{1550 }{81}\nu ^3\nonumber\\ &&+e_t^2
\left(\frac{153766369}{44550}+\left[\frac{15157061}{1701}
+\frac{647 }{8}\pi ^2\right]\nu-\frac{116335
}{54}\nu^2-\frac{96973}{81}\nu^3\right)\nonumber\\
&&+e_t^4\left(-\frac{6561101941}{1663200}+\left[\frac{163935875}{18144} -
\frac{6817 }{256}\pi ^2\right]\nu+\frac{3541255}{1008}\nu^2-\frac{438907}{108}\nu^3\right)\nonumber\\
&&+e_t^6\left(-\frac{10123087}{19800}+\left[-\frac{326603}{2268} 
- \frac{615 }{128}\pi ^2\right]\nu+\frac{2224003}{1008}\nu^2-\frac{283205}{162}\nu^3\right)\nonumber\\
&&+e_t^8\left(-\frac{10305073}{709632}+\frac{417923}{12096}\nu
+\frac{95413}{8064}\nu^2-\frac{146671}{2592}\nu^3\right)\nonumber\\
&&+\sqrt{1-e_t^2} \left[-\frac{379223}{630}+\left[-\frac{48907}{63} + \frac{41
}{6}\pi ^2\right]\nu+\frac{580}{3}\nu^2\right.\nonumber\\ &&+e_t^2
\left(\frac{309083}{315}+\left[-\frac{456250}{63} 
+ \frac{2747 }{96}\pi ^2\right]\nu+1902\nu^2\right)\nonumber\\
&&+e_t^4\left(\frac{13147661}{5040}+\left[-\frac{2267795}{504} 
+ \frac{287 }{96}\pi ^2\right]\nu+\frac{2703}{2}\nu^2\right)\nonumber\\ &&\left.+e_t^6
\left(70-\frac{203}{3}\nu+\frac{77}{3}\nu^2\right)\right]\nonumber\\
&&+\left(\frac{13696}{105}+\frac{98012}{105}e_t^2+\frac{23326}{35}e_t^4
+\frac{2461}{70}e_t^6\right)\,\ln\left[\frac{x}{x_0}\frac{1
+\sqrt{1-e_t^2}}{2(1-e_t^2)}\right]\Biggr\}\,.\label{AMFx3PN}
\end{eqnarray}\end{subequations}
This ADM-coordinates expression, similar to the energy flux case, does not contain the logarithmic terms  $\ln(r/r_0')$, consistent with these being pure gauge dependent terms arising specifically in the SH coordinate
system. However, we notice that there are still some other logarithmic terms involving the constant  $r_0$, in the instantaneous part of the flux, even in ADM coordinates. Actually these terms are parametrized by $x_0$ in~(\ref{AMFx3PN}) where
\begin{equation}\label{x0}
x_0\equiv\frac{G m}{c^2 r_0}\,.
\end{equation}
Recall that $r_0=c\,\tau_0$ is an arbitrary length scale introduced in the general MPM formalism (to regularize Ultra-Violet divergences), which then appears in the definition of the source multipole moments starting explicitly at the 3PN order. This is what leads to the $\ln r_0$ dependence of the instantaneous terms in the angular momentum flux. It is known~\cite{B98mult,B98tail} that the $\ln r_0$ terms will be cancelled by similar terms when we add up all the non-linear multipole interactions (mainly tails) constituting the radiative multipole moments observable at infinity. This has been checked at the 3PN order for compact binaries in the circular orbit case~\cite{BIJ02}, where the contribution due to tails-of-tails effectively removes these unphysical $\ln r_0$ terms. In Ref.~\cite{ABIQ07}, this cancellation was proved in the case of the 3PN energy flux for general non-circular orbits. We shall also recover this necessary cancellation below for the angular momentum flux for general orbits.

As a check of the algebra, we take the circular orbit limit of the averaged instantaneous angular momentum flux, say $\langle{\cal G}_{\rm inst}\rangle_{\odot}$, as given by~(\ref{eq:AvAMFx})--(\ref{AMFxPN}) with $e_t=0$, and get
\begin{eqnarray}
\langle{\cal G}_{\rm inst}\rangle_{\odot} &=&
\frac{32}{5}\,c^2\,m\,\nu^2\,x^{7/2} \Biggl\{ 1 + \left(-\frac{1247}{336} -
\frac{35}{12}\nu \right) x + \left(-\frac{44711}{9072} + \frac{9271}{504}\nu +
\frac{65}{18} \nu^2\right) x^2 \nonumber\\&+&
\left[\frac{1266161801}{9979200}+\frac{1712}{105}\ln\left(\frac{x}{x_0}\right)
+ \left(-\frac{134543}{7776} + \frac{41}{48}\pi^2 \right)\nu -
\frac{94403}{3024}\nu^2 - \frac{775}{324}\nu^3 \right] x^3
\Biggr\}\,.\nonumber\\
\end{eqnarray}
This is in perfect agreement with the averaged instantaneous
energy flux for circular orbits, say $\langle{\cal F}_{\rm inst}\rangle_{\odot}$, in the sense that
\begin{equation}\label{circular}
\langle{\cal F}_{\rm inst}\rangle_{\odot}
=\omega\,\langle{\cal G}_{\rm inst}\rangle_{\odot}\,,
\end{equation}
where the proportionality factor is the binary's mean orbital frequency $\omega=c^3 x^{3/2}/G m$. For the reader's convenience we recall the 3PN energy flux $\langle{\cal F}_{\rm inst}\rangle$ in Eqs.~(\ref{AvEADM})--(\ref{AvEADMa}) below.

We emphasize again that the expressions we provide for the orbital average of the angular momentum flux and later the evolution of orbital elements are in the ADM coordinates; we can easily obtain the corresponding expressions in another coordinate system like the Modified Harmonic (MH) one~\cite{MGS04,ABIQ07} by transforming the eccentricity $e_t\equiv e_t^{\rm ADM}$ used here to another eccentricity such as $e_t^{\rm MH}$. There is no need to transform the parameter $x$ which is a gauge invariant. The relation linking $e_t^{\rm ADM}$ and $e_t^{\rm MH}$ reads (see Eq.~(8.21) in~\cite{ABIQ07})
\begin{eqnarray}\label{ADMMH}
e_t^{\rm ADM}&=&e_t^{\rm MH}\left\{1+\frac{x^2}{1-e_t^2}
\left(\frac{1}{4}+\frac{17}{4}\nu\right)+\frac{x^3}{(1-e_t^2)^2}\left[\frac{1}{2}+\left(\frac{16739}{1680}-\frac{21 }{16}\pi ^2\right) \nu 
-\frac{83 }{24}\nu ^2\right.\right.\nonumber\\
&&\qquad+\left.\left.e_t^2\left(\frac{1}{2}+\frac{249  }{16}\nu
-\frac{241 }{24}\nu ^2\right)\right]\right\}\,.
\end{eqnarray}
Since the transformation begins at 2PN order, the eccentricity in the brackets of the RHS of \eqref{ADMMH} can equivalently be replaced by $e_t^{\rm ADM}$ or $e_t^{\rm MH}$. Furthermore only the instantaneous part needs to be transformed. The hereditary contributions retain the same form up to the 3PN order to which we are concerned at present.
For the explicit form of the angular momentum flux in MH coordinates, see Appendix \ref{appC}.

Note also that the $x$ parametrization that we use can easily be changed to the alternative parametrization by the variable
\begin{equation}
\zeta\equiv\frac{G\,m\,n}{c^3}\,,
\end{equation}
which is often used in the literature  and is also gauge invariant. In ADM
coordinates we have (with $e_t\equiv e_t^{\rm ADM}$)
\begin{eqnarray}\label{xtozetaADM}
x &=& \zeta^{2/3}\left\{1+\frac{2}{1-e_t^2}\zeta^{2/3}+
\frac{1}{(1-e_t^2)^2}\left[12-\frac{14}{3}\nu+e_t^2\left(\frac{17}{2}-\frac{13}{3}\nu\right)\right]\zeta^{4/3}\right.\nonumber\\
&&+\left.\frac{1}{(1-e_t^2)^3}\left[\frac{250}{3}+\left(-\frac{255}{2}+\frac{41 }{16}\pi ^2\right) \nu +\frac{14 }{3}\nu ^2+e_t^2\left(137+\left(-\frac{473}{3}+\frac{41 }{64}\pi ^2\right) \nu +\frac{80 }{3}\nu ^2\right)\right.\right.\nonumber\\
&&+\left.\left.e_t^4\left(13-\frac{55  }{6}\nu+\frac{65 }{12}\nu ^2\right)+\sqrt{1-e_t^2}\left(10-4 \nu +e_t^2(20-8 \nu )\right)\right]\zeta^{2}\right\}.
\end{eqnarray}

\section{Hereditary contributions to the 3PN angular momentum flux}
\label{sec:AMF-hered}
Having obtained all the instantaneous terms in the 3PN angular momentum flux (in averaged form), one must now turn attention to the hereditary contributions. As for the energy flux~\cite{ABIQ07tail} it is not feasible to obtain closed-form results in the time domain for these contributions, and we adapt the strategy set out there by using a (discrete) Fourier decomposition. The details and notation are similar to those in~\cite{ABIQ07tail} so we avoid repeating them and only quote the final results with a few intermediate steps. We however find a new aspect that arises in the case of the angular momentum flux with respect to the energy flux, namely the presence of a contribution due to the memory integral in the angular momentum flux for general systems. We shall prove that this memory contribution yields an interesting zero-frequency effect depending on the past history of the system (see Appendix~\ref{appA}).

\subsection{Tail and tail-of-tail integrals at Newtonian order}
\label{secIVB}
Most of the hereditary contributions will require only relative Newtonian precision. At Newtonian order we can use the Fourier decomposition of the Newtonian (N) source multipole moments $I^{(\mathrm{N})}_L$ and $J^{(\mathrm{N})}_{L-1}$ given by
\begin{subequations}\label{ILJLfourierN}\begin{eqnarray}
I_L^{(\mathrm{N})}(t) &=&
\sum_{p=-\infty}^{+\infty}\,\mathop{{\cal{I}}}_{(p)}{}_{\!\!L}^{(\mathrm{N})}\,e^{\ii
\,p \,\ell},\label{ILfourierN}\\ J_{L-1}^{(\mathrm{N})}(t) &=&
\sum_{p=-\infty}^{+\infty}\,\mathop{\mathcal{J}}_{(p)}
{}_{\!\!\!L-1}^{(\mathrm{N})}\,e^{\ii \,p\,\ell}\,,\label{JLfourierN}
\end{eqnarray}\end{subequations}
with discrete Fourier coefficients ${}_{(p)}{\cal{I}}_L^{(\mathrm{N})}$ and ${}_{(p)}{\cal{J}}_{L-1}^{(\mathrm{N})}$ indexed by the integer $p$. Since the moments are real we have e.g.
${}_{(-p)}{\cal{I}}_L^{(\mathrm{N})}={}_{(p)}{\cal{I}}_L^{(\mathrm{N})*}$ where $*$ is the complex conjugation. Here we denote the mean anomaly by $\ell=n(t-t_\text{P})$, with $n=2\pi/P$ being the mean motion and $P$ the orbital period (we can choose the origin of time so that $t_\text{P}=0$). The
latter Fourier decompositions are simple because the motion is periodic at Newtonian order since there is no relativistic precession, $K=1+\mathcal{O}(c^{-2})$.

We follow exactly the same method as in Ref.~\cite{ABIQ07tail}, and express the tail, tail-of-tail and tail squared terms~\eqref{eq:tail}--\eqref{eq:tail-tail-sq}, i.e. up to the 3PN order, and averaged over the mean anomaly $\ell$, in terms of the Fourier coefficients of the multipole moments. All these terms are Newtonian except the mass-type quadrupolar tail term given by the first term in~\eqref{eq:tail} and which must be evaluated at 1PN. For the Newtonian part of the mass quadrupole tail --- a quadratic interaction $\propto G^2$ between the mass quadrupole moment $I_{ij}$ and the mass monopole $M$ --- we get
\begin{equation}\label{FtailN}
\langle{\cal G}_i^{\rm tail}\rangle_{\rm mass\;quad}^{(\mathrm{N})} = -\ii
\,{8\pi\over5}{G^2 M\over c^{10}}\,\varepsilon_{ijk}\sum_{p=1}^{+\infty}(p\,n)^{6}
\!\mathop{{\cal{I}}}_{(p)}{}_{\!\!ja}^{\!\!(\mathrm{N})}\,
\!\mathop{{\cal{I}}}_{(p)}{}_{\!\!ka}^{\!\!(\mathrm{N})*} \,,
\end{equation}
where the range of $p$'s is set to correspond to positive frequencies only. We add a label (N) to make the distinction with the mass quadrupole tail we compute below at 1PN order. The remaining tail integrals involve the Newtonian mass octupole and current quadrupole moments and read
\begin{subequations}\label{Ftailhigher}\begin{eqnarray} 
\langle{\cal G}_i^{\rm tail}\rangle_{\rm mass\;oct} &=& -\ii\,{4\pi\over63}
  \,{G^2 M\over c^{12}}\,\varepsilon_{ijk}\sum_{p=1}^{+\infty}(p\,n)^{8}
  \!\mathop{{\cal{I}}}_{(p)}{}_{\!\!jab}^{\!\!(\mathrm{N})}\,
  \!\mathop{{\cal{I}}}_{(p)}{}_{\!\!kab}^{\!\!(\mathrm{N})*}\,, \\ \langle{\cal
  G}_i^{\rm tail}\rangle_{\rm curr\;quad} &=& -\ii\,{128\pi\over45}\,{
  G^2 M\over c^{12}}\varepsilon_{ijk}\,\sum_{p=1}^{+\infty}(p\,n)^{6}
  \!\mathop{{\cal{J}}}_{(p)}{}_{\!\!ja}^{\!\!(\mathrm{N})}\,
  \!\mathop{{\cal{J}}}_{(p)}{}_{\!\!ka}^{\!\!(\mathrm{N})*}\,.
\end{eqnarray}\end{subequations}
In Sec.~\ref{sec:plotsheredamf} we shall provide the plots, computed numerically, for  the relevant ``enhancement'' eccentricity-dependent factors associated with Eqs.~\eqref{Ftailhigher}, since they do not admit a closed-form expression.

In addition to the previous quadratic-order tails, we have also at the 3PN order the first cubic non-linear interactions between $I_{ij}$ and two mass monopole factors $M$, namely the so-called tail-of-tail integral and the tail squared one, both being evaluated at Newtonian order. For the tail-of-tail part, averaged over an orbit, we get
\begin{eqnarray}
\label{Ftailtail2}
\langle{\cal G}_i^{\rm tail(tail)}\rangle &=& -\ii\,{8\over5}\,{G^3 M^2\over
c^{11}}\,\varepsilon_{ijk}\sum_{p=1}^{+\infty}(p\,n)^{7}
\!\mathop{{\cal{I}}}_{(p)}{}_{\!\!ja}^{\!\!(\mathrm{N})}\,
\!\mathop{{\cal{I}}}_{(p)}{}_{\!\!ka}^{\!\!(\mathrm{N})*}\nonumber\\
&&\qquad\times\left\{{\pi^2\over6}-2\Bigl(\ln(2p\,n \,\tau_0) +
C\Bigr)^2+{57\over35}\Bigl(\ln(2p\,n\,\tau_0) +
C\Bigr)-{124627\over22050}\right\}\,.
\end{eqnarray}
Here $C=0.577\dots$ denotes the Euler constant. Note that in contrast to the quadratic tails this expression involves some logarithms, and even some squares of logarithms. However some more logarithms are contained in the related contribution of tail squared, namely
\begin{equation}\label{Ftail22}
\langle{\cal G}_i^{\rm (tail)^2}\rangle =
-\ii\,{8\over5}\,{G^3 M^2\over c^{11}}\,\varepsilon_{ijk}\sum_{p=1}^{+\infty}(p\,n)^{7}
\!\mathop{{\cal{I}}}_{(p)}{}_{\!\!ja}^{\!\!(\mathrm{N})}\,
\!\mathop{{\cal{I}}}_{(p)}{}_{\!\!ka}^{\!\!(\mathrm{N})*}
\left\{{\pi^2\over2}+2\biggl(\ln(2p\,n \,\tau_0) + C -
       {11\over12}\biggr)^2\right\}\,, 
\end{equation}
and we can check that in fact the squares of logarithms cancel each other when we add together the two contributions~\eqref{Ftailtail2} and~\eqref{Ftail22}. Such cancellation is known to occur for general sources~\cite{B98tail}. Hence we get for the sum
\begin{align}
\label{Ftailtailres}
\langle{\cal G}_i^{\rm tail(tail)+(tail)^2}\rangle &=
-\ii\,{8\over5}\,{G^3 M^2\over c^{11}}\,\varepsilon_{ijk}\sum_{p=1}^{+\infty}(p\,n)^{7}
\!\mathop{{\cal{I}}}_{(p)}{}_{\!\!ja}^{\!\!(\mathrm{N})}\,
\!\mathop{{\cal{I}}}_{(p)}{}_{\!\!ka}^{\!\!(\mathrm{N})*}
\,\left\{{2\pi^2\over3}-{214\over105}\Bigl(\ln(2p\,n \,\tau_0) +
C\Bigr)-{116761\over29400}\right\}\,.
\end{align}
This result still depends on the arbitrary time scale $\tau_0$. It will be important to trace out the fate of this constant and check that the complete angular momentum flux we obtain at the end is independent of $\tau_0=r_0/c$.

It is worth recalling that the reduction of the above expressions to their simple form is made possible thanks to the following closed-form formulas which are applied to each of the Fourier components of the flux when performing the time average. For any $\sigma$ denoting the product $p\,n$ (which at the initial stage can be positive or negative), we have
\begin{subequations}\label{intAB}\begin{eqnarray}
\label{intA}
\int_{-\infty}^{T_R}\ud V\,e^{-\ii
\,\sigma\,V}\ln\left(\frac{T_R-V}{2\tau_0}\right) &=& -\frac{e^{-\ii
\,\sigma\,T_R}}{\sigma}\left[\frac{\pi}{2}\mathrm{sign}(\sigma) +
\ii\Bigl(\ln(2\vert \sigma\vert \tau_0) + C\Bigr)\right]\,,
\label{eq:tailintegral}\\\label{intB}\int_{-\infty}^{T_R}\ud V\,e^{-\ii
\,\sigma\,V}\ln^2\left(\frac{T_R-V}{2\tau_0}\right) &=& \ii \,\frac{e^{-\ii
\,\sigma\,T_R}}{\sigma}\left\{\frac{\pi^2}{6}
-\left[\frac{\pi}{2}\mathrm{sign}(\sigma) + \ii\Bigl(\ln(2\vert \sigma\vert
\tau_0) + C\Bigr)\right]^2\right\}\,,
\end{eqnarray}\end{subequations}
where $\mathrm{sign}(\sigma)$ and $\vert \sigma\vert$ denote the sign $\pm$ and the absolute value of $\sigma$, and where $C$ is the Euler constant. The rational fractions such as $11/12$ present in the tail integrals are easily taken into account by redefining the constant $\tau_0$ in these formulas as $\tau_0\,e^{-11/12}$ for instance. For dealing with higher non-linear tails (occurring at 4PN order for instance), we would need further integration formulas involving higher powers of the logarithms.
\subsection{The mass quadrupole tail at 1PN order}
\label{secVD}
One hereditary contribution will require the relative 1PN order, namely the leading quadratic-order mass quadrupole tail. In this case the Fourier decomposition is more complicated and one must exploit the ``doubly periodic'' nature of the 1PN dynamics in the two variables $\ell$ and $\lambda\equiv K\,\ell$, where $K=1+k$ is the advance of the periastron per revolution. We can then write the following doubly periodic Fourier decomposition of the mass quadrupole moment at the 1PN order~\cite{ABIQ07tail},
\begin{equation}
I_{ij}(t) = \sum_{p=-\infty}^{+\infty}\,\sum_{m=-2}^{2}
\,\mathop{{\cal{I}}}_{(p,m)}{}_{\!\!ij}\,e^{\ii \,(p+m \,k)
\,\ell}\,.\label{Iijfourier}
\end{equation}
The discrete Fourier coefficients ${}_{(p,m)}{\cal{I}}_{ij}$ depend now on two integers: $p\in\mathbb{Z}$ and the ``magnetic'' number $m\in\mathbb{Z}$ such that $\vert m\vert\leq l=2$. We have ${}_{(-p,-m)}{\cal{I}}_{ij}={}_{(p,m)}{\cal{I}}^*_{ij}$. These Fourier coefficients are valid through 1PN order. Actually one can check that those for which $m=\pm 1$ are zero in the case of the mass quadrupole moment at the 1PN order.

We insert the Fourier decomposition~\eqref{Iijfourier} into the 1PN mass quadrupole tail [first term in Eq.~\eqref{eq:tail}] and perform the average. The result in terms of the doubly-periodic Fourier coefficients at 1PN order reads
\begin{eqnarray}
\label{Ftailfourier}
\langle{\cal G}_i^{\rm mass\;quad}\rangle_{\rm tail} &=&
-\ii\,\frac{4}{5}\,\frac{G^2M}{c^8}\,\varepsilon_{ijk}\!\!\sum_{p,p';m,m'}n^7\,(p+m
k)^2(p'+m' k)^4\,\Bigl[p'-p+(m'-m)
  k\Bigr]\,\mathop{{\cal{I}}}_{(p,m)}{}_{\!\!ja}
\mathop{{\cal{I}}}_{(p',m')}{}_{\!\!ka} \nonumber\\ &\times&\langle
e^{\ii(p+p'+(m+m') k)\ell}\rangle \int_{-\infty}^{T_R}\ud
V\,e^{-\ii\,(p'+m'k)\,n
  \,(T_R-V)}\biggl[\ln\left(\frac{T_R-V}{2\tau_0}\right)+\frac{11}{12}\biggr]\,.
\end{eqnarray}
The summations range from $-\infty$ to $+\infty$ for $p$ and $p'$, and from $-2$ to $2$ for $m$ and $m'$. The last two factors in~\eqref{Ftailfourier}, i.e. the average over $\ell$ of an elementary complex exponential, and the Fourier transform of the tail integral, are both evaluated following~\cite{ABIQ07tail} at first order in the relativistic advance of the periastron $k$ which is a small 1PN quantity, $k=\mathcal{O}(c^{-2})$. The $\ell$-average factor reads
\begin{equation}\label{av} 
\langle e^{\ii\,(p+m\,k)\,\ell}\rangle\equiv
\int_0^{2\pi}\frac{\ud\ell}{2\pi}\,e^{\ii\,(p+m\,k)\,\ell} = \left\{
\begin{array}{ll} \displaystyle \frac{m}{p}\,k & ~~\mathrm{if~} p\neq
0\\[1.5 em] \displaystyle 1+\ii\,\pi \,m \,k & ~~\mathrm{if~}
p=0\end{array}\right\} \,.
\end{equation}
where we neglect some 2PN terms ${\cal O}(c^{-4})$, and we have used the fact that $m\,k\ll 1$ since we are in the limit where $k\rightarrow 0$ (hence $p+m\,k$ is never an integer unless $k=0$). This result depends on whether $p$ is zero or not, and is true for any integer $m$, except that when $m=0$ it becomes exact as there is no remainder term ${\cal O}(c^{-4})$ in this case. The tail integral expanded to first order in $k$ [i.e. up to some remainder ${\cal O}(c^{-4})$] reads
\begin{equation}\label{intk} 
\int_{-\infty}^{T_R}\ud V\,e^{\ii\,(p+m\,k)\,n\,(T_R-V)}\ln\left(\frac{T_R-V}{2\tau_0}\right)
= \left(1-\frac{m \,k}{p}\right) \int_{-\infty}^{T_R}\ud V\,e^{\ii p\,n
\,(T_R-V)}\ln\left(\frac{T_R-V}{2\tau_0}\right) - \ii\frac{m \,k}{p^2 n} \,.
\end{equation}
For the remaining integral in the RHS of~\eqref{intk} we apply the formula~\eqref{intA}.

In this paper we do not give a more explicit form for the Fourier decomposition~\eqref{Ftailfourier}, which together with the two results~\eqref{av} and~\eqref{intk} is well-defined. Indeed~\eqref{Ftailfourier} is given by a very complicated expression which can only be handled using an algebraic computer program. Still it will remain to insert in that expression the explicit 1PN results for the mass quadrupole moment and the total mass $M$ for eccentric binary orbits, and to re-express the series in terms of some elementary eccentricity-dependent ``enhancement'' functions which we shall evaluate numerically.

\subsection{The non-linear memory integral}
\label{memory}
The memory contribution is defined, from Eq.~\eqref{eq:memory}, by
\begin{equation}
{\cal G}_i^{\rm memory} = \frac{4}{35}\,\frac{G^2}{c^{10}}~
\varepsilon_{ijk}\,I_{ja}^{{(3)}}(T_R) \int_{-\infty}^{T_R} \ud V
\,I_{kb}^{{(3)}}(V)~ I_{ab}^{{(3)}}(V)\,.
\label{eq:memory2}
\end{equation}
In principle we should put brackets to indicate a STF projection acting on the indices $k$ and $a$ in the RHS, however the expression is manifestly symmetric with respect to those indices, and automatically trace-free thanks to the presence of $\varepsilon_{ijk}$. The orbital average reads
\begin{equation}
\langle{\cal G}_i^{\rm memory}\rangle =\frac{4}{35}\,
\frac{G^2}{c^{10}}\,\varepsilon_{ijk}\,\int_0^{P} \frac{\ud
T_R}{P}\,I_{ja}^{{(3)}}(T_R) \int_{-\infty}^{T_R} \ud V \,I_{kb}^{{(3)}}(V)~
I_{ab}^{{(3)}}(V)\,.
\label{moymemory2}
\end{equation}
We invert the two integration signs to re-write the latter expression as
\begin{eqnarray}
\langle{\cal G}_i^{\rm memory}\rangle &=&
\frac{4}{35}\,\frac{G^2}{c^{10}}\,\varepsilon_{ijk}\,\int_{-\infty}^{0} \ud
V\,I_{kb}^{{(3)}}(V)~ I_{ab}^{{(3)}}(V)\int_0^{P} \frac{\ud
T_R}{P}\,I_{ja}^{{(3)}}(T_R)\nonumber\\
&+&\frac{4}{35}\,\frac{G^2}{c^{10}}\,\varepsilon_{ijk}\,\int_0^{P} \ud
V\,I_{kb}^{{(3)}}(V)~ I_{ab}^{{(3)}}(V)\int_V^{P} \frac{\ud
T_R}{P}\,I_{ja}^{{(3)}}(T_R)\,.\label{invint}
\end{eqnarray}
The contribution extending over the entire past (i.e. over $-\infty\leq V\leq 0$) involves the orbital average of the third time derivative of the moment, i.e.
\begin{equation}
\langle I_{ij}^{{(3)}}\rangle \equiv \int_0^{P} \frac{\ud
T_R}{P}\,I_{ja}^{{(3)}}(T_R)\,.
\label{avI3}
\end{equation}
Since the quadrupole moment is Newtonian at this level of accuracy, the motion is periodic, and the quadrupole averages to zero. However one is not allowed to replace $\langle I_{ij}^{{(3)}}\rangle=0$ into the first term in the RHS of~\eqref{invint} because the argument neglects the evolution in the remote past of the Keplerian orbital elements by radiation reaction. In Appendix~\ref{appA} we shall study the dependence over the binary's past history and find that it gives a contribution on the current dynamics in the form of a zero-frequency or DC effect. 

On the other hand the memory due to the recent past of the source [second term in the RHS of~\eqref{invint}] is easily seen to be zero on average. Indeed we evaluate the integral in the same way as was done to compute the orbital average of the instantaneous contributions in Sec.~\ref{sec:AMF-AvADM}, and find that it is made of a sum of elementary integrals only of the type~\eqref{int0} that are zero. Therefore we conclude that the memory contribution to the averaged angular momentum flux reduces to the DC term due to the influence of the remote past of the source,
\begin{equation}
\langle{\cal G}_i^{\rm memory}\rangle = \langle{\cal G}_i^{\rm DC}\rangle\,.
\label{moymemory4}
\end{equation}
The DC term has the structure of a Newtonian term and is obtained from a model of past orbital evolution for eccentric orbits in Eq.~\eqref{avDC} of Appendix~\ref{appA}.\footnote{Recently the DC non-linear memory terms in the gravitational-wave polarizations have been computed to post-Newtonian order in the case of quasi-circular binary orbits~\cite{Favata08}.}

However, in this paper we have mostly in view the comparison with numerical simulations such as those in~\cite{HinderPNeccentric08}. Numerical simulations start from initial conditions which are for the moment always set at some very recent instant. This means that the initial eccentricity $e_1$ is comparable to the current one $e_0$ (using the notation of Appendix~\ref{appA}). In this case one can neglect the DC contribution so that
\begin{equation}
\langle{\cal G}_i^{\rm memory}\rangle = 0\,.
\label{moymemory5}
\end{equation}
In the present paper we adopt the result \eqref{moymemory5} appropriate for short-lived binary systems, but keep in mind the possible influence from the past of the non-linear memory DC term computed in Appendix~\ref{appA}.

\subsection{Definition of the eccentricity enhancement factors}
\label{eccfactor}
We shall now closely follow the numerical calculation of the hereditary terms in the energy flux~\cite{ABIQ07tail}. We shall present here only the definitions we use and directly the results we obtain from those definitions. We refer to~\cite{ABIQ07tail} for more details. The source multipole moments (in the center-of-mass frame) at Newtonian order read
\begin{subequations}\begin{eqnarray}
I_L^{(\mathrm{N})}&=&\mu \,s_l(\nu) \,x^{<L>}\,, \\ J_{L-1}^{(\mathrm{N})}
&=&\mu \,s_l(\nu) x^{<L-2}\,\varepsilon^{i_{l-1}>ab}\,x^av^b \,,
\end{eqnarray}\end{subequations}
and involve the following function of the symmetric mass ratio $\nu=\mu/m$,\footnote{Alternative forms for this function can be found in Refs.~\cite{K08,BFIS08}.}
\begin{equation}
s_l(\nu)=X_2^{l-1}+(-)^lX_1^{l-1}\,, \label{sl}
\end{equation}
where $X_1=\frac{1}{2}\left(1+\sqrt{1-4\nu}\right)$ and $X_2=\frac{1}{2}\left(1-\sqrt{1-4\nu}\right)$. Next, we rescale the source moments in an appropriate way and introduce the \textit{dimensionless} moments $\hat{I}_L$ and $\hat{J}_{L-1}$ by
\begin{subequations}\begin{eqnarray}
I_L^{(\mathrm{N})} &\equiv& {\mu\,a^l}\, s_l(\nu)\, \hat{I}_L\,,
\label{ILhat}\\
J_{L-1}^{(\mathrm{N})} &\equiv& {\mu\,a^l\,n}\, s_l(\nu)\, \hat{J}_{L-1}\,,
\end{eqnarray}\end{subequations}
where $a$ is the semi-major axis and $n=2\pi/P$ is the mean motion (such that Kepler's law $n^2 a^3=G\,m$ holds at Newtonian order).

We now define a set of ``enhancement'' functions of the eccentricity $e$ of the orbit (at Newtonian order) by means of the Fourier components of the rescaled moments. Such enhancement functions will exactly parallel similar functions valid in the case of the energy flux~\cite{ABIQ07tail}, and we shall adopt for these exactly the same notation as in~\cite{ABIQ07tail} except that we add a tilde on these functions to distinguish them from the functions parametrizing the energy flux. We thus pose
\begin{subequations}\label{enhancefns}\begin{eqnarray} 
\label{Newtf(e)} \tilde{f}(e) &=& -{\ii \over8}\,\varepsilon_{ijk}
\,\hat{L}_i\sum_{p=1}^{+\infty} p^5
\mathop{\hat{\cal{I}}}_{(p)}{}_{\!\!ja}^{(\mathrm{N})}
\mathop{\hat{\cal{I}}}_{(p)}{}_{\!\!ka}^{(\mathrm{N})*}\,,  \\\label{phie}
\tilde{\varphi}(e) &=& -{\ii \over16}\,\varepsilon_{ijk}
\,\hat{L}_i\sum_{p=1}^{+\infty} p^6 \mathop{\hat{\cal{I}}}_{(p)}{}_{\!\!ja}
\mathop{\hat{\cal{I}}}_{(p)}{}_{\!\!ka}^{*}\,,  \\ \tilde{\beta}(e) &=&
-{20\,\ii\over 16403}\,\varepsilon_{ijk} \,\hat{L}_i\sum_{p=1}^{+\infty} p^8
\mathop{\hat{\cal{I}}}_{(p)}{}_{\!\!jab}
\mathop{\hat{\cal{I}}}_{(p)}{}_{\!\!kab}^{*}\,,
\label{betae}\\
\tilde{\gamma}(e) &=& -8\,\ii\,\,\varepsilon_{ijk}
\,\hat{L}_i\sum_{p=1}^{+\infty} p^6\mathop{\hat{\cal{J}}}_{(p)}{}_{\!\!ja}
\mathop{\hat{\cal{J}}}_{(p)}{}_{\!\!ka}^{*}\,,\label{gammae}\\\tilde{F}(e)
&=&-{\ii\over32}\,\varepsilon_{ijk} \,\hat{L}_i\sum_{p=1}^{+\infty} p^7
\mathop{\hat{\cal{I}}}_{(p)}{}_{\!\!ja}
\mathop{\hat{\cal{I}}}_{(p)}{}_{\!\!ka}^{*}\,,\\\tilde{\chi}(e) &=&
-{\ii\over32}\,\varepsilon_{ijk} \,\hat{L}_i\sum_{p=1}^{+\infty}
p^7\ln\left(\frac{p}{2}\right) \,\mathop{\hat{\cal{I}}}_{(p)}{}_{\!\!ja}
\mathop{\hat{\cal{I}}}_{(p)}{}_{\!\!ka}^{*}\,.\label{chie}
\end{eqnarray}\end{subequations}
Like for the definitions adopted in~\cite{ABIQ07tail} all the latter tilde functions \textit{but one} are chosen in such a way that they \textit{tend to one} in the circular orbit limit, when $e\rightarrow 0$. The notable exception is $\tilde{\chi}(e)$ which \textit{vanishes} in this limit. This is easily checked since in the circular orbit limit (and at Newtonian order) the quadrupole moment possesses only the harmonic for which $p=2$, and consequently the log-term in $\tilde{\chi}(e)$ --- Eq.~\eqref{chie} --- becomes zero. Most of these functions will not admit any algebraic closed-form expression, and we shall leave them in the form of Fourier series to be evaluated numerically. However, as we shall now see, two functions can be computed algebraically, namely $\tilde{f}(e)$ and $\tilde{F}(e)$.\footnote{We reserve Latin names for algebraic closed-form functions, and Greek names for numerically generated ones.}

The first function parametrizes the Newtonian part of the averaged angular momentum flux,
\begin{equation}
\label{JfluxN:avf(e)}\langle {\cal G}^{(\mathrm{N})}\rangle = 
\frac{32}{5}c^2\,\nu^2\,m\,x^{7/2}\,\tilde{f}(e)\,,
\end{equation}
where $x$ is defined by \eqref{x} and reduces at Newtonian order to $G\,m/(a c^2)$. Thus, $\tilde{f}(e)$ is the Peters \& Mathews~\cite{PM63,Pe64} function, which admits an algebraically closed-form expression which is used in the timing of the binary pulsar PSR~1913+16~\cite{TW82}, and given by
\begin{equation} \label{fe}
\tilde{f}(e) = \frac{1+\frac{7}{8}e^2}{(1-e^2)^{2}}\,,
\end{equation}
in agreement with the Newtonian part of our earlier
result~\eqref{eq:AvAMFx}--\eqref{AMFxPN}.

On the other hand the function $\tilde{F}(e)$ is the analogue of the function $F(e)$ in~\cite{ABIQ07tail} and will partly parametrize the tails-of-tails, where it will have in factor a contribution depending on the arbitrary constant scale $r_0$. That dependence on $r_0$ is the same as in the instantaneous part of the flux, as given by Eqs.~\eqref{eq:AvAMFx}--\eqref{AMFxPN}. Such a specific dependence of the hereditary terms on $r_0$ will just be appropriate to exactly cancel out the $\ln r_0$ in the total angular momentum flux. The function $\tilde{F}(e)$ is given by 
\begin{equation}\label{Fetilde}
\tilde{F}(e) = \frac{1+{229\over32}\,e^2+{327\over64}\,e^4+{69\over256}\,e^6}
{(1-e^2)^{5}}\,.
\end{equation}
The analogous function $F(e)$ in the energy flux is recalled in Eq.~\eqref{Fe} below.

From the previous definitions we can express the quadratic tails at Newtonian order as
\begin{subequations}\begin{eqnarray}
\label{Nmqt}\langle{\cal G}^{\rm tail}\rangle_{\rm
mass\;quad}^{(\mathrm{N})} &=&
\frac{32}{5}c^2\,\nu^2\,m\,x^{7/2}\,\Bigl[
4\pi\,x^{3/2}\,\tilde{\varphi}(e)\Bigr]\,,\\\langle{\cal G}^{\rm
tail}\rangle_{\rm mass\;oct} &=&
\frac{32}{5}c^2\,\nu^2\,m\,x^{7/2}\,\Biggl[\frac{16403}{2016}\,\pi\,(1-4\,\nu)\,
x^{5/2}\,\tilde{\beta}(e)\Biggr]\,, \label{Fbeta}\\ \langle{\cal G}^{\rm
tail}\rangle_{\rm curr\;quad} &=&
\frac{32}{5}c^2\,\nu^2\,m\,x^{7/2}\,\biggl[\frac{\pi}{18}\,(1-4\,\nu)\,x^{5/2}\,
\tilde{\gamma}(e)\biggr]\,.  \label{Fgamma}
\end{eqnarray}\end{subequations} 
In the Newtonian mass-quadrupole tail~\eqref{Nmqt} we recognize in particular the coefficient $4\pi$ computed analytically in Ref.~\cite{RS97} (recall that $\tilde{\varphi}(0)=1$). The function $\tilde{\varphi}(e)$ has already been computed numerically from its Fourier series~\eqref{phie} in Ref.~\cite{RS97}. Notice that these formulas and similar formulas below take exactly the same form (i.e. with the same coefficients) as the corresponding formulas valid in the case of the energy flux~\cite{ABIQ07tail}. The reason of course is that in the circular-orbit limit $e\rightarrow 0$ the energy and angular momentum fluxes are proportional, and related by Eq.~\eqref{circular}. However, for non-zero eccentricities, the enhancement functions will differ from the corresponding functions in the energy flux, and this is why we add a tilde on them.

In a similar way, with the above definitions we get the sum of contributions from tails-of-tails and squared-tails as
\begin{align}\label{FFPsi}
&\langle{\cal G}^{\rm tail(tail)+(tail)^2}\rangle =\nonumber\\
&\qquad {32\over5}c^2\nu^2\,m \,x^{13/2} \Biggl\{ \left[ -{116761\over3675} +
{16\over3} \pi^2 -{1712\over105}C -
{1712\over105}\ln\left(4\,\omega\,r_0\right)\right] \tilde{F}(e)
-{1712\over105}\,\tilde{\chi}(e) \Biggr\}\,.
\end{align}
The circular-orbit limit of this result is immediately read off and seen to agree with the previous result in Refs.~\cite{B98tail,BIJ02}.

Finally we provide the mass quadrupole tail at relative 1PN order. Its computation is much more involved because the Fourier series~\eqref{Ftailfourier} contains several summations, and depends on intermediate results~\eqref{Ftailfourier} and~\eqref{intk}. The computation is based on the known 1PN relativistic corrections in the mass quadrupole moment $I_{ij}$ and the total mass $M$, which are given in Eqs.~(5.16)--(5.17) in~\cite{ABIQ07tail}. The result is of the type
\begin{equation}\label{1PNmq}
\langle{\cal G}^{\rm tail}\rangle_{\rm mass\;quad} =
\frac{32}{5}c^2\,\nu^2\,m\,x^{5} \Biggl\{
4\pi\,\tilde{\varphi}(e_t)+\pi\,x\,\Biggl[
-\frac{428}{21}\,\tilde{\alpha}(e_t) + \frac{178}{21}
\nu\,\tilde{\theta}(e_t)\Biggr]\Biggr\}\,,
\end{equation}
which defines two new enhancement functions $\tilde{\alpha}$ and $\tilde{\theta}$ which are the analogues of the functions $\alpha$ and $\theta$ in the energy flux~\cite{ABIQ07tail}. Since $\tilde{\alpha}$ and $\tilde{\theta}$ are given by some very complicated Fourier series (handled with Mathematica) --- instead of relatively simple ones like in~\eqref{enhancefns} --- we shall directly compute them numerically using the same method as in~\cite{ABIQ07tail}. Notice that since we are at the 1PN order we must be specific about which definition of eccentricity we use; here we adopt the \textit{time eccentricity} denoted $e_t$ which enters the generalized Kepler equation at 1PN order~\cite{DD85}. On the other hand, the 1PN corrections arising from the parameter $x$ [see~\eqref{x}] are evidently crucial in the result (they include the 1PN periastron advance $k$).

For the final presentation it is convenient to redefine in a minor way our elementary enhancement functions. Let us choose (still paralleling Ref.~\cite{ABIQ07tail})
\begin{subequations}\label{redef}\begin{eqnarray} \tilde{\psi}(e) 
&\equiv& \frac{13\,696}{8191}\,\tilde{\alpha}(e)
-\frac{16\,403}{24\,573}\,\tilde{\beta}(e) -
\frac{112}{24\,573}\,\tilde{\gamma}(e)\,, \\ \tilde{\zeta}(e) &\equiv&
-\frac{1424}{4081}\,\tilde{\theta}(e)
+\frac{16\,403}{12\,243}\,\tilde{\beta}(e)
+\frac{16}{1749}\,\tilde{\gamma}(e)\,, \\ \tilde{\kappa}(e) &\equiv&
\tilde{F}(e) +\frac{59\,920}{11\,6761} \tilde{\chi}(e)\,.
\end{eqnarray}\end{subequations}
Considering thus the 1.5PN and 2.5PN terms, composed of tails, and the 3PN terms, composed of the tails-of-tails and the squared-tails, we get the final form of the total hereditary contribution to the averaged angular momentum flux~\eqref{eq:hered}: $\langle{\cal G}_i^{\rm hered}\rangle = \hat{L}_i\,\langle{\cal G}_{\rm hered}\rangle$ with
\begin{align}\label{Gtailfinal}
\langle{\cal G}_{\rm hered}\rangle &=
\frac{32}{5}c^2\,\nu^2\,m\,x^{7/2} \Biggl\{
4\pi\,x^{3/2}\,\tilde{\varphi}(e_t)+\pi\,x^{5/2}
\left[-\frac{8191}{672}\,\tilde{\psi}(e_t)
-\frac{583}{24}\nu\,\tilde{\zeta}(e_t)\right]\nonumber\\ &~+x^3\left[
-\frac{116\,761}{3675}\,\tilde{\kappa}(e_t) +\left[ \frac{16}{3} \,\pi^2
-\frac{1712}{105}\,C -
\frac{1712}{105}\ln\left(4\omega\,r_0\right)\right]
\tilde{F}(e_t)\right]\Biggr\}\,.
\end{align}
For circular orbits this angular momentum flux is in agreement with the energy flux $\langle{\cal F}_{\rm hered}\rangle_{\odot}$ computed in~\cite{ABIQ07}, since we have $\langle{\cal F}_{\rm hered}\rangle_{\odot} =\omega\,\langle{\cal G}_{\rm hered}\rangle_{\odot}$.  

\subsection{Computation of the enhancement functions}
\label{sec:plotsheredamf}
The eccentricity-dependent enhancement functions we use in~\eqref{Gtailfinal}, computed numerically, are now provided in the form of plots. The numerical computation has been described in the case of the energy flux~\cite{ABIQ07tail} and we adapt the same for the angular momentum flux. 

Essentially, the fitting procedure to obtain the Fourier coefficients of the Newtonian source moments (or their periodic part in the mean anomaly $\ell$ for 1PN accurate moments) in terms of $\ell$ can be implemented either starting with the basic multipole moments themselves or the leading time derivatives appearing in each of the terms in the angular momentum flux. The latter method is known to improve the numerical convergence of the final sum because one deals with lower order time-derivatives of the basic functions. However,  here we have followed the former method which has the advantage that the fitting function is much more simple for the basic source moments than for their multi-time-derivatives and thus takes less time and is also less prone to errors. Proceeding in this way also provides another check on the energy flux calculation as we have reproduced the results of Ref.~\cite{ABIQ07tail} using this alternative choice. At the Newtonian order it is in fact more efficient to make use of the well-known Fourier decomposition of the Keplerian motion to compute the Fourier coefficients. Using this we can derive the components of the multipole moments (at Newtonian order) as series of combinations of Bessel functions. It is then simple to compute numerically the associated Newtonian enhancement functions [namely the functions $\tilde{\varphi}(e)$, $\tilde{\beta}(e)$, $\tilde{\gamma}(e)$ and $\tilde{\chi}(e)$]. On the other hand, for the Newtonian tail terms, we could proceed exactly in the same way as for the 1PN term, following the various steps and evaluating numerically the functions. We have verified that both methods agree well. The above procedure is quite general, and provides a method which could be extended to higher PN orders.

\begin{figure}[t]
\vspace{0.5cm}
\centering 
\includegraphics[width=3.2in]{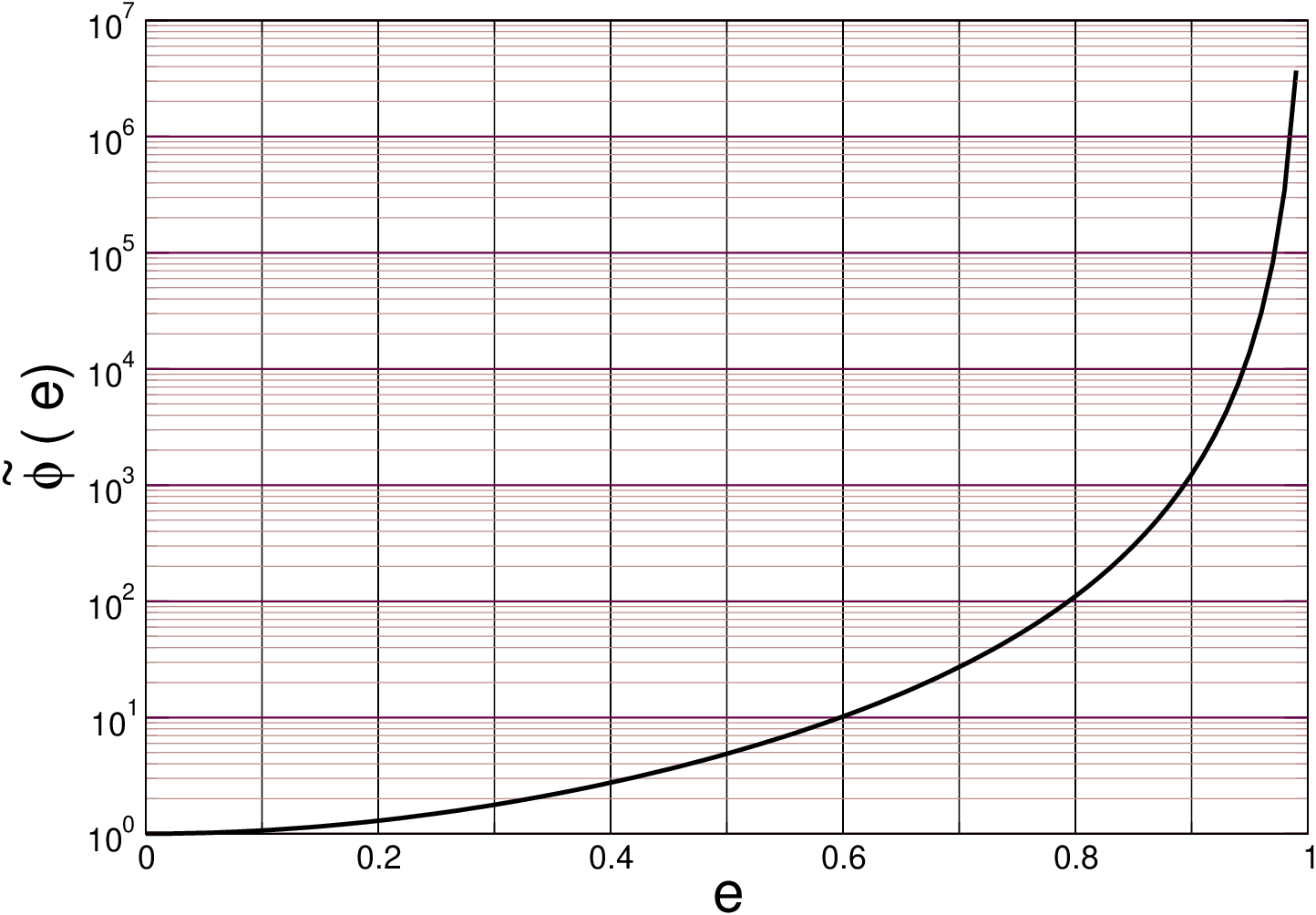}
\caption{Enhancement function $\tilde{\varphi}(e)$ in the angular momentum flux at 1.5PN order.}
\label{fig1}
\end{figure}

\begin{figure}[t]
\vspace{0.5cm}
\centering 
\includegraphics[width=3.2in]{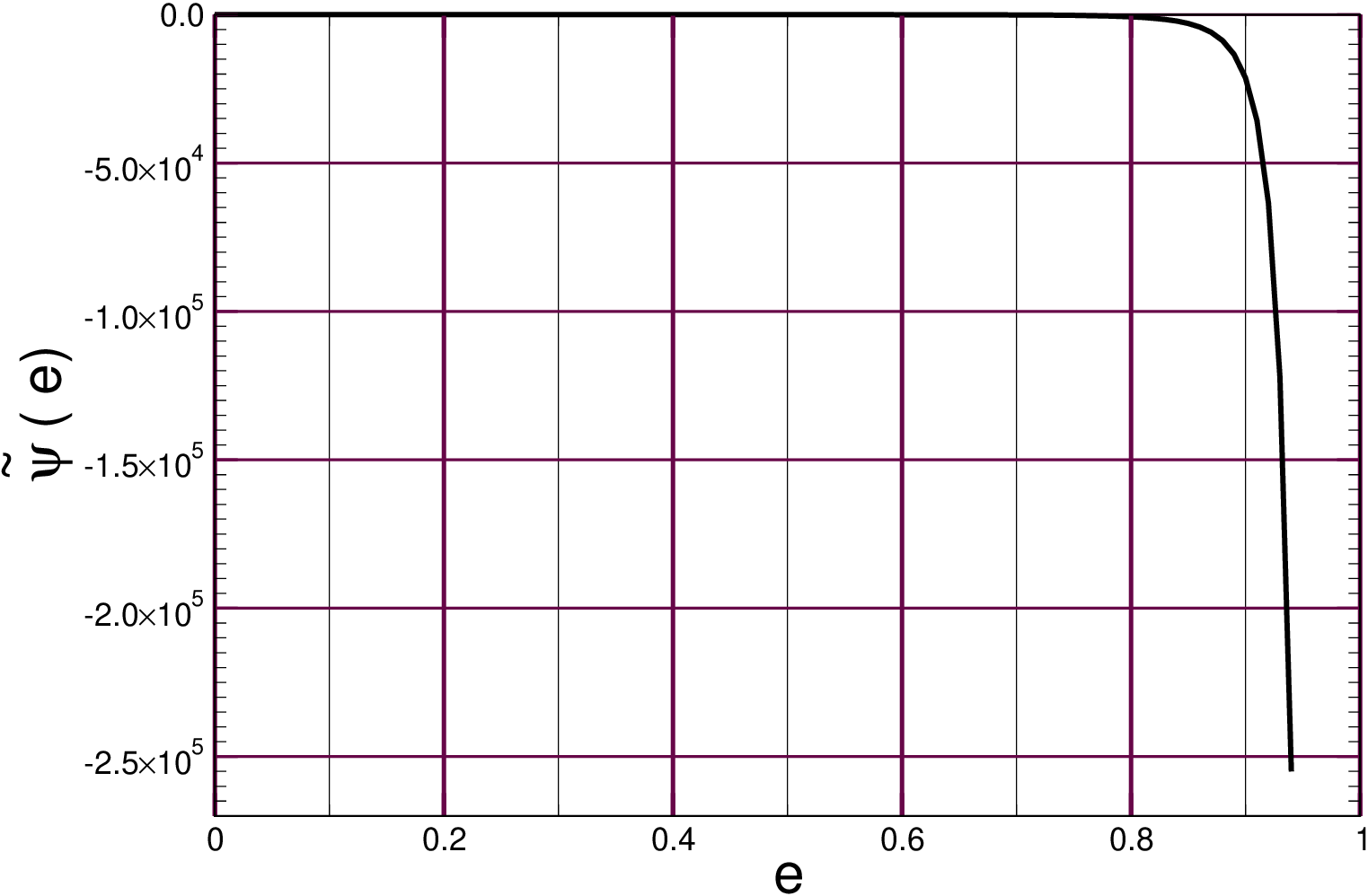}
\hskip 0.2 true cm 
\includegraphics[width=3.1in]{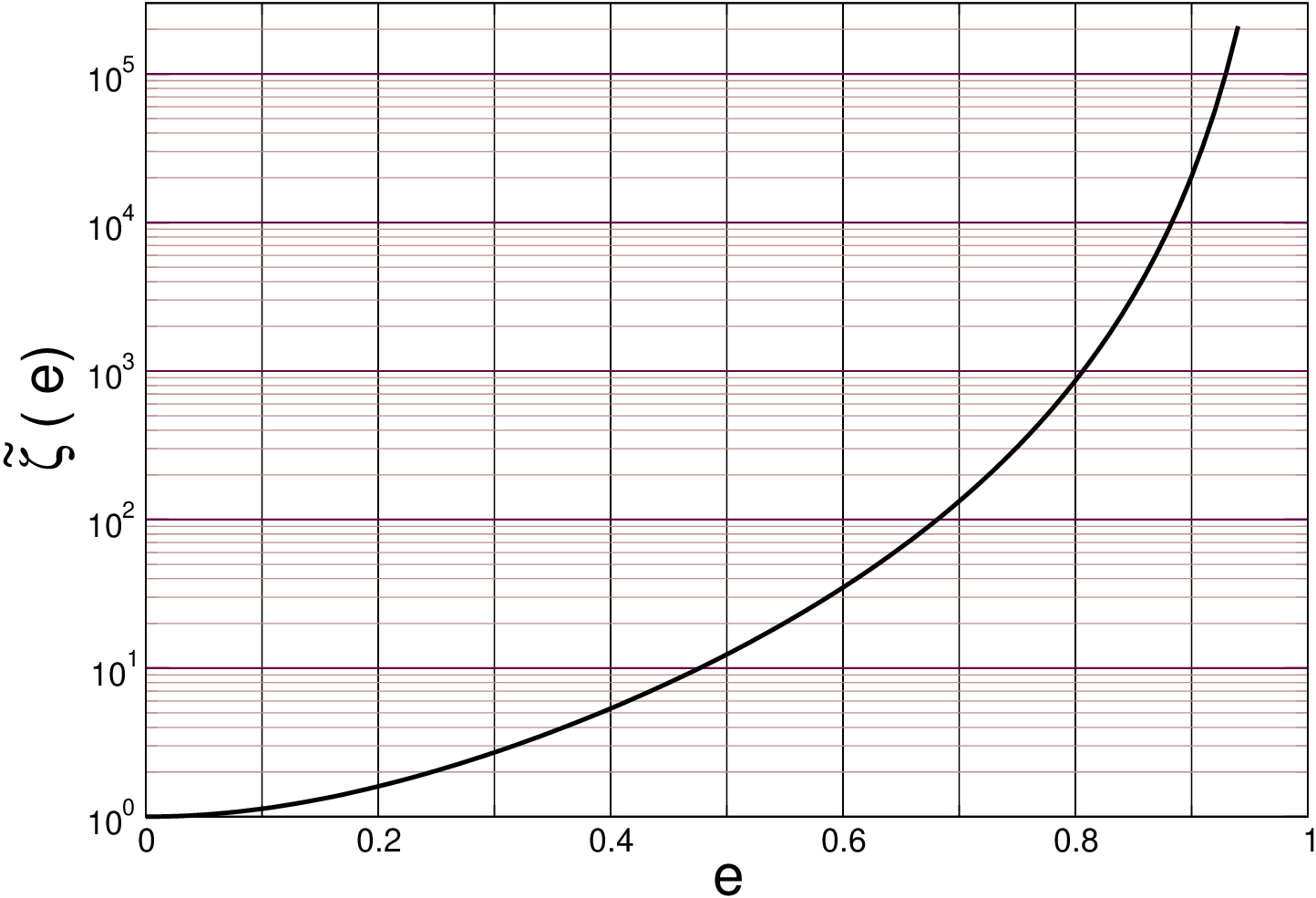} 
\caption{Enhancement functions $\tilde{\psi}(e)$ and $\tilde{\zeta}(e)$ in the angular momentum flux at 2.5PN order.}
\label{fig2}
\end{figure}

\begin{figure}[t]
\vspace{0.5cm}
\centering 
\includegraphics[width=3.2in]{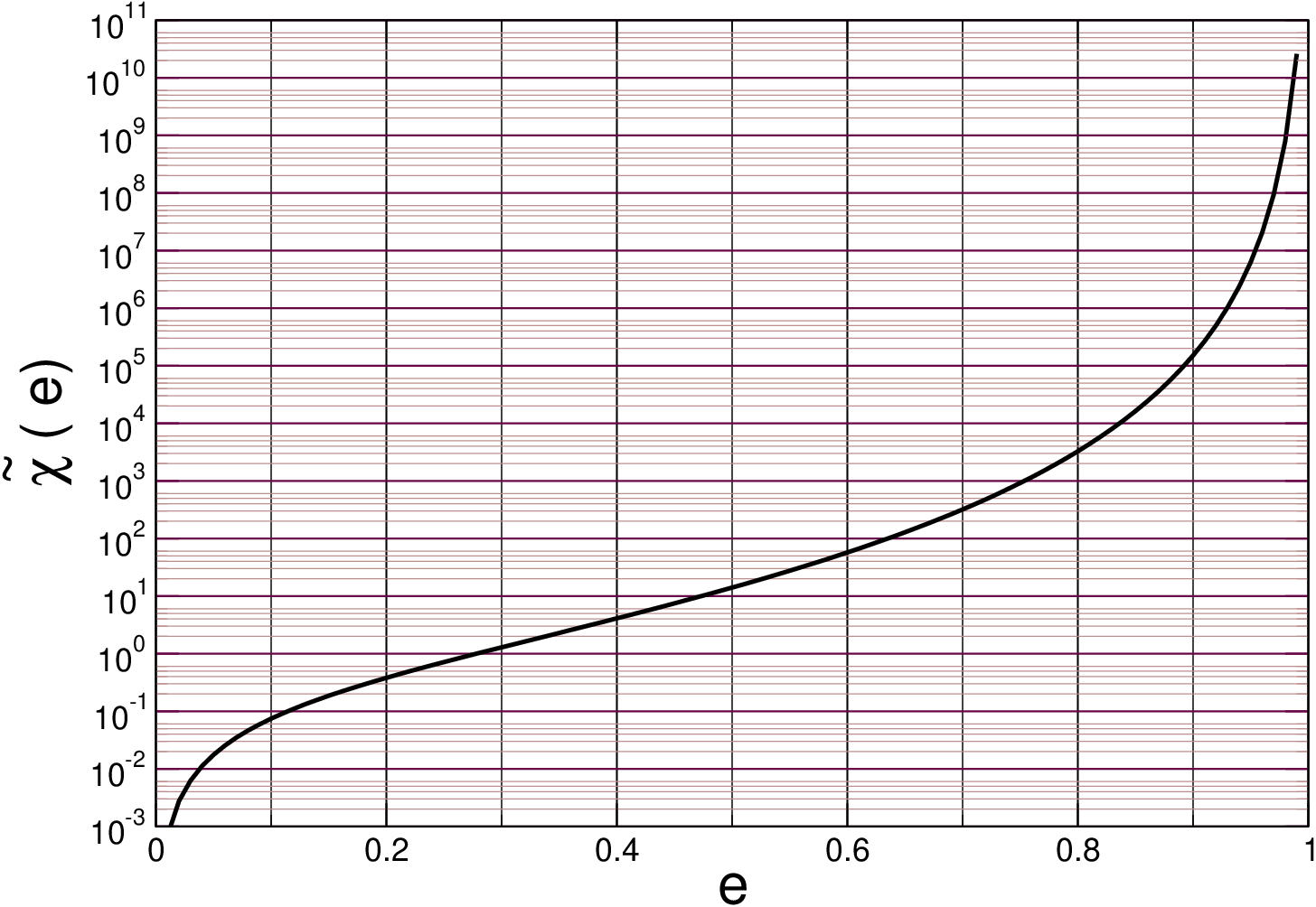}
%\hskip 0.2 true cm 
\includegraphics[width=3.2in]{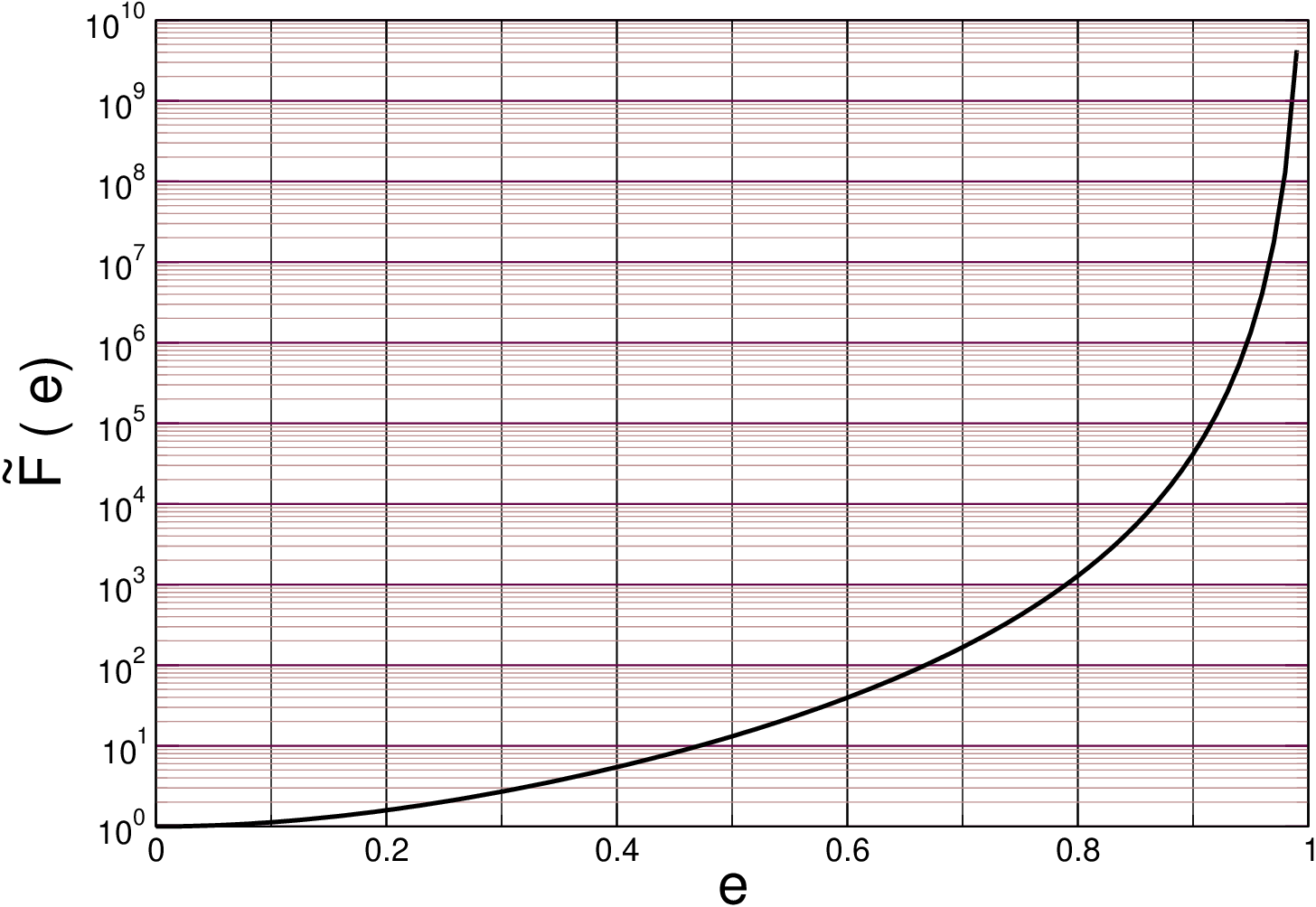} 
\caption{Enhancement functions $\tilde{\chi}(e)$ and $\tilde{F}(e)$ in the angular momentum flux at 3PN order.}
\label{fig3}
\end{figure}

We present in Figs.~\ref{fig1}--\ref{fig3} the functions which permit to define the hereditary part of the angular momentum flux \eqref{Gtailfinal}. We recall that these functions are such that they reduce to one in the circular-orbit limit $e\rightarrow 0$.
To facilitate the comparison with the results of numerical relativity~\cite{HinderPNeccentric08} or use in data-analysis applications
 we provide also some numerical tables for these functions in Appendix~\ref{appB}.

\section{Evolution of orbital elements under 3PN radiation reaction}
\label{sec:orbelem}
The most important application of the 3PN angular momentum flux is to calculate, using also the energy flux~\cite{ABIQ07}, how the binary's orbital elements evolve under 3PN gravitational radiation reaction. 
We shall compute the time evolution of the mean motion $n$, the periastron precession $k$, the mean orbital frequency $\omega=n\,K$, the semi-major axis $a_r$, and the time eccentricity $e_t$.

\subsection{General method}
We start with the 3PN expressions for $n$, $k$, $\omega$, $a_r$ and $e_t$ in terms of the conserved energy $E$ and conserved angular momentum $J$ of the orbit~\cite{MGS04}. In an attempt to simplify some expressions we often employ, in place of the energy $E$ and angular momentum $J$, the dimensionless variables~\cite{ABIQ07}
\begin{subequations}
\begin{align}
\varepsilon &= -\frac{2\, E}{c^2}\,,\\ j &= -\frac{2\,E\,  J^2}{(G m)^2}\,.
\end{align}\label{epsj}\end{subequations}
Since $\varepsilon={\cal O}(c^{-2})$ this energy variable can be viewed as a book-keeping parameter labeling the successive PN orders. Recall that the semi-major axis $a_r$ and the eccentricity $e_t$ depend on the coordinate system, but that the mean motion $n$ and periastron precession $k$ are \textit{gauge invariant}, i.e. take the same expressions in ADM and, say, modified  harmonic coordinates. Of course this is also true of the mean orbital frequency $\omega$ and hence of the parameter $x$. The expressions we give for $e_t$ and $a_r$ are valid in ADM coordinates. We have 
\begin{subequations}\label{3PNQK}\begin{align}
n =&\frac{c^3}{G\,m} \varepsilon^{3/2}\bigg\{ 1+\frac{\varepsilon}{8}\, (
-15+\nu ) +\frac{\varepsilon^{2}}{128} \biggl[ 555 +30\,\nu +11\,\nu^{2} +
  \frac{192}{j^{1/2}} ( -5+2\,\nu ) \biggr] \nonumber\\& +
\frac{\varepsilon^{3}}{3072} \biggl[ -29385 -4995\,\nu-315\,\nu^{2}+135
  \,\nu^{3} \nonumber\\&\quad - \frac{16}{j^{3/2}} \bigg(
  10080+\left(-13952+123\,\pi^{2} \right)\,\nu+1440\,\nu^{2}\bigg) +
  \frac{5760}{j^{1/2}} (17 -9\,\nu+2\,\nu^{2} ) \biggr] \bigg\} \,,
\label{n}\\ k =& 3\frac{\varepsilon}{j}+ \frac{\varepsilon^{2}}{4} \biggl [
  \frac{3}{j} ( -5+2\,\nu ) + \frac{15}{j^2} ( 7 -2\,\nu ) \biggr]
+\frac{\varepsilon^{3}}{128} \biggl[ \frac{24}{j} ( 5 -5\nu+ 4\nu^2)
\nonumber\\&\quad - \frac{1}{j^2} \biggl( 10080 
+\left(-13952 +123 \,\pi^{2}\right)\nu+1440\,\nu^{2} \biggr)\nonumber\\&\quad \quad+ \frac{5}{j^3}
\biggl(7392+\left(-8000 + 123\,\pi^{2}\right)\nu + 336\,\nu^{2} \biggr) \biggr]\,, \\
\omega = &\frac{c^3}{G\,m}\varepsilon^{3/2}\biggl\{ \biggl[1+\frac{\varepsilon}{8}\, (
  -15+\nu +{24\over j})\biggr]+\frac{\varepsilon^2}{128}\biggl[555+30 \nu +11 \nu ^2+\frac{192} {\sqrt{j}}(-5+2 \nu ) +\frac{240}{j}\left(-5+\nu\right)\nonumber\\
&\quad-\frac{480} {j^2}(-7+2 \nu )\biggr]+\frac{\varepsilon^3}{3072}\biggl[
45 \left(-653-111 \nu -7 \nu ^2+3 \nu ^3\right)+\frac{5760}{{j^{1/2}}} \left(17-9 \nu +2 \nu ^2\right)\nonumber\\
&\quad+\frac{72}{j} \left(895-150 \nu +51 \nu ^2\right)+\frac{1}{j^{3/2}}\left(-230400-16 \left(-15680+123 \pi ^2\right) \nu -23040 \nu
   ^2\right)\nonumber\\
&\quad+\frac{1}{j^2}\left(-393120-24 \left(-16172+123 \pi ^2\right) \nu -37440 \nu
   ^2\right)\nonumber\\
&\quad+\frac{1}{j^3}\left(887040+120 \left(-8000+123 \pi ^2\right) \nu +40320 \nu^2\right)\biggr]\biggr\}\\
 a_r =&
\frac{G\,m}{c^2}\frac{1}{\varepsilon}\bigg\{ 1+\frac{\varepsilon}{4} ( -7+\nu ) +
\frac{\varepsilon^{2}}{16}\,\bigg[ 1+10\,\nu+{\nu}^{2} +\frac{1}{j}
(-68+44\,\nu) \bigg] \nonumber\\& + \frac {\varepsilon^{3}}{192}\, \biggl[
3-9\,\nu-6\,{\nu}^{2} +3\,{\nu}^{3}+\frac{1}{j} \biggl( 864+ \left( -2212-3\,{\pi}^{2} \right) \nu+432\,{\nu}^{2}\biggr)\nonumber\\ &\quad + \frac{1}
{j^2} \biggl( -6432+ \left( 13488-240\,{\pi}^{2} \right) \nu -768\,{\nu}^{2}\biggr)
\biggr] \bigg\}\,,\\ e_t =&\Biggl[1-j+ \frac{\varepsilon}{4}\, \bigg\{
-8+8\,\nu -j ( -17+7\,\nu ) \bigg\} \nonumber\\& + \frac{\varepsilon^{2}}{8}
\bigg\{8+4\,\nu +20\,{\nu}^{2} - j( 112-47\,\nu +16\,\nu^{2}) -24\,j^{1/2}\, (
-5+2\,\nu ) \nonumber\\ &\quad+\frac{4}{j} (17 - 11\,\nu ) -\frac{24}{j^{1/2}}
\, ( 5 -2\,\nu ) \bigg\}\nonumber\\ &\quad + \frac{\varepsilon^{3}}{192}
\bigg\{ 24\, ( -2+5\,\nu ) (-23+10\,\nu+ 4\,\nu^{2} ) -15\,j \biggl
(-528+200\,\nu-77\,\nu^{2} + 24\,\nu^{3} \biggr) \nonumber\\&\quad
-72\,j^{1/2}( 265-193\,\nu +46\,\nu^{2} ) - \frac{2}{j} \bigg( 6732
+\left( -12508 +117\,{\pi }^{2}\right)\nu+2004\,{\nu}^{2}\bigg) \nonumber\\&\quad +
\frac{2}{j^{1/2}} \bigg( 16380+\left(-19964+123\,\pi^{2}\right)\nu+3240\,\nu^{2}
\bigg) \nonumber\\&\quad -\frac{2}{j^{3/2}} \bigg(
10080+\left(-13952+123\,\pi^{2}\right)\nu+1440\,\nu^{2} \bigg) \nonumber\\&\quad +
\frac{96}{j^2} \bigg( 134+\left( -281+5\,{\pi }^{2}\right)\nu+16\,{\nu}^{2} \bigg)
\bigg\}\Biggr]^{1/2}\,.
\end{align}\end{subequations}

The procedure to compute the evolution of the orbital elements under gravitational radiation-reaction is straightforward but lengthy. 
Differentiating the orbital elements with respect to time, and using the heuristic balance equations, we equate the decreases of energy and angular momentum to the corresponding averaged fluxes, and obtain the (secular) rate of change of the orbital elements. This extends earlier analyses at previous PN orders: Newtonian~\cite{Pe64}, 1PN order~\cite{BS89,JunkS92}, 1.5PN order~\cite{BS93,RS97} and 2PN~\cite{GI97,DGI04}. Taking the example of the mean motion we have
\begin{equation}\label{timeder}
\frac{\ud n}{\ud t} = \frac{\partial n}{\partial E}\,\frac{\ud E}{\ud t} +
\frac{\partial n}{\partial J}\,\frac{\ud J}{\ud t}\,.
\end{equation}
The usual (heuristically derived) balance equations for energy $\mu\,\langle\ud E/\ud t\rangle = -\langle{\cal F}\rangle$ and angular momentum $\mu\,\langle\ud J/\ud t\rangle = -\langle{\cal G}\rangle$, where the fluxes are known up to 3PN order, give the 3PN evolution equation averaged over one orbit,
\begin{equation}\label{evoleq}
\langle\frac{\ud n}{\ud t}\rangle = - \frac{1}{\mu}\left[\frac{\partial n}{\partial
E}\,\langle{\cal F}\rangle + \frac{\partial n}{\partial J}\,\langle{\cal
G}\rangle\right]\,.
\end{equation}
For eccentric orbits we recall that this gives only the slow \textit{secular} evolution under gravitational radiation reaction. The complete evolution includes superimposed on this a fast but smaller periodic oscillation on the orbital time scale which can be conveniently computed using a two-scale decomposition following~\cite{DGI04,KG06}. 

To express the final result of (\ref{evoleq}) in terms of $x$ and $e_t$,
the variables in terms of which we have represented the energy and angular
momentum fluxes, we will require the expressions for $\varepsilon$ and $j$
in terms of $x$ and $e_t$. We have,
\begin{subequations}\label{varepsj}\begin{align}
\varepsilon=&x\left\{1+
\frac{x}{1-e_t^2}\left[-{3\over4}-{\nu\over12}
+e_t^2\left(-{5\over4}+{\nu\over12}\right) \right]
+ \frac{x^2}{(1-e_t^2)^2}\left[
-\frac{67}{8} +\frac{35  }{8}\nu -\frac{1}{24}\nu ^2
\right.\right.\nonumber\\&\left.
+{e_t}^2 \left( -\frac{19}{4} +\frac{21  }{4}\nu +\frac{1}{12}\nu ^2
\right)
+{e_t}^4 \left( \frac{5}{8} -\frac{5  }{8}\nu -\frac{1}{24}\nu ^2
\right)+
(1-e_t^2)^{3/2}(5-2\nu)
\right]
\nonumber\\&
\left.
+\frac{x^3}{(1-e_t^2)^{3}}
\left[
-\frac{835}{64}
+ \left({18319\over192}-{41\over16}\pi^2\right) \nu -\frac{169 }{32}\nu ^2
-\frac{35 }{5184}\nu ^3
\right.\right.\nonumber\\&
+{e_t}^2 \left( -\frac{3703}{64}
 + \left({21235\over192}-{41\over64}\pi^2\right) \nu 
-\frac{7733 }{288}\nu ^2+ \frac{35 }{1728}\nu ^3 \right)
\nonumber\\&
+{e_t}^4 \left( \frac{103}{64} -\frac{547  }{192}\nu -\frac{1355 }{288}\nu ^2
-\frac{35 }{1728}\nu ^3 \right)
+{e_t}^6 \left( \frac{185}{192} +\frac{75  }{64}\nu +\frac{25 }{288}\nu ^2
+\frac{35 }{5184}\nu ^3 
\right) 
\nonumber\\&
+\sqrt{1-e_t^2}\left(\frac{5}{2} +\left(-\frac{641}{18}+\frac{41 }{96}\pi ^2\right) \nu +\frac{11 }{3}\nu ^2
+{e_t}^2 \left( -35 +\left(\frac{394}{9}-\frac{41 }{96}\pi ^2\right) \nu 
-\frac{1}{3}\nu ^2 \right)
\right.\nonumber\\&\left.\left.\left.
+{e_t}^4 \left( \frac{5}{2} +\frac{23  }{6}\nu -\frac{10 }{3}\nu ^2
\right)
\right)
\right]
 \right\}\,,
\\j=&(1-e_t^2)\left\{1+\frac{x}{1-e_t^2}\left[
\frac{9}{4} +\frac{1 }{4}\nu
+{e_t}^2 \left( -\frac{17}{4}+ \frac{7  }{4}\nu \right)
\right]+\frac{x^2}{(1-e_t^2)^2}\left[
\frac{27}{8} -\frac{19  }{8}\nu +\frac{1}{24}\nu ^2
\right.\right.\nonumber\\&\left.
+{e_t}^2 \left( -\frac{1}{4} +\frac{53  }{12}\nu -\frac{5 }{4}\nu ^2 \right)
+{e_t}^4 \left( \frac{75}{8} -\frac{277  }{24}\nu+ \frac{29 }{24}\nu ^2 \right)
+\sqrt{1-e_t^2}\,e_t^2\,\left(-15+6\nu\right) \right]
\nonumber\\&
+\frac{x^3}{(1-e_t^2)^3}\left[
-\frac{747}{64} +\left(-\frac{5797}{192}+\frac{41 }{32}\pi ^2\right) \nu +\frac{167 }{96}\nu ^2 -\frac{1}{192}\nu ^3
\right.\nonumber\\&
+{e_t}^2 \left( \frac{5281}{64} +\left(-\frac{8955}{64}+\frac{39 }{32}\pi ^2\right) \nu +\frac{1751 }{96}\nu ^2 +\frac{67 }{192}\nu ^3 \right)
+{e_t}^4 \left( \frac{1023}{64} -\frac{3701  }{64}\nu +\frac{947 }{32}\nu ^2 -\frac{131 }{192}\nu ^3 \right)
\nonumber\\&
+{e_t}^6 \left( -\frac{757}{64} +\frac{7381  }{192}\nu -\frac{1207 }{96}\nu ^2+ \frac{65 }{192}\nu ^3 \right)
+\sqrt{1-e_t^2}\left(
\frac{45}{4} -\frac{13  }{4}\nu -\frac{1}{2}\nu ^2
\right.\nonumber\\&\left.\left.\left.
+{e_t}^2 \left( -\frac{505}{4} + \left({2227\over12}-{41\over32} \pi ^2\right) \nu -\frac{47 }{2}\nu ^2
\right)
+{e_t}^4 \left(55-40\nu +3 \nu ^2\right)
\right)
\right]
\right\}.
\end{align}
\end{subequations}

\subsection{Recalls for the energy flux}
To proceed further we also need the energy flux and so we recapitulate here the results from~\cite{ABIQ07tail,ABIQ07}. The instantaneous part of the energy flux reads
\begin{equation}\label{AvEADM}
\langle\,\mathcal{F}_\mathrm{inst}\rangle =\frac{32 }{5 }\,\frac{c^5}{G}\,\nu^2\,x^5\,\biggl(\mathcal{I}_\mathrm{N} + x\,\mathcal{I}_{\rm 1PN} +
x^2\,\mathcal{I}_\mathrm{2PN} + x^3\,\mathcal{I}_{\rm 3PN} \biggr)\,,
\end{equation}
where the coefficients (in ADM coordinates) are
\begin{subequations}\label{AvEADMa}\begin{align}
\label{AvEADM0} \mathcal{I}_\mathrm{N} =& \frac{1}{(1-e_t^2)^{7/2}}\left\{1
+\frac{73}{24}\,e_t^2+\frac{37}{96}\,e_t^4\right\}\,,\\
%==================================
\label{AvEADM1} \mathcal{I}_\mathrm{1PN} =&{\frac{1}{(1-e_t^2)^{9/2}}
\left\{-\frac{1247}{336}-\frac{35}{12} \nu
+e_t^2\left(\frac{10475}{672}-\frac{1081}{36} \nu
\right)\right.}\nonumber\\&
{\left.+e_t^4\left(\frac{10043}{384}-\frac{311}{12} \nu
\right)+e_t^6\left(\frac{2179}{1792}-\frac{851}{576} \nu
\right)\right\}}\,,\\
%==================================
\label{AvEADM2} \mathcal{I}_\mathrm{2PN} =& 
\frac{1}{(1-e_t^2)^{11/2}}\left\{-\frac{203471}{9072}
+\frac{12799}{504} \nu +\frac{65}{18}
\nu^2\right.\nonumber\\&+e_t^2\left(-\frac{3866543}{18144}+\frac{4691}{2016}
\nu +\frac{5935}{54} \nu ^2\right)\nonumber\\&
{+e_t^4\left(-\frac{369751}{24192}-\frac{3039083}{8064} \nu
+\frac{247805}{864} \nu ^2\right)}\nonumber\\&
{+e_t^6\left(\frac{1302443}{16128}-\frac{215077}{1344} \nu
+\frac{185305}{1728} \nu ^2\right)}\nonumber\\&
{+e_t^8\left(\frac{86567}{64512}-\frac{9769}{4608} \nu
+\frac{21275}{6912} \nu ^2\right)}\nonumber\\&
{+\sqrt{1-e_t^2}\left[\frac{35}{2}-7 \nu
+e_t^2\left(\frac{6425}{48}-\frac{1285}{24} \nu
\right)\right.}\nonumber\\&
{\left.\left.+e_t^4\left(\frac{5065}{64}-\frac{1013}{32} \nu
\right)+e_t^6\left(\frac{185}{96}-\frac{37}{48} \nu
\right)\right]\right\}}\,,\\
%==================================
\label{AvEADM3} \mathcal{I}_\mathrm{3PN} =&
 \frac{1}{(1-e_t^2)^{13/2}}
\left\{\frac{2193295679}{9979200}+\left[\frac{8009293 }{54432}
-\frac{41  }{64}\pi ^2\right]\nu-\frac{209063 }{3024}\nu ^2-\frac{775 }{324}\nu ^3
\right.\nonumber\\& {+e_t^2\left(
\frac{2912411147}{2851200}+\left[\frac{249108317
}{108864}+\frac{31255}{1536}\pi^2\right]\nu-\frac{3525469 }{6048}\nu
^2-\frac{53696 }{243}\nu ^3 \right)} \nonumber\\& {+e_t^4\left(
-\frac{4520777971}{13305600}+\left[\frac{473750339 }{108864}
-\frac{7459  }{1024}\pi ^2\right]\nu+\frac{697997 }{576}\nu ^2-\frac{10816087}{7776} \nu ^3
\right)}\nonumber\\& {+e_t^6\left(
\frac{3630046753}{26611200}+\left[-\frac{8775247}{145152}
-\frac{78285  }{4096}\pi ^2\right]\nu+\frac{31147213 }{12096}\nu ^2-\frac{983251 }{648} \nu ^3
\right)}\nonumber\\& {+e_t^8\left(
\frac{21293656301}{141926400}+\left[-\frac{36646949}{129024}-\frac{4059 \pi
^2}{4096}\right]\nu+\frac{85830865 }{193536}\nu ^2-\frac{4586539 }{15552}\nu
^3 \right)}\nonumber\\&
{+e_t^{10}\left(-\frac{8977637}{11354112}+\frac{9287}{48384} \nu
+\frac{8977}{55296} \nu ^2-\frac{567617}{124416} \nu ^3\right)}\nonumber\\&
+\sqrt{1-e_t^2}\left[ -\frac{14047483}{151200}+\left[-\frac{165761 }{1008}
+\frac{287  }{192}\pi ^2\right]\nu+\frac{455 }{12}\nu ^2\right.\nonumber\\&
+e_t^2\left(\frac{36863231}{100800}+\left[-\frac{14935421}{6048} +\frac{52685
}{4608}\pi ^2\right]\nu+\frac{43559 }{72}\nu ^2\right) \nonumber\\&
+e_t^4\left(\frac{759524951}{403200}+\left[-\frac{31082483}{8064} +\frac{41533
}{6144}\pi ^2\right]\nu+\frac{303985 }{288}\nu ^2\right)\nonumber\\&
+e_t^6\left(\frac{1399661203}{2419200}+\left[-\frac{40922933 }{48384}
+\frac{1517 }{9216}\pi ^2\right]\nu+\frac{73357 }{288}\nu
^2\right)\nonumber\\& +\left.e_t^8\left(\frac{185}{48}-\frac{1073 }{288}\nu
+\frac{407 }{288}\nu ^2\right)\right]\nonumber\\& \left. +
\left(\frac{1712}{105}+\frac{14552}{63} e_t^2+\frac{553297}{1260}
e_t^4+\frac{187357}{1260} e_t^6+\frac{10593}{2240} e_t^8\right)
\ln\left[\frac{x}{x_0}\frac{1+\sqrt{1-e_t^2}}{2(1-e_t^2)}\right]\right\}\,.
\end{align}
\end{subequations}
The hereditary part of the energy flux is given by
\begin{eqnarray}\label{Ftailfinal} 
\langle{\cal{F}}_\mathrm{hered}\rangle &=& \frac{32}{5}\,\frac{c^5}{G}\,\nu^2\,x^5 \Biggl\{
4\pi\,x^{3/2}\,\varphi(e_t)+\pi\,x^{5/2} \left[-\frac{8191}{672}\,\psi(e_t)
-\frac{583}{24}\nu\,\zeta(e_t)\right]\nonumber\\ &&+x^3\left[
-\frac{116761}{3675}\,\kappa(e_t) +\left[ \frac{16}{3} \,\pi^2
-\frac{1712}{105}\,C - \frac{1712}{105}\ln\left(\frac{4\omega\,r_0}{c}\right)\right]
F(e_t)\right]\Biggr\}\,.
\end{eqnarray}
With the exception of $F$ that can be given in a closed analytic form,
\begin{equation}\label{Fe}
F(e) = \frac{1+\frac{85}{6}e^2+\frac{5171}{192}e^4+\frac{1751}{192}e^6
+\frac{297}{1024}e^8}{(1-e^2)^{13/2}}\,,
\end{equation} 
the untilded enhancement functions $\varphi$, $\psi$, $\dots$ (differing for non-zero eccentricities from the tilded ones $\tilde{\varphi}$, $\tilde{\psi}$, $\dots$ in the angular momentum flux) are computed numerically in~\cite{ABIQ07tail}; in the Appendix~\ref{appB} below we provide the numerical tables of these functions.

\subsection{Instantaneous contributions}\label{secinst}
It is clear that since the evolution of the orbital elements is linear in $\langle{\cal F}\rangle$ and $\langle{\cal G}\rangle$ one can separate out the contributions due to the instantaneous and hereditary components in the fluxes.

Starting with the instantaneous terms, we find for the evolution of  the orbital elements $\{n,k,\omega,a_r,e_t\}$\footnote{Of course this is a redundant set of orbital elements; for instance $\omega=n\,(1+k)$.} a PN structure with coefficients depending on $E$ and $J$ or, alternatively, on the frequency-related parameter $x$ and the time eccentricity $e_t$. Choosing an eccentricity parameter like $e_t$ as one of the basic independent variables has the advantage of yielding more usual-looking formulas, e.g. by recovering at the lowest order the Peters-Mathews enhancement function. However it has the disadvantage of depending on the employed gauge. For some purposes it is better to use a pair of gauge invariant variables like $(E,J)$ or its variant $(\varepsilon,j)$. Another interesting choice for a pair of gauge-independent variables is $(x,k)$ or alternatively $(x,\iota)$ with $\iota=3x/k$ as was used in~\cite{ABIQ07}. Here we present the results in terms of the pair of parameters $(x,e_t)$; if necessary it is straightforward to transform the results into a different set of parameters like $(x,k)$ or $(\varepsilon,j)$.

We start with the time evolution of the mean motion $n=2\pi/P$. Since $n$ depends on the angular momentum $J$ only from the 2PN level [see~\eqref{n}], the angular momentum flux $\langle{\cal G}\rangle$ will only be needed in this case up to 1PN order (while the energy flux is needed with full 3PN accuracy). We present the result in the form
\begin{eqnarray}
\langle\frac{\ud n}{\ud t}\rangle_{\rm
inst}&=&\frac{c^6\,\nu}{G^2\,m^2}\,x^{11/2}\,\biggl[\mathcal{N}_{\rm
N}+x\,\mathcal{N}_{\rm 1PN}+x^2\,\mathcal{N}_{\rm 2PN}+x^3\,\mathcal{N}_{\rm
3PN}\biggr]\,,
\end{eqnarray}
where the PN coefficients explicitly read
\begin{subequations}
\begin{eqnarray} 
\mathcal{N}_{\rm N}&=&\frac{1}{\left(1-e_t^2\right)^{7/2}}\left\{\frac{96}{5}
+\frac{292}{5}e_t^2+\frac{37 }{5}e_t^4\right\}\,,\\ %%%%%%%%%%%%%%%%%%%%%%%%%%%%
\mathcal{N}_{\rm
1PN}&=&\frac{1}{\left(1-e_t^2\right)^{9/2}}\left\{ -\frac{4846}{35}
-\frac{264}{5}\nu+e_t^2\left(\frac{5001}{35}-570\nu\right)\right.\nonumber\\
&&+\left.e_t^4\left(\frac{2489}{4}-\frac{5061}{10}\nu\right)
+e_t^6\left(\frac{11717}{280}-\frac{148}{5}\nu\right)\right\}\,,\\
%%%%%%%%%%%%%%%%%%%%%%%%%%%%
\mathcal{N}_{\rm 2PN}&=&
\frac{1}{\left(1-e_t^2\right)^{11/2}}\left\{-\frac{1159}{945}
+\frac{15265}{21}\nu+\frac{944}{15}\nu^2\right.\nonumber\\
&&+e_t^2\left(-\frac{975868}{189}+\frac{10817}{5}\nu
+\frac{182387}{90}\nu^2\right)\nonumber\\
&&+e_t^4\left(-\frac{211193}{90}-\frac{955709}{140}\nu
+\frac{396443}{72}\nu^2\right)\nonumber\\
&&+e_t^6\left(\frac{4751023}{1680}-\frac{203131}{48}\nu
+\frac{192943}{90}\nu^2\right)\nonumber\\
&&+e_t^8\left(\frac{391457}{3360}-\frac{6037}{56}\nu
+\frac{2923}{45}\nu^2\right)\nonumber\\ 
&&+\sqrt{1-e_t^2}
\left[48-\frac{96}{5}\nu+e_t^2\left(2134-\frac{4268}{5}\nu\right)
+e_t^4\left(2193-\frac{4386}{5}\nu\right)\right.\nonumber\\
&&+\left.\left.e_t^6\left(\frac{175}{2}-35\nu\right)\right]\right\}\,,\\
%%%%%%%%%%%%%%%%%%%%%%%5
\mathcal{N}_{\rm
  3PN}&=&\frac{1}{\left(1-e_t^2\right)^{13/2}}\left\{\frac{4915859933}{1039500}
  +\left[\frac{1463719}{2268} 
- \frac{369 }{10}\pi ^2\right]\nu-\frac{711931}{420}\nu^2 -\frac{1121}{27}\nu^3\right.\nonumber\\
  &&+e_t^2\left(\frac{76740432133}{2079000} +\left[\frac{140649817}{3240} +
  \frac{24777 }{80}\pi ^2\right]\nu-\frac{17171137}{840}\nu^2
  -\frac{1287385}{324}\nu^3\right)\nonumber\\
  &&+e_t^4\left(-\frac{136998957827}{8316000} +\left[\frac{11146580197}{90720}
  - \frac{26887 }{160}\pi ^2\right]\nu+\frac{18119597}{3360}\nu^2
  -\frac{33769597}{1296}\nu^3\right)\nonumber\\
  &&+e_t^6\left(-\frac{88115571763}{5544000} +\left[\frac{4653403}{4032} -
  \frac{59093 }{160}\pi ^2\right]\nu+\frac{123833019}{2240}\nu^2
  -\frac{3200965}{108}\nu^3\right)\nonumber\\
  &&+e_t^8\left(\frac{59327921801}{7392000}+\left[-\frac{9471607}{672} -
  \frac{12177 }{640}\pi ^2\right]\nu+\frac{2260735}{168}\nu^2
  -\frac{982645}{162}\nu^3\right)\nonumber\\
  &&+e_t^{10}\left(\frac{33332681}{197120}-\frac{1874543}{10080}\nu
  +\frac{109733}{840}\nu^2 -\frac{8288}{81}\nu^3\right)\nonumber\\
  &&+\sqrt{1-e_t^2}\left[-\frac{2667319}{1125}+\left[\frac{56242}{105} -
  \frac{41 }{10}\pi ^2\right]\nu+\frac{632}{5}\nu^2\right.\nonumber\\
  &&+e_t^2\left(-\frac{2673296}{375}+\left[-\frac{10074037}{315} + \frac{45961
  }{240}\pi ^2\right]\nu+\frac{125278}{15}\nu^2\right)\nonumber\\
  &&+e_t^4\left(\frac{700397951}{21000}+\left[-\frac{4767517}{60} + \frac{6191
  }{32}\pi ^2\right]\nu+\frac{317273}{15}\nu^2\right)\nonumber\\
  &&+e_t^6\left(\frac{708573457}{31500}+\left[-\frac{6849319}{252} + \frac{287
  }{960}\pi ^2\right]\nu+\frac{232177}{30}\nu^2\right)\nonumber\\
  &&+\left.e_t^8\left(\frac{56403}{112}-\frac{427733}{840}\nu
  +\frac{4739}{30}\nu^2\right)\right]\nonumber\\
  &&+\left.\left(\frac{54784}{175} + \frac{465664 }{105}e_t^2 + \frac{4426376
  \ }{525}e_t^4 + \frac{1498856 }{525}e_t^6 + \frac{31779
  }{350}e_t^8\right)\right.\nonumber\\
&&\left.\times\ln\left[\frac{x}{x_0}\frac{1
  +\sqrt{1-e_t^2}}{2(1-e_t^2)}\right]\right\}\,.
\end{eqnarray}\end{subequations}
Next, the evolution of the periastron precession is
\begin{equation}
\langle \frac{\ud k}{\ud t}\rangle_{\rm inst} =
\frac{c^3\,\nu}{G\,m}\,x^4\,\biggl[x\,\mathcal{K}_{\rm
1PN}+x^2\,\mathcal{K}_{\rm 2PN}+x^3\,\mathcal{K}_{\rm 3PN}\biggr]
\end{equation}
(notice that the expansion starts at 1PN order), where
\begin{subequations}\begin{eqnarray}
\mathcal{K}_{\rm 1PN}&=&\frac{1}{(1-e_t^2)^{7/2}}\left\{
\frac{192}{5}+\frac{168 e_t^2}{5}\right\}\,,\\ \mathcal{K}_{\rm 2PN}&=&
\frac{1}{{(1-e_t^2)}^{9/2}}\left\{
\frac{9124}{35}-\frac{1424}{5}\nu+e_t^2\left(\frac{28512}{35}
-\frac{3804}{5}\nu\right)\right.\nonumber\\
&&+\left.e_t^4\left(\frac{10314}{35}-\frac{1017}{5}\nu\right)
\right\},\\ \mathcal{K}_{\rm
3PN}&=&\frac{1}{(1-e_t^2)^{11/2}}\left\{\frac{232082}{189}+\left[-\frac{131366}{21}
+ \frac{738 }{5}\pi ^2\right]\nu+\frac{13312}{15}\nu^2\right.\nonumber\\
&&+e_t^2\left(\frac{2842199}{630}+\left[-\frac{1659934}{105} 
+ \frac{1271 }{10}\pi ^2\right]\nu+\frac{29879}{5}\nu^2\right)\nonumber\\
&&+e_t^4\left(\frac{1640713}{252}+\left[-\frac{1304524}{105} 
+ \frac{5371 }{320}\pi ^2\right]\nu+\frac{54133}{10}\nu^2\right)\nonumber\\
&&+e_t^6\left(\frac{1850407}{1680}-\frac{388799}{280}\nu
+\frac{19573}{30}\nu^2\right)\nonumber\\
&&+\left.\sqrt{1-e_t^2}\left[672-\frac{1344 }{5}\nu +\left(2436-\frac{4872
}{5}\nu \right) e_t^2+\left(672-\frac{1344 }{5}\nu \right)
e_t^4\right]\right\}\,.
\end{eqnarray}\end{subequations}
Note that there is no dependence on the scale $x_0$ in this expression; the reason is that the 3PN coefficient is actually 2PN relatively to the dominant term. From these above results we immediately deduce the time evolution of the mean orbital frequency $\omega=K n$ as
\begin{equation}
\langle\frac{\ud \omega}{\ud t}\rangle_{\rm
inst}=\frac{c^6\nu}{G^2\,m^2}\,x^{11/2}\,\biggl[\mathcal{O}_{\rm
N}+x\,\mathcal{O}_{\rm 1PN}+x^2\,\mathcal{O}_{\rm 2PN}+x^3\,\mathcal{O}_{\rm
3PN}\biggr]\,,
\end{equation}
where
\begin{subequations}\begin{eqnarray} 
\mathcal{O}_{\rm N}&=&\frac{1}{\left(1-e_t^2\right)^{7/2}}\left\{\frac{96}{5}
+\frac{292}{5}e_t^2+\frac{37 }{5}e_t^4\right\}\,,\\ 
%%%%%%%%%%%%%%%%%%%%%%%%%%%%
\mathcal{O}_{\rm 1PN}&=& \frac{1}{\left(1-e_t^2\right)^{9/2}}\Biggl\{-\frac{1486}{35}-\frac{264}{5}\nu +e_t^2\left(\frac{2193}{7}-570\nu \right)+e_t^4\left(\frac{12217}{20}-\frac{5061}{10}\nu \right)\nonumber\\
&&+e_t^6\left(\frac{11717}{280}-\frac{148}{5}\nu \right)\Biggr\}\\
%%%%%%%%%%%%%%%%%%%%%%%%%%%%
\mathcal{O}_{\rm 2PN}&=& \frac{1}{\left(1-e_t^2\right)^{11/2}}\Biggl\{-\frac{11257}{945}+\frac{15677}{105}\nu +\frac{944}{15}\nu ^2\nonumber\\
&&+e_t^2\left(-\frac{2960801}{945}-\frac{2781}{5}\nu+\frac{182387}{90}\nu^2\right)\nonumber\\
&&+e_t^4\left(-\frac{68647}{1260}-\frac{1150631}{140}\nu +\frac{396443}{72}\nu ^2\right)\nonumber\\
&&+e_t^6\left(\frac{925073}{336}-\frac{199939}{48}\nu +\frac{192943}{90}\nu^2\right)\nonumber\\
&&+e_t^8\left(\frac{391457}{3360}-\frac{6037}{56}\nu
+\frac{2923}{45}\nu^2\right)\nonumber\\ 
&&+\sqrt{1-e_t^2}\left[\left(48-\frac{96}{5}\nu\right)+e_t^2\left(2134-\frac{4268}{5}\nu \right)+e_t^4\left(2193-\frac{4386}{5}\nu\right)\right.\nonumber\\
&&+\left.e_t^6\left(\frac{175}{2}-35\nu\right)\right]\Biggr\}\\
%%%%%%%%%%%%%%%%%%%%%%%
\mathcal{O}_{\rm 3PN}&=&\frac{1}{\left(1-e_t^2\right)^{13/2}}\Biggl\{\frac{614389219}{148500}+\left[-\frac{57265081}{11340} + \frac{369 }{2}\pi ^2\right]\nu-\frac{16073}{140}\nu ^2-\frac{1121}{27}\nu ^3\nonumber\\
&&+e_t^2\left(\frac{19769277811}{693000}+\left[\frac{66358561}{3240} + \frac{42571 }{80}\pi ^2\right]\nu-\frac{3161701}{840}\nu ^2-\frac{1287385}{324}\nu ^3\right)\nonumber\\
&&+e_t^4\left(-\frac{3983966927}{8316000}+\left[\frac{6451690597}{90720} - \frac{12403 }{64}\pi ^2\right]\nu+\frac{34877019}{1120}\nu ^2-\frac{33769597}{1296}\nu ^3\right)\nonumber\\
&&+e_t^6\left(-\frac{4548320963}{5544000}+\left[-\frac{59823689}{4032} - \frac{242563 }{640}\pi ^2\right]\nu+\frac{411401857}{6720}\nu ^2-\frac{3200965}{108}\nu ^3\right)\nonumber\\
&&+e_t^8\left(\frac{19593451667}{2464000}+\left[-\frac{6614711}{480} - \frac{12177 }{640}\pi ^2\right]\nu+\frac{92762}{7}\nu^2-\frac{982645}{162}\nu ^3\right)\nonumber\\
&&+e_t^{10}\left(\frac{33332681}{197120}-\frac{1874543}{10080}\nu +\frac{109733}{840}\nu ^2-\frac{8288}{81}\nu ^3\right)\nonumber\\
&&+\sqrt{1-e_t^2}\left[-\frac{1425319}{1125}+\left[\frac{9874}{105} 
- \frac{41} {10}\pi ^2\right]\nu+\frac{632}{5}\nu^2\right.\nonumber\\
&&+e_t^2\left(\frac{933454}{375}+\left[-\frac{2257181}{63} 
+ \frac{45961 }{240}\pi ^2\right]\nu+\frac{125278}{15}\nu ^2\right)\nonumber\\
&&+e_t^4\left(\frac{840635951}{21000}+\left[-\frac{4927789}{60} 
+ \frac{6191 }{32}\pi ^2\right]\nu+\frac{317273}{15}\nu^2\right)\nonumber\\
&&+e_t^6\left(\frac{702667207}{31500}+\left[-\frac{6830419}{252} 
+ \frac{287 }{960}\pi ^2\right]\nu+\frac{232177}{30}\nu^2\right)\nonumber\\
&&+\left.e_t^8\left(\frac{56403}{112}-\frac{427733}{840}\nu+\frac{4739}{30}\nu^2\right)\right]\nonumber\\
&&+\left(\frac{54784}{175} + \frac{465664 }{105}e_t^2 + \frac{4426376 }{525}e_t^4 + \frac{1498856 }{525}e_t^6 + \frac{31779
  }{350}e_t^8\right)\nonumber\\
&&\times\ln\left[\frac{x}{x_0}\frac{1
  +\sqrt{1-e_t^2}}{2(1-e_t^2)}\right]\Biggr\}\,.
\end{eqnarray}\end{subequations}
For the semi-major axis $a_r$, again we need the 3PN energy flux but only the 1PN angular momentum flux, and get
\begin{eqnarray}
\langle\frac{\ud a_r}{\ud t}\rangle_{\rm inst}&=&\nu\,c\,x^3\,\biggl[
\mathcal{A}_{\rm N}+x\,\mathcal{A}_{\rm 1PN}+x^2\,\mathcal{A}_{\rm
2PN}+x^3\,\mathcal{A}_{\rm 3PN}\biggr]\,,
\end{eqnarray}
with
\begin{subequations}\begin{eqnarray}
\mathcal{A}_{\rm N}&=&\frac{1}{\left(1-e_t^2\right)^{7/2}}\left\{
-\frac{64}{5}-\frac{584}{15}e_t^2-\frac{74}{15}e_t^4\right\}\,,\\
%%%%%%%%%%%%%%%%%%%%%%%%%%%%%%%%%%%%%%%%%%%%%
\mathcal{A}_{\rm 1PN}&=&\frac{1}{\left(1-e_t^2\right)^{9/2}}
\left\{\frac{2972}{105}+\frac{176}{5}\nu+e_t^2\left(-\frac{30442}{105}
+380\nu\right)\right.\nonumber\\
&&+\left.e_t^4\left(-\frac{879}{2}+\frac{1687}{5}\nu\right)+e_t^6\left(
-\frac{11717}{420}+\frac{296}{15}\nu\right)\right\}\,,\\
%%%%%%%%%%%%%%%%%%%%%%%%%%%%%%%%%%%%%%%%%%%%%
\mathcal{A}_{\rm 2PN}&=&\,\frac{1}{(1-e_t^2)^{11/2}}\,
\left\{\frac{194882}{2835}-\frac{34882}{315}\nu-\frac{608}{15}\nu^2\right.\nonumber
\\&&+e_t^2\left(\frac{4813252}{2835}+\frac{46121}{45}\nu
-\frac{20243}{15}\nu^2\right)\nonumber\\
&&+e_t^4\left(-\frac{402869}{270}+\frac{1670329}{252}\nu
-\frac{73549}{20}\nu^2\right)\nonumber\\
&&+e_t^6\left(-\frac{5413417}{2520}+\frac{1092403}{360}\nu
-\frac{64169}{45}\nu^2\right)\nonumber\\
&&+e_t^8\left(-\frac{366593}{5040}+\frac{9703}{126}\nu
-\frac{1924}{45}\nu^2\right)\nonumber\\ &&+\sqrt{1-e_t^2}
\left[-96+\frac{192 \,\nu }{5}+e_t^2\,\left(-1452+\frac{2904}{5}\nu\right)
  +e_t^4\left(-1353+\frac{2706}{5}\nu\right)\right.\nonumber\\
&&\left.\left.+e_t^6\,\left( -74
  +\frac{148}{5}\nu\right)\right]\right\}\,,\\
%%%%%%%%%%%%%%%%%%%%%%%%%%%%%%%%%%%%%%%%%%% 
\mathcal{A}_{\rm
  3PN}&=&\frac{1}{\left(1-e_t^2\right)^{13/2}}\left\{
-\frac{4121173183}{1559250}+\left[\frac{3347773}{1701} 
- \frac{449 }{5}\pi ^2\right]\nu+\frac{285629}{1890}\nu^2 +\frac{122}{5}\nu^3\right.\nonumber\\
&&+e_t^2\,\left(-\frac{17602259111}{1039500}+\left[-\frac{81401371}{8505} -
  \frac{17607 }{40}\pi ^2\right]\nu+\frac{720193}{3780}\nu^2
+\frac{78437}{30}\nu^3\right)\nonumber\\
&&+e_t^4\left(-\frac{63269331823}{12474000}+\left[-\frac{38423549}{1701} +
  \frac{20609 }{480}\pi ^2\right]\nu-\frac{95502193}{3024}\nu^2
+\frac{2089273}{120}\nu^3\right)\nonumber\\
&&+e_t^6\left(-\frac{98870259137}{8316000}+\left[ \frac{111259109}{3780} +
  \frac{67053 }{320}\pi ^2\right]\nu-\frac{491278049}{10080}\nu^2
+\frac{1781461}{90}\nu^3\right)\nonumber\\
&&+e_t^8\left(-\frac{72878500601}{11088000}+\left[\frac{6292747}{560} +
  \frac{12271 }{960}\pi ^2\right]\nu-\frac{25030639}{2520}\nu^2
+\frac{180428}{45}\nu^3\right)\nonumber\\
&&+e_t^{10}\left(-\frac{81086491}{887040}+\frac{109847}{864}\nu
-\frac{66209}{630}\nu^2 +\frac{592}{9}\nu^3\right)\nonumber\\&+&\sqrt{1-e_t^2}
\left[\frac{17249966}{23625}+\left[\frac{21500}{21} 
- \frac{41 }{5}\pi ^2\right]\nu-240\nu^2\right.\nonumber\\
  &&+e_t^2\left(-\frac{39170981}{7875}+\left[\frac{2741416}{105} - \frac{4961
      }{40}\pi ^2\right]\nu-\frac{29964}{5}\nu^2\right)\nonumber\\
  &&+e_t^4\left(-\frac{975995201}{31500}+\left[\frac{5741363}{105} -
    \frac{18491 }{160}\pi ^2\right]\nu-\frac{68913}{5}\nu^2\right)\nonumber\\
  &&+e_t^6\left(-\frac{2779019203}{189000}+\left[\frac{22404341}{1260} -
    \frac{1517 }{240}\pi ^2\right]\nu-\frac{24347}{5}\nu^2\right)\nonumber\\
  &&+\left.e_t^8\left(-\frac{17933}{42}+\frac{47459}{105}\nu
  -\frac{666}{5}\nu^2\right)\right]\nonumber\\
&&-\left(\frac{109568}{525}+\frac{931328 }{315}e_t^2+\frac{8852752}{1575}
e_t^4+\ \frac{2997712 }{1575}e_t^6+\frac{10593 }{175}e_t^8\right)\nonumber\\
&&\times\,\ln\left[\frac{x}{x_0}\frac{1+\sqrt{1-e_t^2}}{2(1-e_t^2)}\right]
\Biggr\}\,.
\end{eqnarray}\end{subequations}
The evolution of the eccentricity $e_t$ is the only one to require both the energy flux $\langle{\cal F}\rangle$ and angular momentum flux $\langle{\cal G}\rangle$ with full 3PN accuracy. We obtain
\begin{equation}\label{evolecc}
\langle\frac{\ud e_t}{\ud t}\rangle_{\rm
inst}=-\frac{c^3}{G\,m}\,e_t \nu\,x^4\,\biggl[ \mathcal{E}_{\rm
N}+x\,\mathcal{E}_{\rm 1PN}+x^2\,\mathcal{E}_{\rm 2PN}+x^3\,\mathcal{E}_{\rm
3PN}\biggr]\,,
\end{equation}
where
\begin{subequations}\label{evoleccPN}
\begin{eqnarray}
\mathcal{E}_{\rm
N}&=&\frac{1}{\left(1-e_t^2\right)^{5/2}}\left\{\frac{304}{15}+\frac{121
e_t^2}{15}\right\}\,,\\
%%%%%%%%%%%%%%%%%%%%%%%%%%%%%%%%%%%%%%%
\mathcal{E}_{\rm
1PN}&=&\frac{1}{\left(1-e_t^2\right)^{7/2}}
\left\{-\frac{939}{35}-\frac{4084}{45}\nu+e_t^2\,\left(\frac{29917}{105}
-\frac{7753}{30}\nu\right)\right.\nonumber\\&&+\left.e_t^4\,\left(\frac{13929}{280}
-\frac{1664}{45}\nu\right)\right\}\,,\\ 
%%%%%%%%%%%%%%%%%%%%%%%%%%%%%%%%%%%%%%%%%%
\mathcal{E}_{\rm
2PN}&=&\frac{1}{\left(1-e_t^2\right)^{9/2}}\left\{-\frac{961973}{1890}
+\frac{70967}{210}\nu+\frac{752}{5}\nu^2+e_t^2\left(-\frac{3180307}{2520}
-\frac{1541059}{840}\nu+\frac{64433}{40}\nu^2\right)\right.\nonumber\\
&&+e_t^4\,\left(\frac{23222071}{15120}-\frac{13402843}{5040}\nu
+\frac{127411}{90}\nu^2\right)
+e_t^6\,\left(\frac{420727}{3360}-\frac{362071}{2520}\nu
+\frac{821}{9}\nu^2\right)\nonumber\\ &&+\left.\sqrt{1-e_t^2}
\left[\frac{1336}{3}-\frac{2672}{15}\nu+e_t^2\,\left(\frac{2321}{2}
-\frac{2321}{5}\nu\right)
+e_t^4\left(\frac{565}{6}-\frac{113}{3}\nu\right)\right]\right\}\,,\\
%%%%%%%%%%%%%%%%%%%%%%%%%%%%%%%%%%%%%%%%%
\mathcal{E}_{\rm 3PN}&=&\frac{1}{\left(1-e_t^2\right)^{11/2}}\biggl\{\frac{54177075619}{6237000}
+\left[\frac{7198067}{22680} 
+ \frac{1283 }{10}\pi ^2\right]\nu-\frac{3000281}{2520}\nu^2-\frac{61001}{486}\nu^3\nonumber\\
&&+e_t^2\,\left(\frac{6346360709}{891000}+\left[\frac{9569213}{360} +
\frac{54001 }{960}\pi ^2\right]\nu+\frac{12478601}{15120}\nu^2-\frac{86910509}{19440}
\nu^3\right)\nonumber\\
&&+e_t^4\,\left(-\frac{126288160777}{16632000}+\left[\frac{418129451}{181440}
- \frac{254903 }{1920}\pi ^2\right]\nu+\frac{478808759}{20160}\nu^2-\frac{2223241}{180}\nu^3
\right)\nonumber\\
&&+e_t^6\,\left(\frac{5845342193}{1232000}+\left[-\frac{98425673}{10080} -
\frac{6519 }{640}\pi ^2\right]\nu+\frac{6538757}{630}\nu^2
-\frac{11792069}{2430}\nu^3\right)\nonumber\\
&&+e_t^8\left(\frac{302322169}{1774080}-\frac{1921387}{10080}\nu
+\frac{41179}{216}\nu^2 -\frac{193396}{1215}\nu^3\right)\nonumber\\&&
+\sqrt{1-e_t^2} \left[-\frac{22713049}{15750}+\left[-\frac{5526991}{945} +
\frac{8323 }{180}\pi ^2\right]\nu+\frac{54332}{45}\nu^2\right.\nonumber\\
&&+e_t^2\left(\frac{89395687}{7875}+\left[-\frac{38295557}{1260} + \frac{94177
}{960}\pi ^2\right]\nu+\frac{681989}{90}\nu^2\right)\nonumber\\
&&+e_t^4\left(\frac{5321445613}{378000}+\left[-\frac{26478311}{1512} +
\frac{2501 }{2880}\pi ^2\right]\nu+\frac{225106}{45}\nu^2\right)\nonumber\\
&&+e_t^6\left(\frac{186961}{336}-\frac{289691}{504}\nu
+\frac{3197}{18}\nu^2\right)\Biggr]+\frac{730168
}{23625}\,\frac{1}{1+\sqrt{1-e_t^2}}\nonumber\\ &&+{304\over 15}
\left(\frac{82283}{1995}
+ \frac{297674 }{1995}e_t^2 + \frac{1147147 \ }{15960}e_t^4 + \frac{61311
}{21280}e_t^6\right)\,\ln\left[\frac{x}{x_0}\frac{1
+\sqrt{1-e_t^2}}{2(1-e_t^2)}\right]\Biggr\}\,.
\end{eqnarray}\end{subequations}
The leading order Newtonian term is in agreement with the work of Peters~\cite{Pe64}. The match to earlier results for the evolution of orbital elements includes also 1PN~\cite{BS89} and 2PN~\cite{GI97} orders.
The explicit expressions for the evolution of the orbital elements in modified
harmonic coordinates are provided in Appendix \ref{appC}.

\subsection{Hereditary contributions}\label{sechered}
The hereditary contribution to the flux begins at relative 1.5PN order and consequently the 1PN quasi-Keplerian representation, given by the truncation of Eqs.~\eqref{3PNQK} at the 1PN order, suffices for this analysis. Namely,
\begin{subequations}\label{1PNQK}\begin{align}
n =&\frac{c^3}{G\,m}\,\varepsilon^{3/2} \biggl[1+\frac{\varepsilon}{8}\, (
  -15+\nu ) \biggr] \,,\label{nN}\\
K =& 1 + 3\frac{\varepsilon}{j}\,, \\
\w =& \frac{c^3}{G\,m}\,\varepsilon^{3/2} \biggl[1+\frac{\varepsilon}{8}\, \Bigl(
  -15+\nu +{24\over j}\Bigr) \biggr] \,,\\
e_t =&\sqrt{1-j}\,\biggl[1+ \frac{\varepsilon}{8(1-j)}\, \Bigl(-8+8\,\nu +j (
  17-7\,\nu ) \Bigr)\biggr] \,,\\ a_r =& \frac{G\,m}{c^2}\,\frac{1}{\varepsilon}\biggl\{
  1+\frac{\varepsilon}{4} ( -7+\nu ) \biggr\}\,.
\end{align}\end{subequations}
We also need the 1PN accurate expressions for $\varepsilon$ and $j$, which we express in terms of $e_t$ and $x$:
\begin{subequations}\label{EJ}
\begin{align} 
\varepsilon=&x\,\left[1 +x\,\left({5\over4}-{\nu\over12}-{2\over
1-e_t^2}\right)\right]\,,\\
j=& (1-e_t^2)\,\left[1+\frac{x}{4}\left(17-7\nu-8{1-\nu\over
1-e_t^2}\right)\right]\,.
\end{align}\end{subequations} 

The results for the hereditary parts $\langle{\cal F}_{\rm hered}\rangle$ and $\langle{\cal G}_{\rm hered}\rangle$ depended on some numerically-computed enhancements functions of the eccentricity: $\varphi$, $\psi$, $\zeta$, $\dots$ in the energy flux, and $\tilde{\varphi}$, $\tilde{\psi}$, $\tilde{\zeta}$, $\dots$ in the angular momentum flux. Obviously the evolution of orbital elements will involve some linear combinations of $\varphi$, $\psi$, $\dots$ and their tilde analogues [see e.g.~\eqref{evoleq}]. We thus have to introduce some new enhancement functions to parametrize the time evolutions of the various orbital elements $\{n, k, \omega, a_r, e_t\}$. We pose\footnote{To avoid heavy notation and since we deal with only the time eccentricity $e_t$, the corresponding enhancement functions $\varphi_{e}$, $\psi_{e}$, $\zeta_{e}$, $\kappa_{e}$ and $F_{e}$ are labelled by $e$ rather than $e_t$.}
\begin{subequations}\label{enhancenk}
\begin{eqnarray}
\psi_{n}(e)&=&
{1344\over17\,599}\,{7-5e^2\over1-e^2}\,\varphi(e)
+{8191\over17\,599}\,\psi(e)\,,\\
\zeta_{n}(e)&=&
{583\over567}\,\zeta(e)-{16\over567}\,\varphi(e)\,,\\
\varphi_{k}(e)&=&
{\tilde{\varphi}(e)\over(1-e^2)^{3/2}}\,,\\
\psi_{\w}(e)&=&
{1344\over4159}{1\over(1-e^2)^{3/2}}\,\left[\sqrt{1-e^2}\,\biggl(1-5e^2\biggr)\,\varphi(e)-4\,\tilde{\varphi}(e)\right]
+{8191\over4159}\,\psi(e)\,,\\
\zeta_{\w}(e)&=&
{583\over567}\,\zeta(e)-{16\over567}\,\varphi(e)\,,\\
\psi_{a}(e)&=& -
{1344\over4159}\,{3+5e^2\over1-e^2}\,\varphi(e)+{8191\over4159}\,\psi(e) \,,\\
\zeta_{a}(e)&=&
 {583\over567}\,\zeta(e)-{16\over567}\,\varphi(e)\,,\\
\varphi_{e}(e)&=&
 {192\over985}{\sqrt{1-e^2}\over e^2}\,\biggl[\sqrt{1-e^2}\,\varphi(e)-\tilde{\varphi}(e)\biggr]\,,\\
\psi_{e}(e)&=&
\frac{18\,816}{55\,691}{1\over
e^2\sqrt{1-e^2}}\biggl[\sqrt{1-e^2}\,\biggl(1-{11\over7}\,e^2\biggr)\,\varphi(e)
-\biggl(1-{3\over7}\,e^2\biggr)\,\tilde{\varphi}(e)\biggr]\nonumber\\ &+&
{16\,382\over55\,691}{\sqrt{1-e^2}\over
e^2}\,\biggl[\sqrt{1-e^2}\,\psi(e)-\tilde{\psi}(e)\biggr]\,,\\
 \zeta_{e}(e)&=&
{924\over19\,067}{1\over e^2\sqrt{1-e^2}}\left[-(1-e^2)^{3/2}\,\varphi(e)
+\biggl(1-{5\over11}\,e^2\biggr)\,\tilde{\varphi}(e)\right]\nonumber\\
&+&{12\,243\over76\,268}{\sqrt{1-e^2}\over
e^2}\,\biggl[\sqrt{1-e^2}\,\zeta(e)-\tilde{\zeta}(e)\biggr]\,,\\
\kappa_{e}(e)&=&
{\sqrt{1-e^2}\over
e^2}\,\biggl[\sqrt{1-e^2}\,\kappa(e)-\tilde{\kappa}(e)\biggr]\left(\frac{769}{96}
-\frac{3\,059\,665}{700\,566}\ln 2+\frac{8\,190\,315}{1\,868\,176}\ln 3\right)^{-1}\!\!.
\end{eqnarray}\end{subequations} 
Note that $\zeta_{n}(e)=\zeta_{a}(e)=\zeta_{\w}(e)$. We have also a function known analytically,
\begin{eqnarray}\label{Fee}
F_{e}(e)&=&\frac{96}{769}{\sqrt{1-e^2}\over
e^2}\,\biggl[\sqrt{1-e^2}\,F(e)-\tilde{F}(e)\biggr]={1+{2782\over769}\,e^2
+{10\,721\over6152}\,e^4+{1719\over24\,608}\,e^6\over (1-e^2)^{11/2}}\,.
\end{eqnarray}
All these new functions of eccentricity reduce to one in the circular orbit limit $e\rightarrow0$.

We now list the relevant 1PN accurate expressions for the hereditary part of the evolution of the orbital elements, as computed from the hereditary energy and angular momentum fluxes~\eqref{Ftailfinal} and~\eqref{Gtailfinal}. However we notice that, for what concerns the hereditary part, the evolutions of $n$ and $a_r$ depend only of the energy flux~\eqref{Ftailfinal} up to the accuracy we need (which is 1.5PN relative order); similarly the evolution of $k$ depends only on the angular momentum flux. We find
\begin{subequations}\label{ar-n-evo}
\begin{eqnarray}
\langle {\ud n\over \ud t}\rangle_{\rm hered}&=& {96\over5}{c^6\over G^2}{\nu \over
m^2}\,x^{11/2}\Biggl\{4\pi\,x^{3/2}\,\varphi(e_t)+\pi\,x^{5/2}
\biggl[-{17599\over672}\,\psi_{n}(e_t)
-{189\over8}\,\nu\,\zeta_{n}(e_t)\biggr] \nonumber\\
&+&x^3\left(-\frac{116\,761}{3675}\,\kappa(e_t) +\left[ \frac{16}{3} \,\pi^2
-\frac{1712}{105}\,C
-\frac{1712}{105}\ln\left(\frac{4\omega\,r_0}{c}\right)\right]F(e_t)\right)\Biggr\}\,,\\
\langle {\ud k\over \ud t}\rangle_{\rm hered}&=& {192\over5}\,{c^3\over G}\,{\nu\over
m}\,x^5\Biggl\{4\pi\,x^{3/2}\,\varphi_{k}(e_t)\Biggr\},\\
\langle{\ud \w \over \ud t}\rangle_{\rm hered}&=& {96\over5}{c^6\over G^2}{\nu \over
m^2}\,x^{11/2}\Biggl\{4\pi\,x^{3/2}\,\varphi(e_t)+\pi\,x^{5/2}
\biggl[-{4159\over672}\,\psi_{\w}(e_t)
-{189\over8}\,\nu\,\zeta_{\w}(e_t)\biggr] \nonumber\\
&+&x^3\left(-\frac{116\,761}{3675}\,\kappa(e_t) +\left[ \frac{16}{3} \,\pi^2
-\frac{1712}{105}\,C
-\frac{1712}{105}\ln\left(\frac{4\omega\,r_0}{c}\right)\right]F(e_t)\right)\Biggr\}\,,\\
\langle {\ud a_r\over \ud t}\rangle_{\rm hered}&=&
{64\over5}\nu\,c\,x^3\Biggl\{-4\pi\,x^{3/2}\,
\varphi(e_t)+\pi\,x^{5/2}\biggl[{4159\over672}\,\psi_{a}(e_t)
+{189\over8}\,\nu\,\zeta_{a}(e_t)\biggr]\nonumber\\
&+&x^3\left(\frac{116\,761}{3675}\,\kappa(e_t) -\left[ \frac{16}{3} \,\pi^2
-\frac{1712}{105}\,C
-\frac{1712}{105}\ln\left(\frac{4\omega\,r_0}{c}\right)\right]F(e_t)\right)\Biggr\}\,.
\end{eqnarray}\end{subequations} 
The hereditary part of the evolution of $e_t$ depends on both the energy and angular momentum fluxes at 1.5PN order, and we find
\begin{eqnarray}\label{et-evo}
\langle {\ud e_t\over \ud t}\rangle_{\rm hered}&=&
{32\over5}{c^3 \over G}\,e_{t}\,{\nu\over
m}\,x^4\Biggl\{-\frac{985}{48}\pi\,x^{3/2}\,\varphi_{e}(e_t)+\pi\,x^{5/2}\biggl[
\frac{55691}{1344}\psi_{e}(e_t)+\frac{19067}{126}\nu\,\zeta_{e}(e_t)\biggr]\nonumber
\\ &+&x^3\left(\left[\frac{89\,789\,209}{352\,800}
-\frac{87\,419}{630}\ln2+\frac{78\,003}{560}\ln3\right]
\kappa_{e}(e_t)\right.\nonumber\\&&\qquad\left.-\frac{769}{96}\left[
\frac{16}{3} \,\pi^2 -\frac{1712}{105}\,C
-\frac{1712}{105}\ln\left(\frac{4\omega\,r_0}{c}\right)\right]F_{e}(e_t)\right)\Biggr\}\,.
\end{eqnarray}
All the numerical functions of eccentricity are provided in Figs.~\ref{fig4}--\ref{fig10}. Numerical tables are given in Appendix~\ref{appB} (see Tables I-IV).
\begin{figure}[t]
\vspace{0.5cm}
\centering 
\includegraphics[width=3.2in]{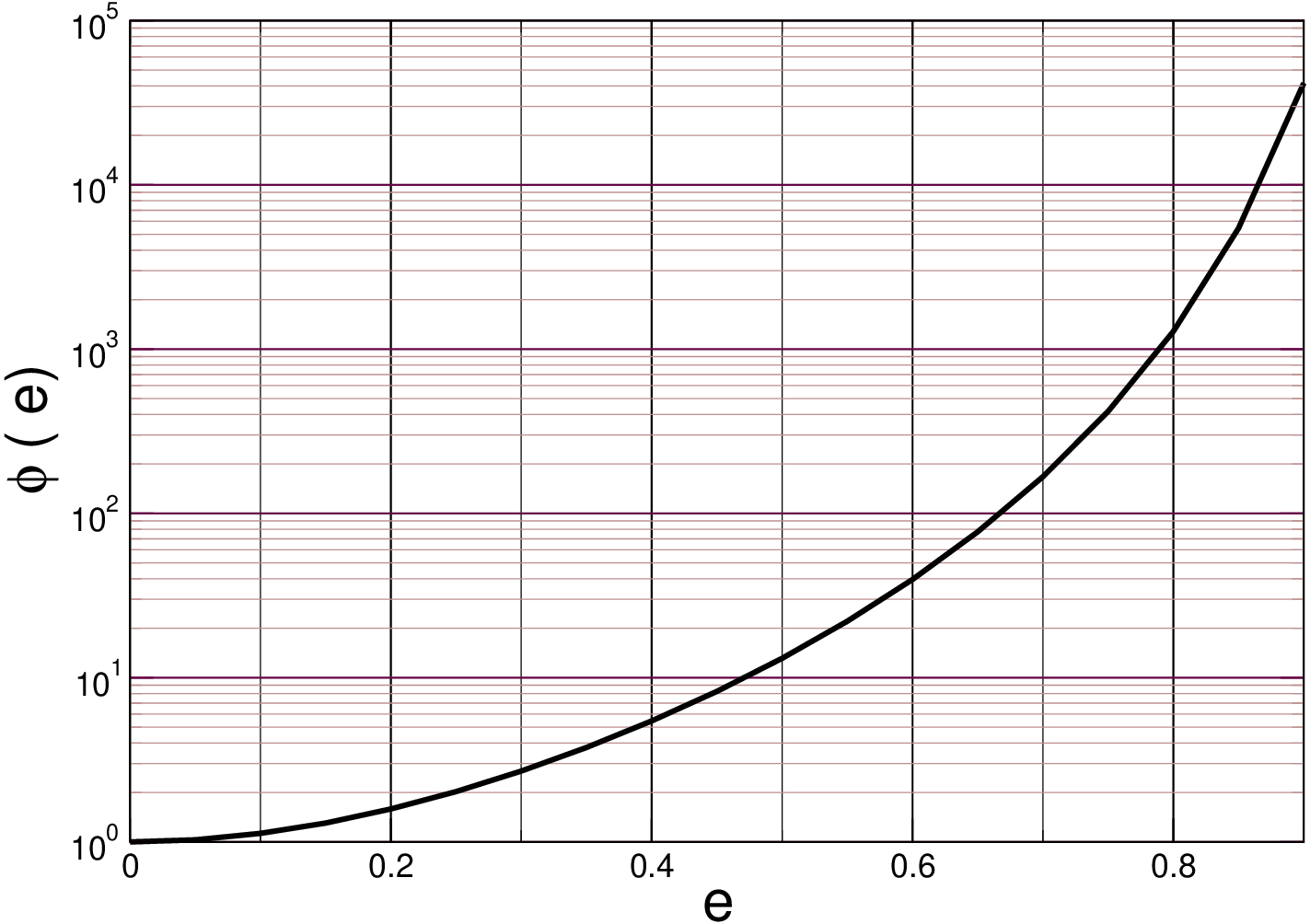}
\hskip 0.092 true cm 
\includegraphics[width=3.2in]{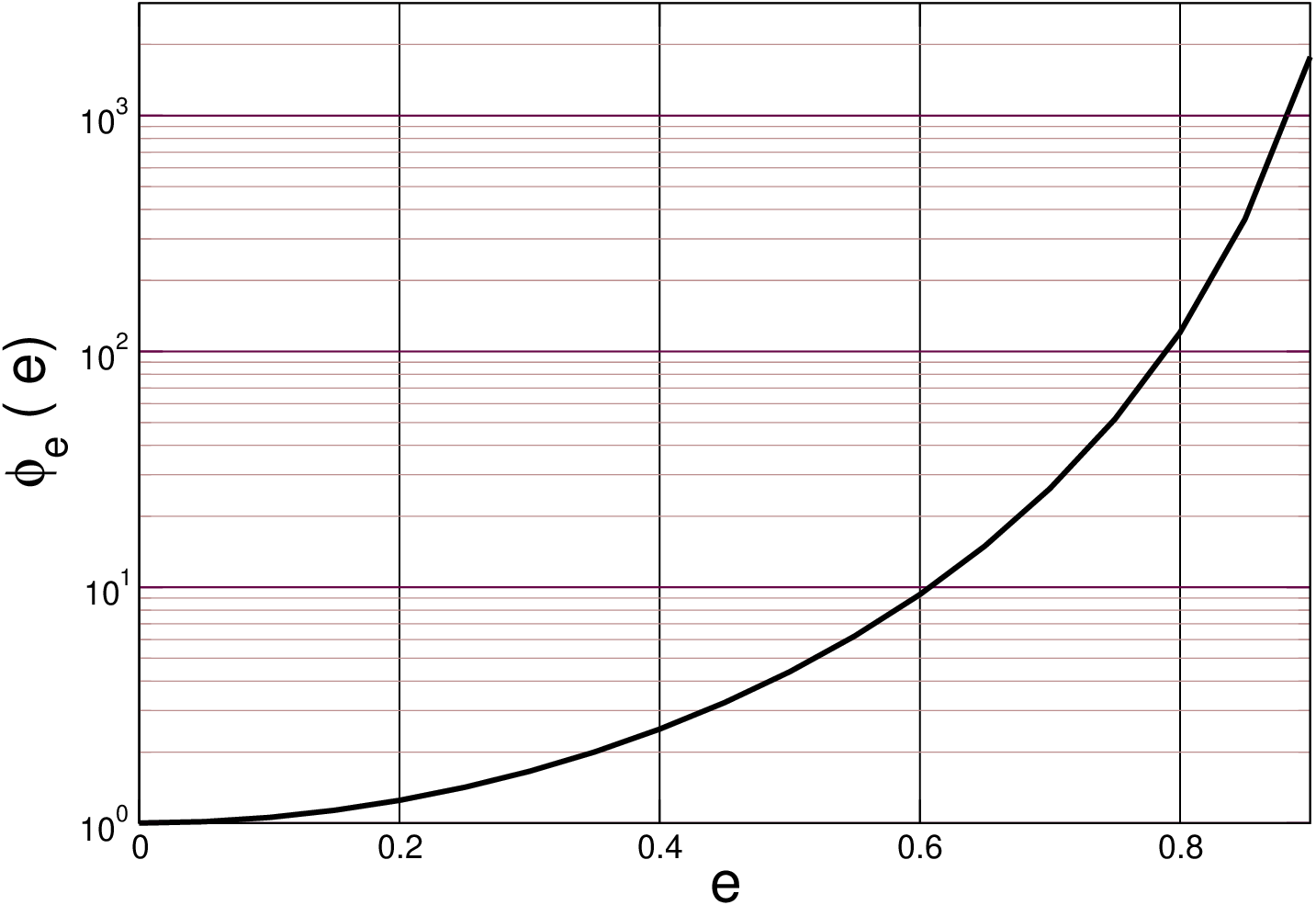}
\caption{Enhancement functions $\varphi(e)$ and $\varphi_{e}(e)$ in the evolutions of $n$ and  $e_t$ at 1.5PN order.}
\label{fig4}
\end{figure}
\begin{figure}[t]
\vspace{0.5cm}
\centering 
\includegraphics[width=3.2in]{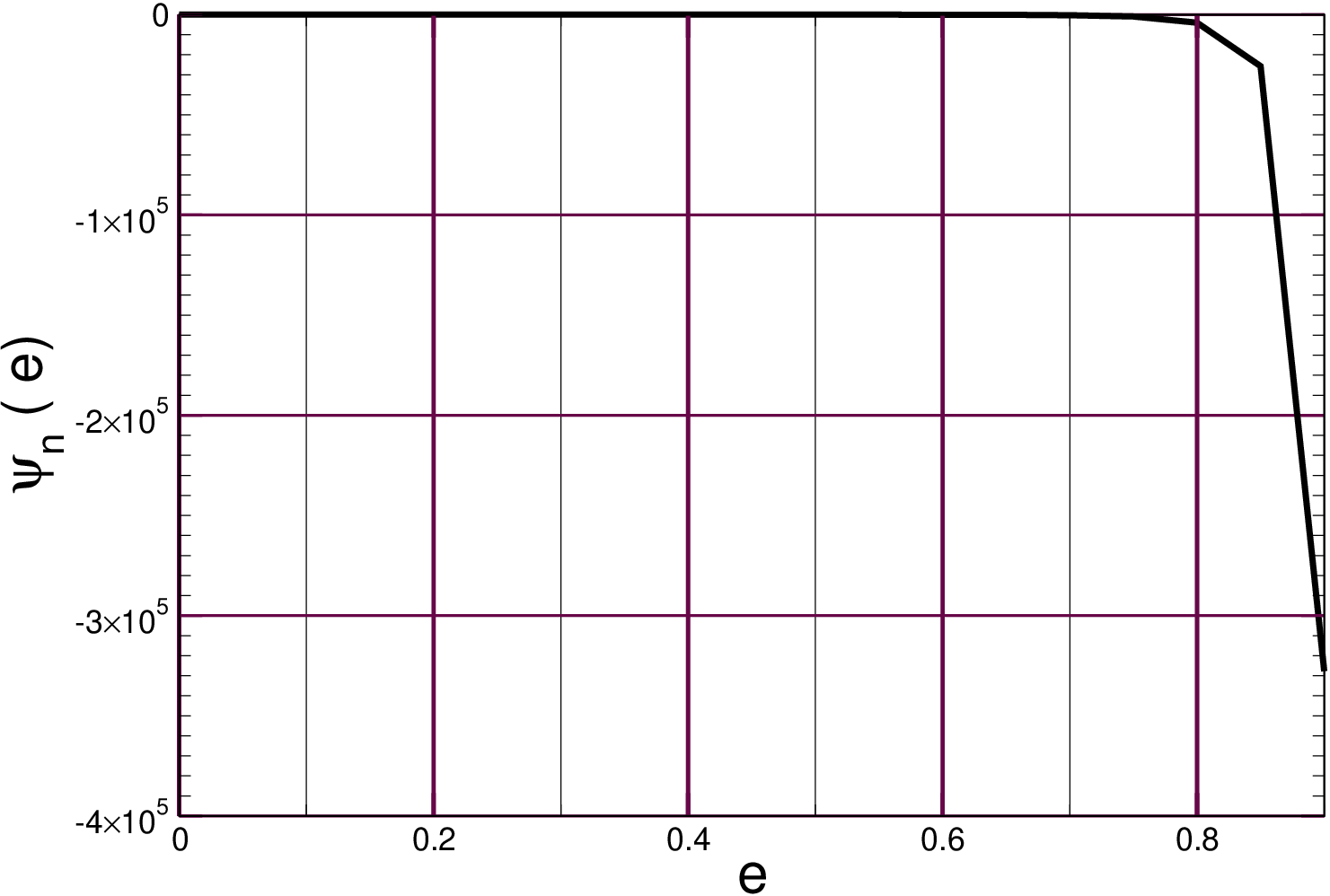}
\hskip 0.2 true cm 
\includegraphics[width=3.1in]{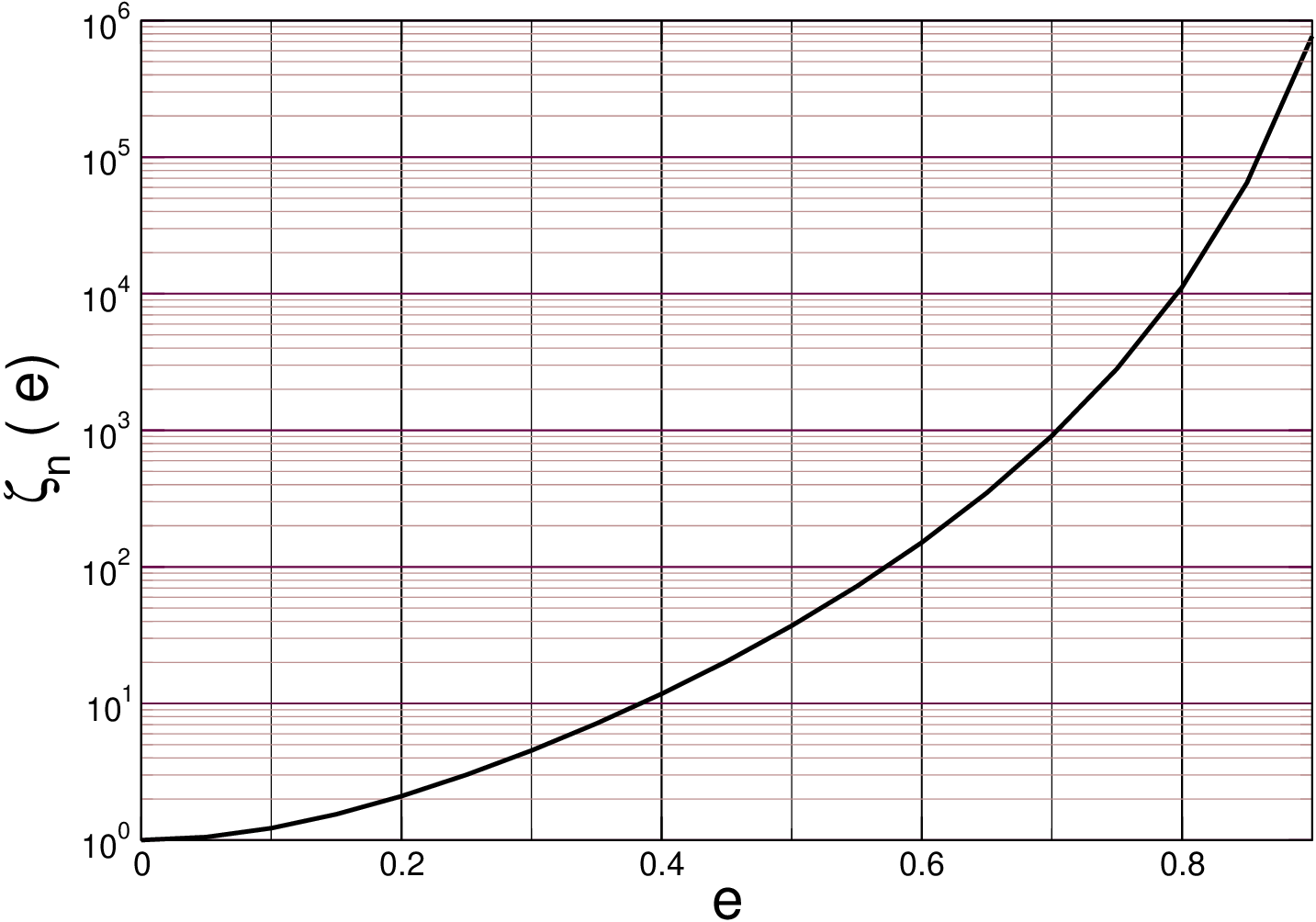} 
\caption{Enhancement functions $\psi_{n}(e)$ and $\zeta_{n}(e)$ in the evolution of $n$ at 2.5PN order. Note the similarity of $\psi_{n}(e)$ with $\tilde{\psi}(e)$ in Fig.~\ref{fig2}.}
\label{fig5}
\end{figure}
\begin{figure}[t]
\vspace{0.5cm}
\centering 
\includegraphics[width=3.2in]{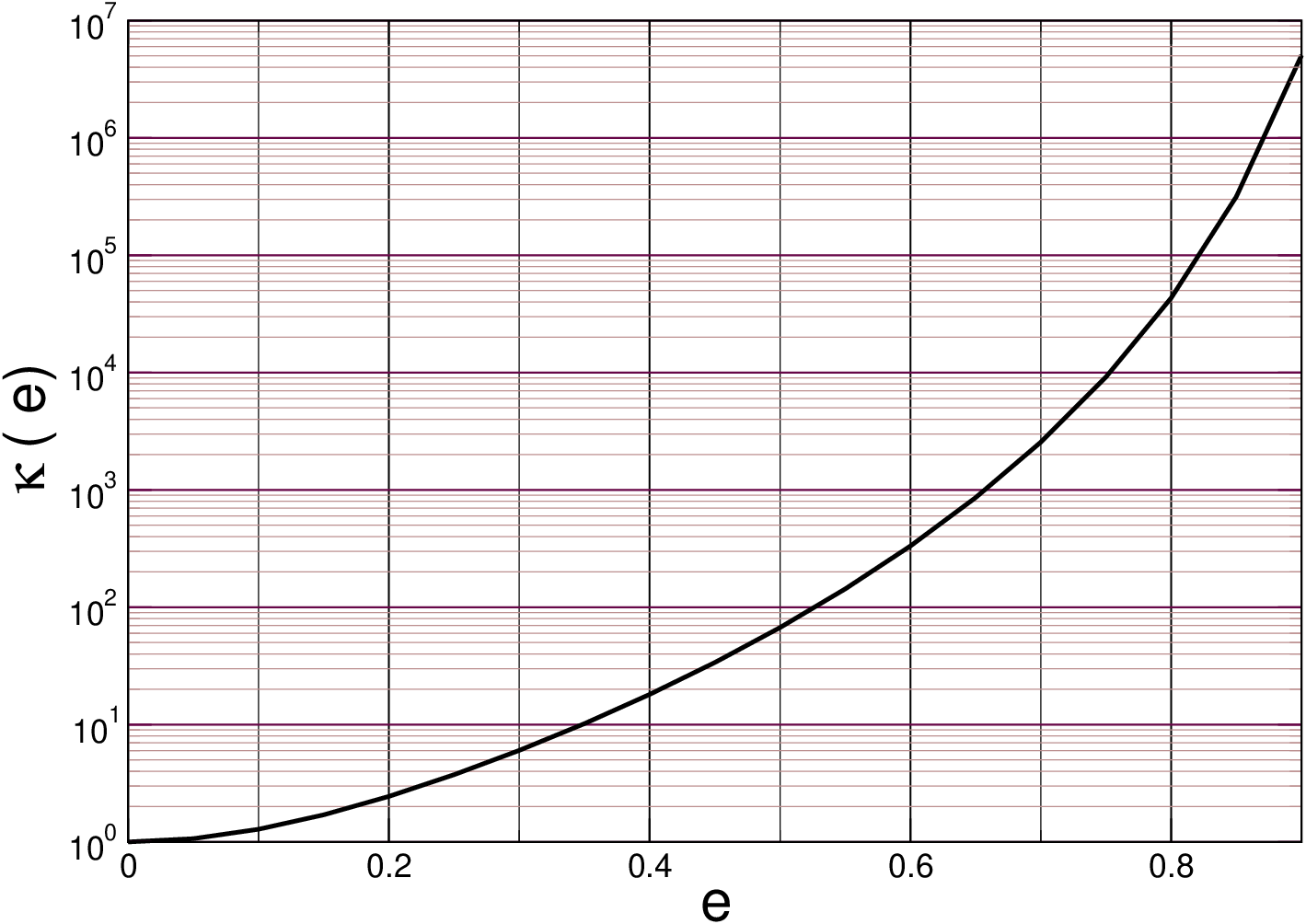}
\hskip 0.095 true cm 
\includegraphics[width=3.2in]{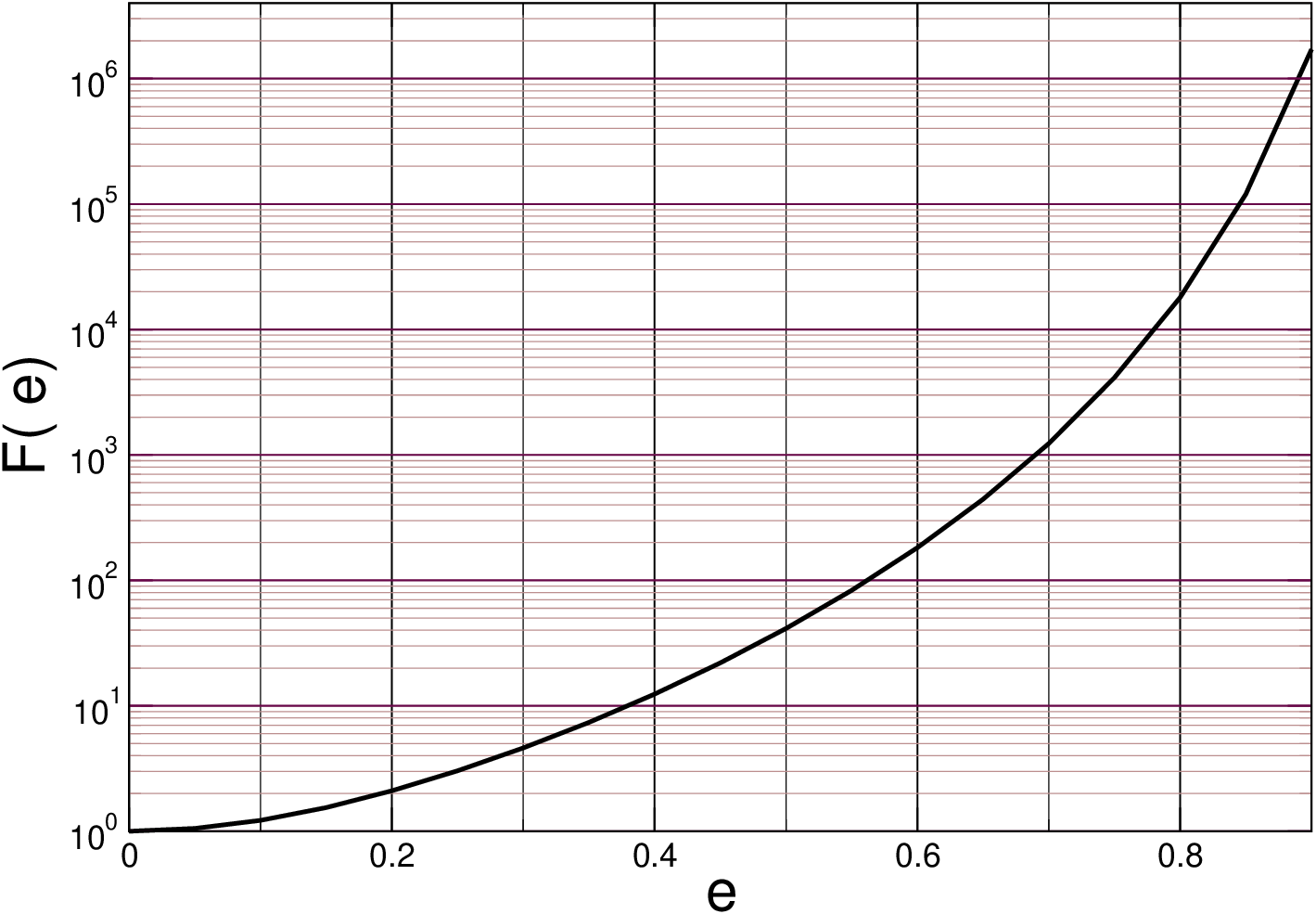} 
\caption{Enhancement functions $\kappa(e)$ and $F(e)$ in the evolution of $n$ at 3PN order.} 
\label{fig6}
\end{figure}
\begin{figure}[t]
\vspace{0.5cm}
\centering
\includegraphics[width=3.2in]{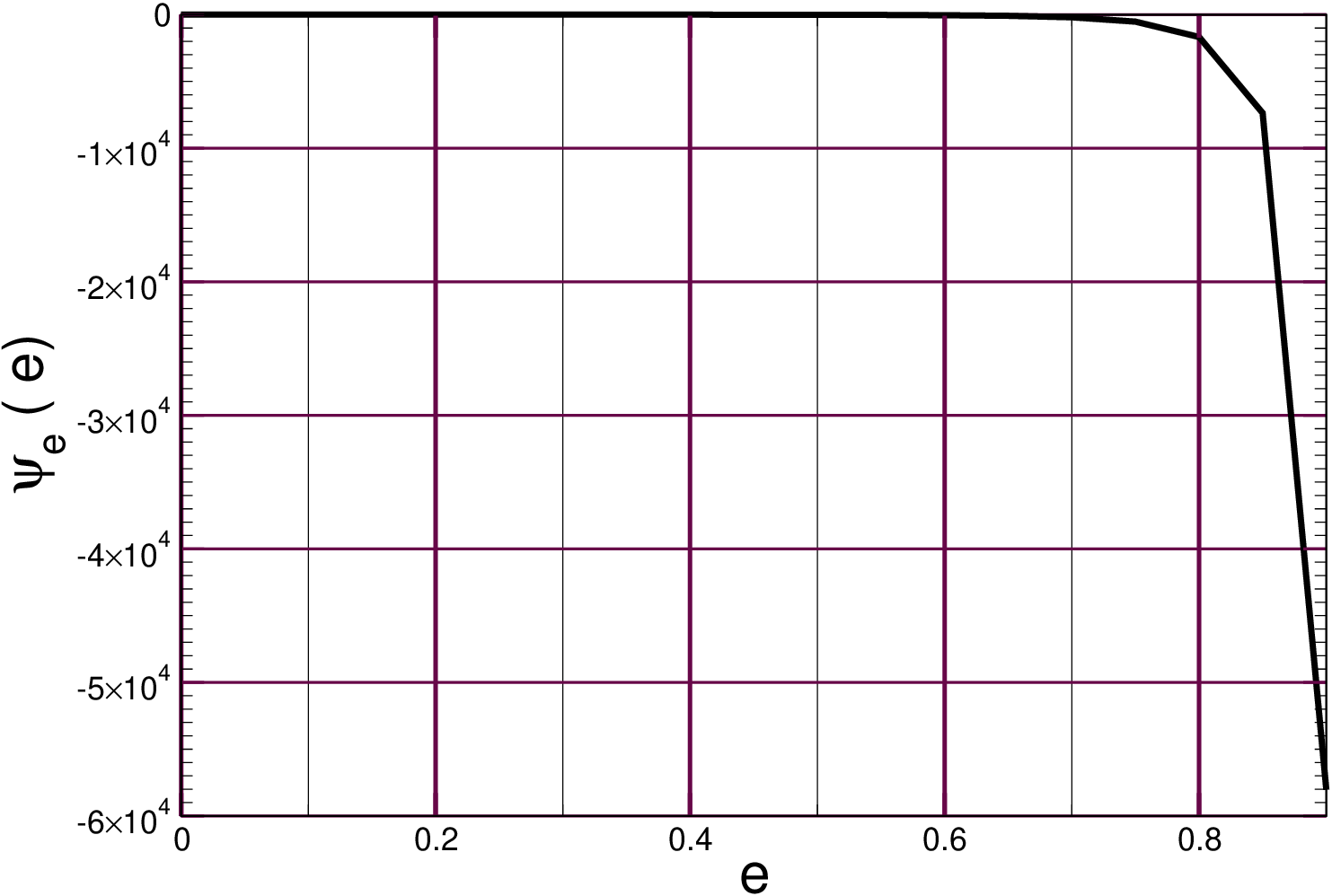}
\hskip 0.2 true cm 
\includegraphics[width=3.1in]{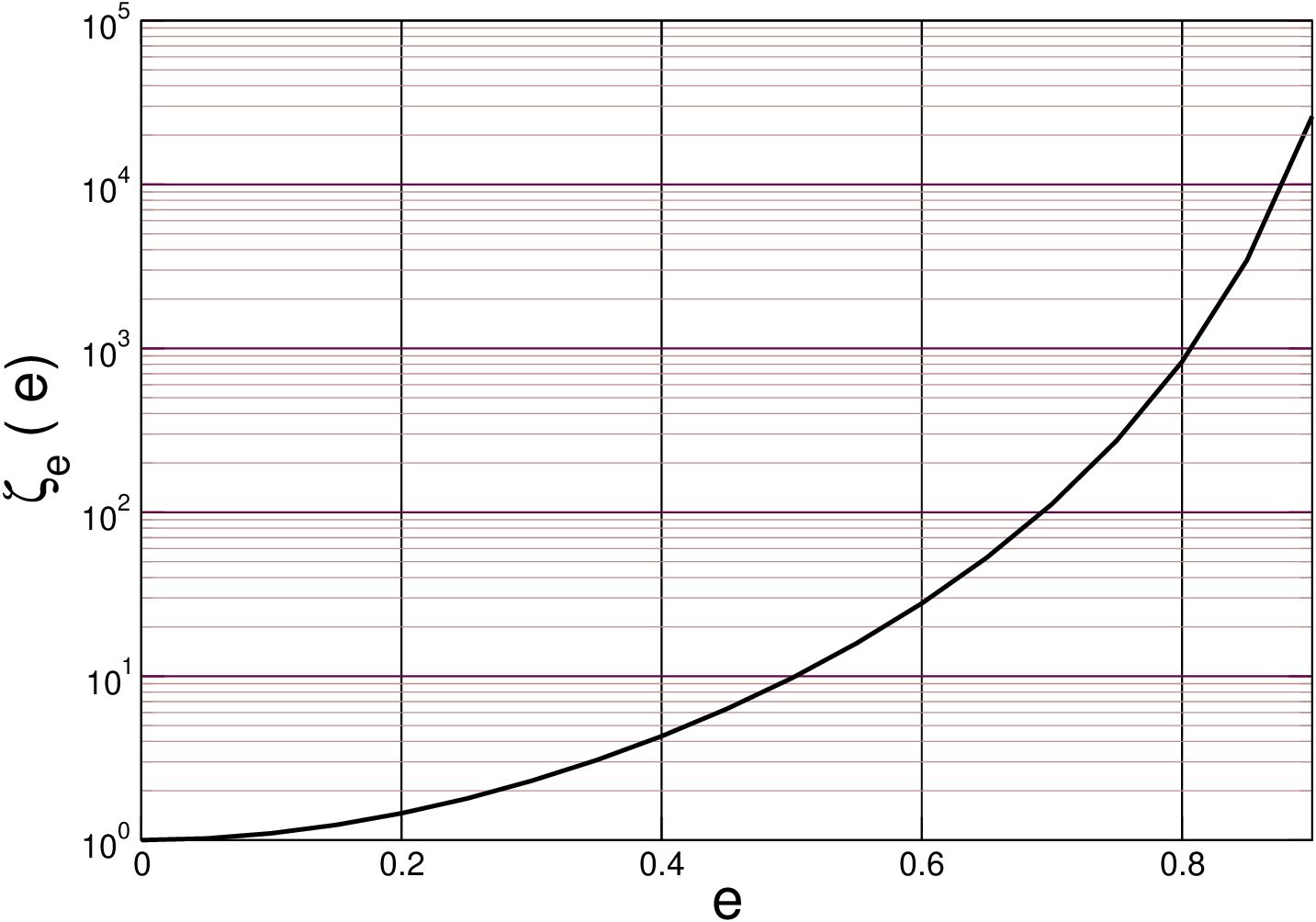} \vskip 1cm
\caption{Enhancement functions $\psi_{e}(e)$ and $\zeta_{e}(e)$ in the evolution of $e_t$ at 2.5PN order. Note the similarity of $\psi_{e}(e)$ with $\tilde{\psi}(e)$ in Fig.~\ref{fig2}. } 
\label{fig7}
\end{figure}
\begin{figure}[t]
\vspace{0.5cm}
\centering
\includegraphics[width=3.2in]{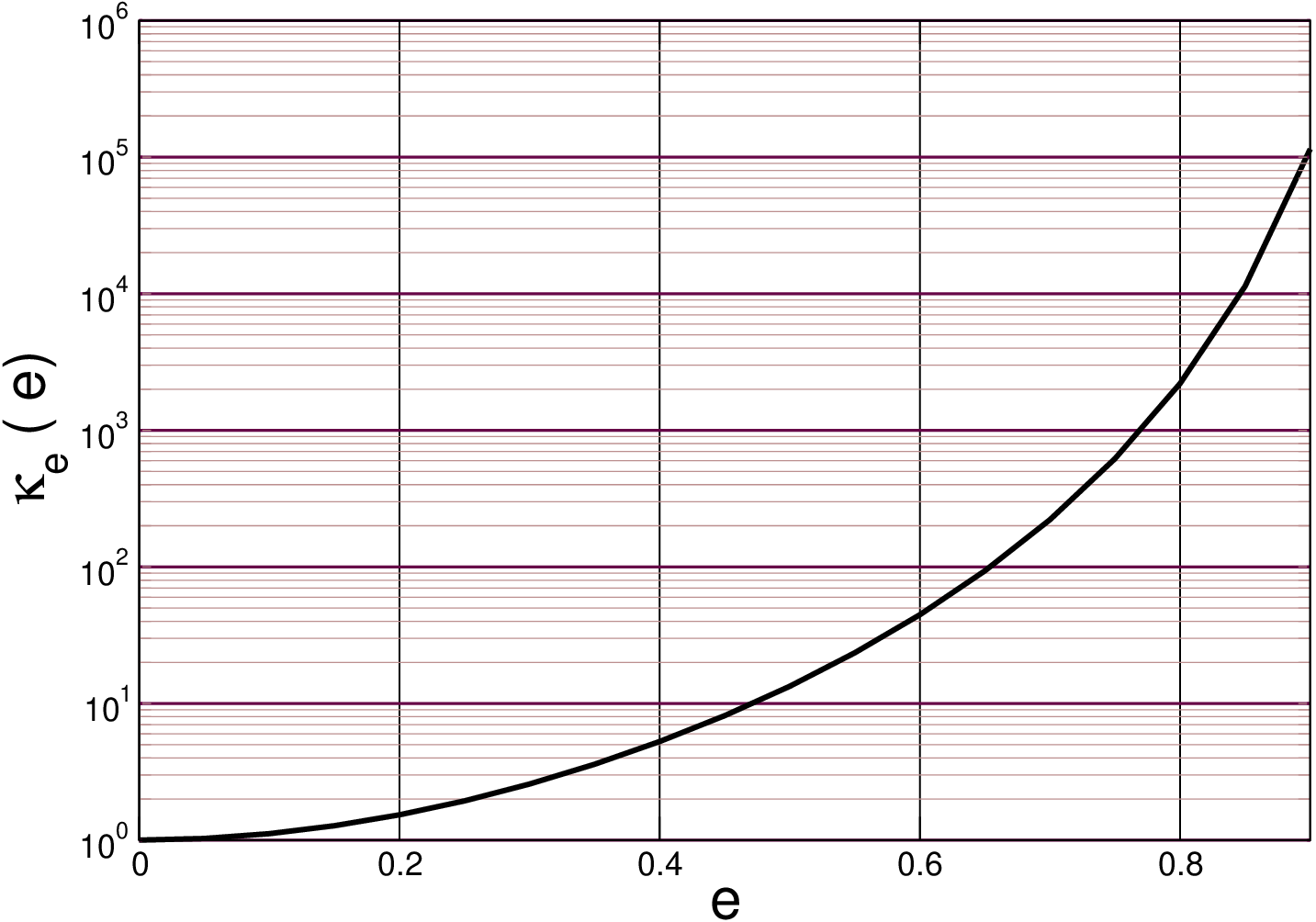}
\hskip 0.02 true cm 
\includegraphics[width=3.2in]{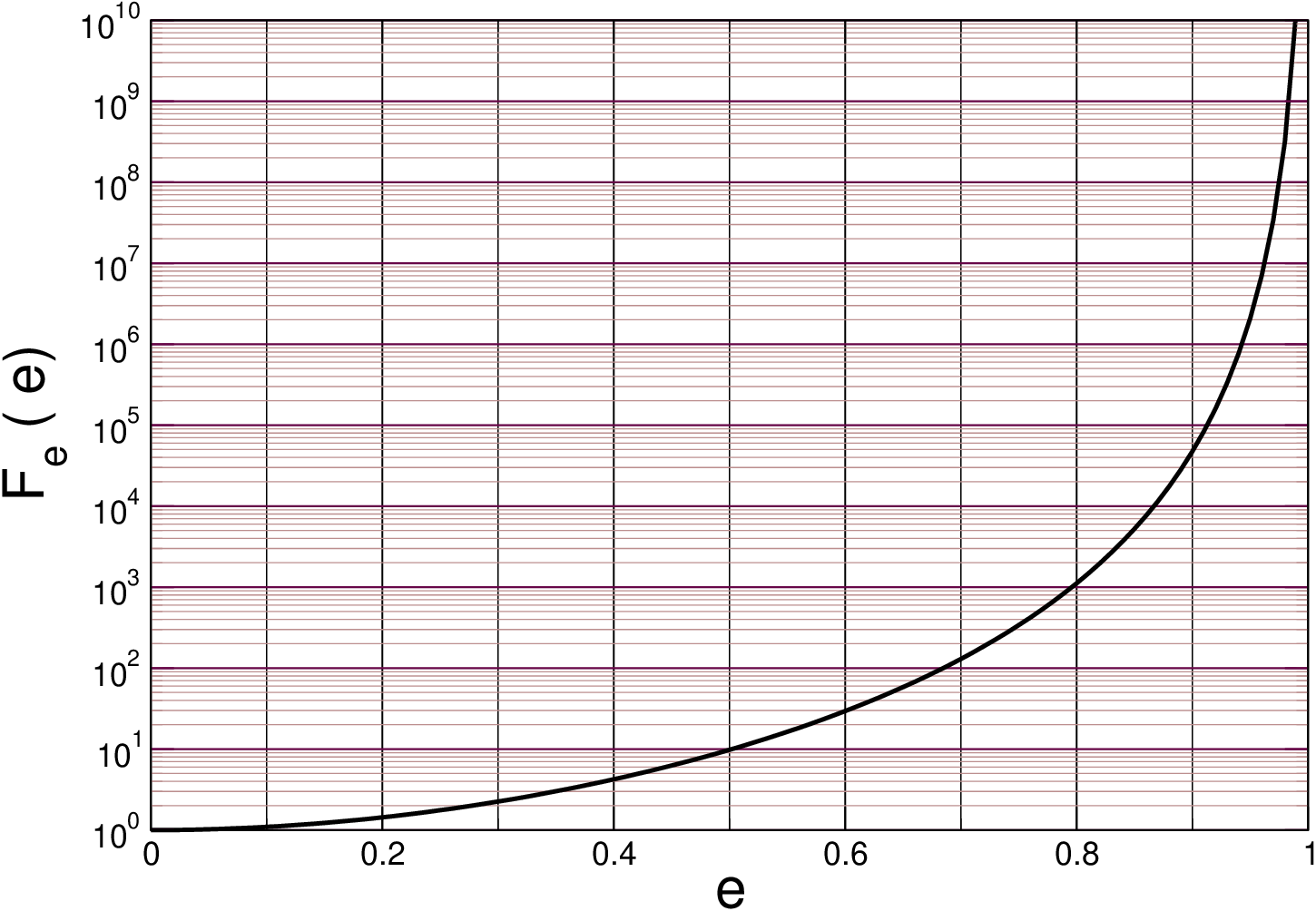} 
\caption{Enhancement functions $\kappa_{e}(e)$ and $F_{e}(e)$ in the evolution of $e_t$ at 3PN order.}
\label{fig8}
\end{figure}
\begin{figure}[t]
\vspace{0.5cm}
\centering 
\includegraphics[width=3.2in]{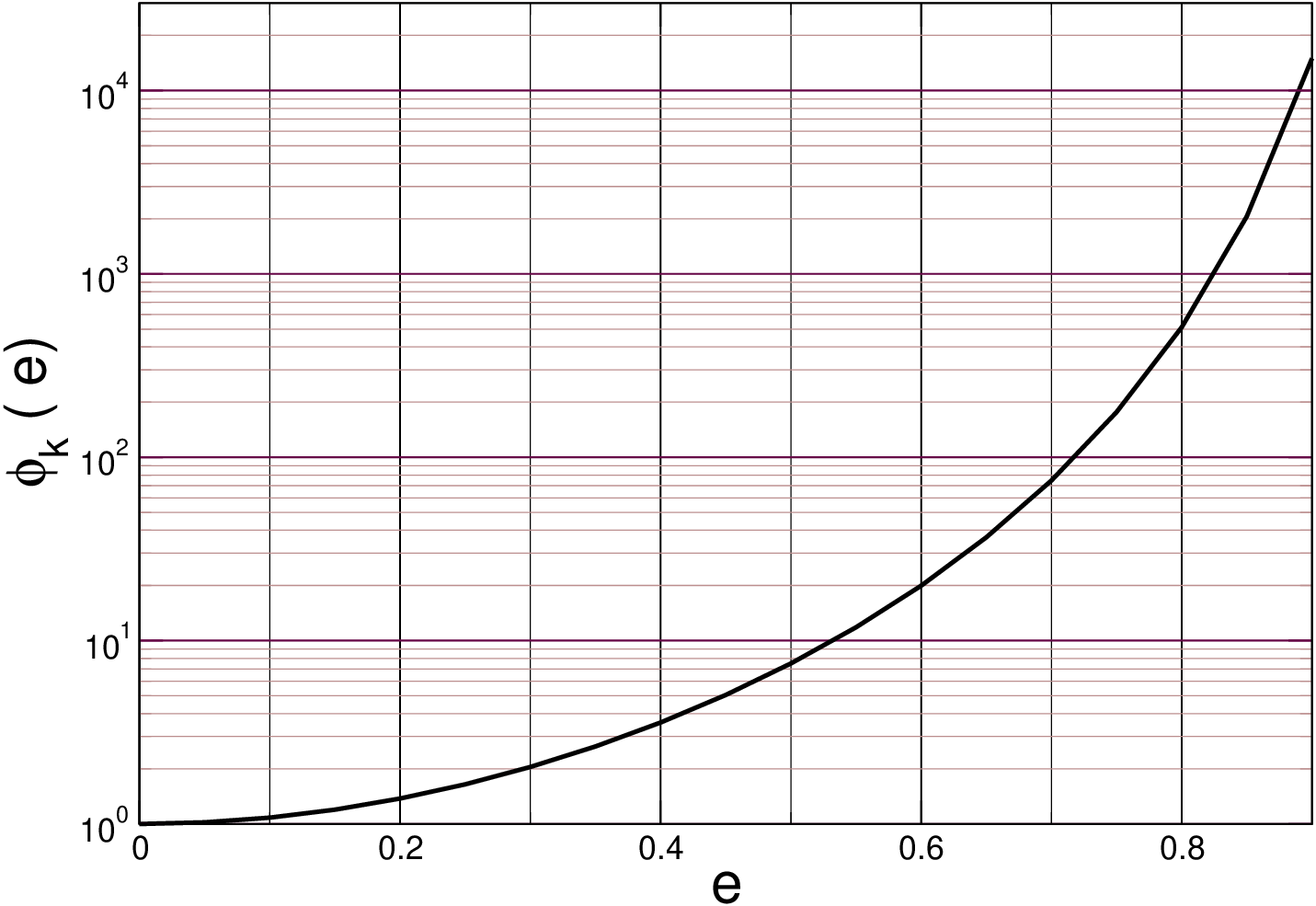}
\hskip 0.2 true cm 
\includegraphics[width=3.1in]{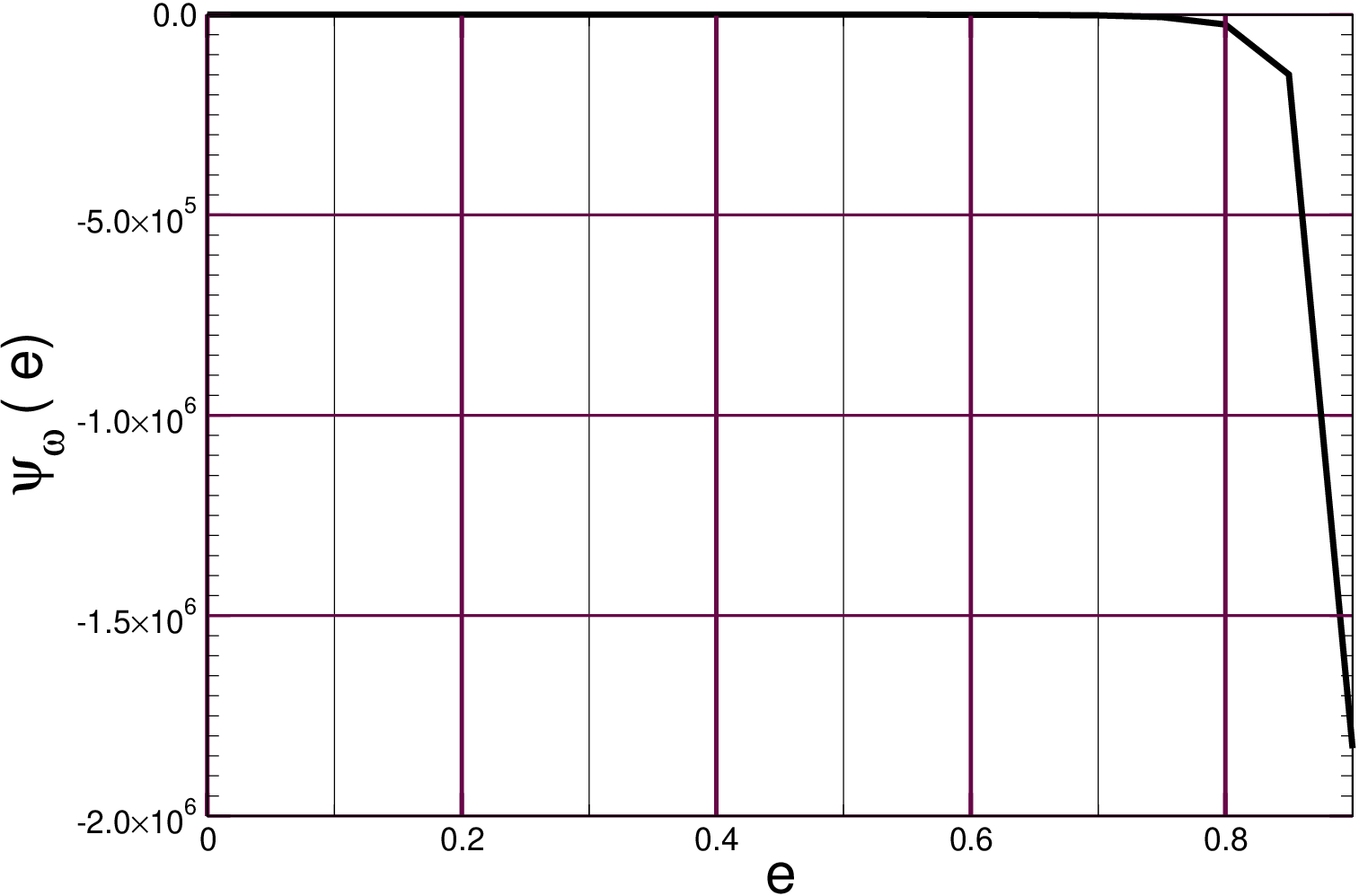}
\caption{Enhancement functions $\varphi_{k}(e)$ and $\psi_{\w}(e)$ in the evolution of $k$ and $\w$ at 1.5PN and 2.5PN orders. Note the similarity of $\psi_{w}(e)$ with $\tilde{\psi}(e)$ in Fig.~\ref{fig2}.}
\label{fig9}
\end{figure}
\begin{figure}[t]
\vspace{0.5cm}
\centering
\includegraphics[width=3.2in]{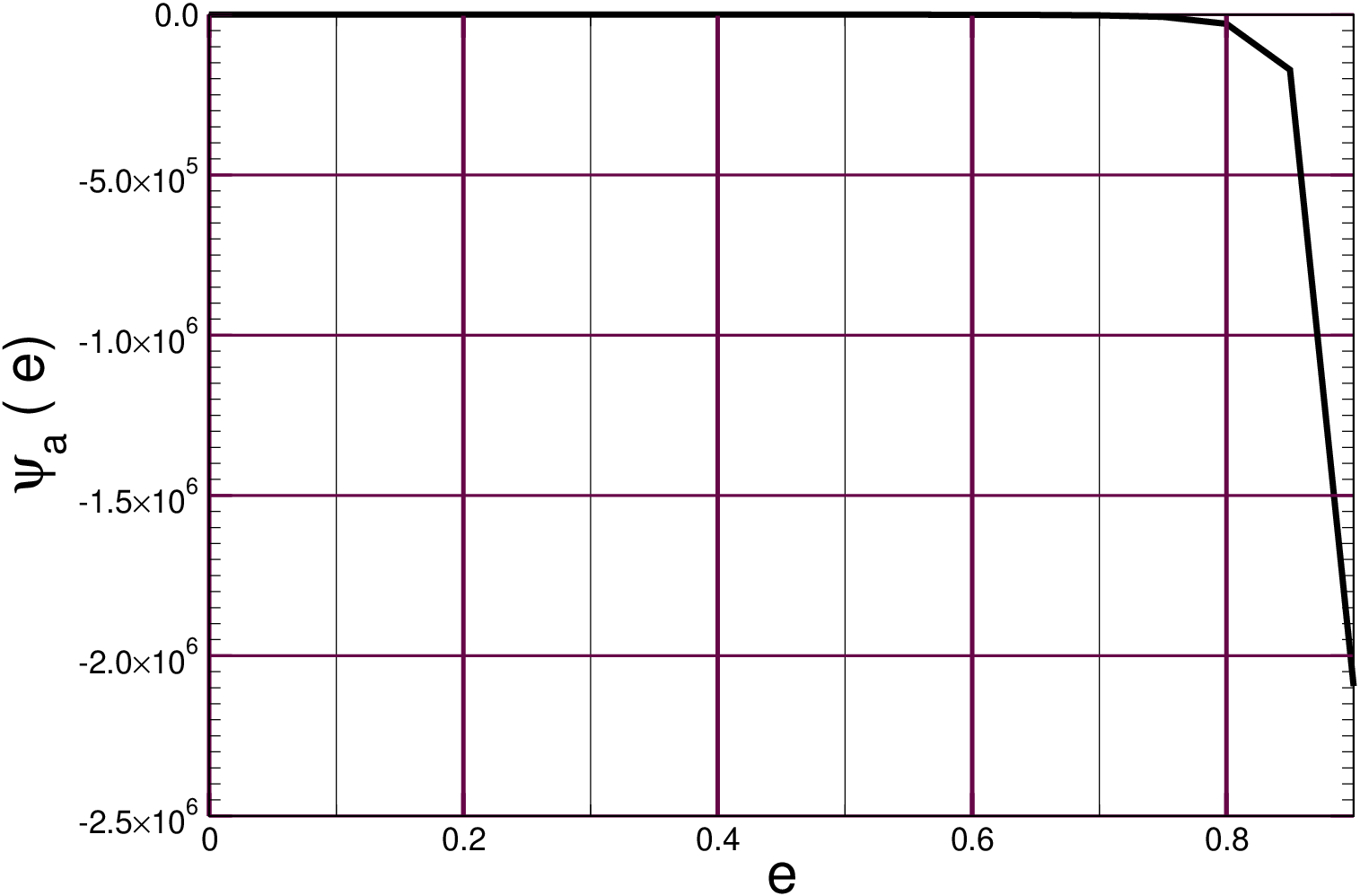} 
\caption{Enhancement function $\psi_{a}(e)$ in the evolution of $a_r$ at 2.5PN order. The other function $\zeta_a(e)$ at 2.5PN order is the same as $\zeta_n(e)$ and hence is not plotted. Similarly at 3PN order it involves the same functions $\kappa(e)$ and $F(e)$ as for $n$. Note the similarity of $\psi_{a}(e)$ with $\tilde{\psi}(e)$ in Fig.~\ref{fig2}.}
\label{fig10}
\end{figure}

\section{Limiting forms for small eccentricity}
\label{sec:small-ecc}
In some cases one may have prior information about the smallness of the eccentricity and in this case one may only need the leading corrections when the eccentricity parameter $e_t\rightarrow 0$. The main problem we face is to treat the hereditary parts because we could compute them only numerically for general eccentricities. However it was shown in~\cite{ABIQ07tail} how to obtain \textit{analytically} the leading corrections [neglecting $\mathcal{O}(e_t^4)$] for the enhancement factors in the case of the energy flux. However, here we need to push the accuracy of these results to the next order [neglecting $\mathcal{O}(e_t^6)$] in order to evaluate the leading-order correction in the enhancements functions present in the evolution of the orbital eccentricity, Eq.~\eqref{et-evo}; this can be checked from the explicit expressions of these functions as given 
in~\eqref{enhancenk}. We find
\begin{subequations}\label{phipsiexp}\begin{eqnarray} 
\label{phiexp} \varphi\left(e\right)&=& 1+\frac{2335
}{192}\,e^2+\frac{42955}{768}\,e^4+\mathcal{O}\left(e^6\right)\,,\\
\label{psiexp} \psi\left(e\right)&=& 1-\frac{22988
}{8191}\,\,e^2-\frac{36508643}{524224}\,e^4+\mathcal{O}\left(e^6\right)\,,\\ \zeta\left(e\right)&=& 1+
\frac{1011565} {48972} \,e^2+\frac{106573021}{783552}\,e^4+\mathcal{O}\left(e^6\right)\,,\\
\kappa\left(e\right)&=& 1+\left(\frac{62}{3}
-\frac{4613840}{350283}\ln 2+\frac{24570945}{1868176}\ln 3\right)\,e^2\nonumber\\
&&+\left(\frac{9177}{64}+\frac{271636085}{1401132}\ln 2-\frac{466847955}{7472704}\ln 3\right)\,e^4
+\mathcal{O}\left(e^6\right)\,,\\
F\left(e\right) &=& 1+\frac{62}{3}\,e^2+\frac{9177}{64}\,e^4
+\mathcal{O}\left(e^6\right)\,.
\end{eqnarray}\end{subequations}
Similar results are obtained in the case of the angular momentum flux, i.e. for the tilded enhancement factors, which read
\begin{subequations}\label{phipsitildeexp}\begin{eqnarray} 
\tilde{\varphi}\left(e\right)&=& 1+\frac{209}{32}\,e^2+\frac{2415}{128}\,e^4
+\mathcal{O}\left(e^6\right)\,,\\ \tilde{\psi}\left(e\right)&=&
1-\frac{17\,416}{8191}\,e^2 -\frac{14199197}{524224}e^4+\mathcal{O}\left(e^6\right)\,,\\
\tilde{\zeta}\left(e\right)&=& 1+ \frac{102\,371} {8162}
\,e^2+\frac{14250725}{261184}e^4+\mathcal{O}\left(e^6\right)\,,\\ 
\tilde{\kappa}\left(e\right)&=&
1+\left(\frac{389}{32} -\frac{2\,056\,005}{233\,522}\ln 2+\frac{8\,190\,315}{934\,088}\ln 3\right)\,e^2\nonumber\\
&&+\left(\frac{3577}{64}+\frac{50149295 }{467044}\,\ln 2-\frac{155615985}{3736352}\ln 3\right)e^4
+\mathcal{O}\left(e^6\right)\,,\\
\tilde{F}\left(e\right) &=&
1+\frac{389}{32}\,e^2 +\frac{3577}{64}e^4+\mathcal{O}\left(e^6\right)\,.
\end{eqnarray}\end{subequations}
Using~\eqref{phipsiexp} and~\eqref{phipsitildeexp} we then get the leading-order corrections in the functions parametrizing the evolution of orbital elements and defined by Eqs.~\eqref{enhancenk}--\eqref{Fee},
\begin{subequations}\label{enhancenke0}
\begin{eqnarray}
\psi_{n}(e)&=&1+\frac{94\,115}{17\,599}\,e^2 +\mathcal{O}\left(e^4\right)\,,\\
\zeta_{n}(e)&=&1+\frac{9215}{441}\,e^2 +\mathcal{O}\left(e^4\right)\,,\\
\varphi_{k}(e)&=&1+\frac{257}{32}\,e^2 +\mathcal{O}\left(e^4\right)\,,\\
\psi_{\w}(e)&=&1-\frac{55\,195}{4159}\,e^2 +\mathcal{O}\left(e^4\right)\,,\\
\zeta_{\w}(e)&=&1+\frac{9215}{441}\,e^2 +\mathcal{O}\left(e^4\right)\,,\\
\varphi_{e}(e)&=&1+\frac{21\,729}{3940}\,e^2 +\mathcal{O}\left(e^4\right)\,,\\
\psi_{e}(e)&=&1-\frac{5\,448\,991}{891\,056}\,e^2 +\mathcal{O}\left(e^4\right)\,,\\
\zeta_{e}(e)&=&1+\frac{23\,431\,113}{2\,440\,576}\,e^2 +\mathcal{O}\left(e^4\right)\,,\\
\kappa_{e}(e)&=&1+\left(\frac{1\,637\,339\,503+2\,135\,608\,720\,\ln2-663\,415\,515\,\ln3}{179\,578\,418-97\,909\,280\,\ln2 +98\,283\,780\,\ln3} \right)\,e^2 +\mathcal{O}\left(e^4\right)\,,\\
F_{e}(e)&=&1+\frac{14\,023}{1538}\,e^2 +\mathcal{O}\left(e^4\right)\,\\
\psi_{a}(e)&=&1-\frac{82\,775}{4159}\,e^2 +\mathcal{O}\left(e^4\right)\,,\\
\zeta_{a}(e)&=&1+\frac{9215}{441}\,e^2 +\mathcal{O}\left(e^4\right)\,.
\end{eqnarray}\end{subequations}
To finish, we provide the complete results (composed of both instantaneous and hereditary contributions) valid up to first order in $e_t^2$. The total fluxes of energy and angular momentum read\footnote{Consistent with the accuracy which was needed in Eqs.~\eqref{phipsiexp}--\eqref{phipsitildeexp}, we recall that the evolution of the orbital eccentricity to the level $e_t^2$ [as given by \eqref{detdt} below] requires these expressions to be accurate up to $e_t^4$.} 
\begin{subequations}\begin{align}
\langle\,{\cal F}\,\rangle &= \frac{32
}{5}\,\frac{c^5}{G}\,\nu^2\,x^5\,\left\{1-\left(\frac{1247}{336}+ \frac{35}{12}\nu
\right)x+4\,\pi\, x^{3/2}+\left(-\frac{44711}{9072} + \frac{9271}{504} \nu + \frac{65}{18} \nu
^2\right)x^2\right.\nonumber\\
&+\left(-\frac{8191  }{672}-\frac{583}{24}\nu\right)\pi\,x^{5/2}+\left(\frac{6643739519}{69854400}+\frac{16}{3}\,\pi^2-\frac{1712}{105}\,C-\frac{856}{105}\,\ln \left(16\,x\right)\right.\nonumber\\
&+\left.\left(-\frac{134543}{7776}+\frac{41 }{48}\pi ^2\right)\nu-\frac{94403}{3024}\nu^2-\frac{775}{324}\nu^3\right)x^3\nonumber\\
&+e_t^2\left[\frac{157}{24}+\left(-\frac{187}{168} - \frac{3107}{72} \nu
\right)x+\frac{2335}{48}\,\pi x^{3/2}+\left(-\frac{348079}{3024} + \frac{35923}{672} \nu +
\frac{14015}{108} \nu ^2\right)\,x^2\right.\nonumber\\
&+\left(\frac{821 \pi }{24}-\frac{1011565 \pi}{2016}\,\nu\right)\,x^{5/2}\nonumber\\
&+\left(\frac{112742385071}{69854400}+\frac{992
   }{9}\,\pi^2+\left(-\frac{1918465}{9072}+\frac{84295 }{2304}\pi ^2\right)\,\nu-\frac{11189}{56}\,\nu^2-\frac{459793}{1944}\,\nu^3\right.\nonumber\\
&-\left.\left.\left.\frac{106144}{315}\,C+\frac{18832}{45}\,\ln 2-\frac{234009 }{560}\ln 3-\frac{53072}{315}\,\ln \left(16\,x\right)\right)x^3\right]\right\}+
\mathcal{O}\left(e_t^4\right)\,,\label{Fsmallet}\\\nonumber\\
%%%%%%%%%%%%%%%%%%%%%%%%%%%%%%%%%%%%%%%%%%%%%
\langle\,{\cal G}\,\rangle
&=\frac{32}{5}\,c^2 \,m \,\nu ^2 \,x^{7/2} \,\left\{1-\left(\frac{1247}{336}
+ \frac{35}{12}\nu \right)x+ 4\pi\,x^{3/2}+\left(-\frac{44\,711}{9072} + \frac{9271}{504} \nu +
\frac{65}{18} \nu ^2\right)x^2\right.\nonumber\\
&+\pi\,\left(-\frac{8191}{672}-\frac{583}{24}\nu\right)x^{5/2}+\left(\frac{6\,643\,739\,519}{69\,854\,400}+\frac{16}{3} \,\pi^2 -\frac{1712}{105}\,C + \left(-\frac{134\,543}{7776} + \frac{41\pi \ ^2}{48}\right) \nu\right.\nonumber\\
&\left.-\frac{94\,403}{3024} \nu ^2 - \frac{775}{324} \nu ^3
- \frac{856}{105}\ln16x \right)x^3
+e_t^2\left[{\frac{23}{8}}+\left(-\frac{361}{168} -\frac{545}{24} \nu
\right)x+{209\over8}\,\pi\,x^{3/2}\right.\nonumber\\
&+\left({-\frac{1\,510\,015}{18\,144}+\frac{13\,975}{288} \nu +\frac{2839}{36}
\nu ^2}\right)x^2+\pi\,\left({311\over12}-{102\,371\over336}\,\nu\right)x^{5/2}\nonumber\\
&+\left(\frac{59\,524\,389\,803}{69\,854\,400}+{19\,581\over70}\ln2-{78\,003\over280}\ln3\,+{389\over32}\left(\frac{16}{3} \,\pi^2 -\frac{1712}{105}\,C \right)\right.\nonumber\\
&\left.\left.\left.+\left( \frac{9\,285\,217}{54\,432} +\frac{13\,463 }{768}\pi ^2 \right) \nu -\frac{75\,493}{378} \nu ^2
-\frac{104\,723}{648} \nu ^3-\frac{41\,623}{420}\ln16x\right)x^3\right]\right\} +\mathcal{O}\left(e_t^4\right)\label{Gsmallet}\,.
\end{align}\end{subequations}
We have compared the $e_t^2$ terms of the angular momentum flux expression above with the result of black-hole perturbation theory given in Eq.~(181) of Ref.~\cite{TSLivRev03}. Similar to the case of the energy flux~\cite{ABIQ07}, we find that our result in the test-mass limit $\nu\rightarrow 0$ matches with the perturbation result with a transformation of eccentricity given by:
\begin{equation}\label{ete} e_t^2=
e^2\left(1 - 6 x + \frac{9}{2}\,x^2 - 10 x^3\right)\,.
\end{equation}
Here the eccentricity $e_t$ is the one appearing in \eqref{Gsmallet}, i.e.  in ADM coordinates, while $e$ is the eccentricity used in perturbation theory. The relation \eqref{ete} is the same as the one we found for the case of the energy flux, as it must surely be: indeed see Eq.~(9.6) in \cite{ABIQ07} where we must take into account the change between the modified harmonic (MH) coordinates used there and the ADM coordinates used here.\footnote{From Eq.~\eqref{ADMMH} we have $e_t^\mathrm{MH}=e_t^\mathrm{ADM}\bigl[1-\frac{1}{4}x^2-\frac{1}{2}x^3+\mathcal{O}(\nu)\bigr]$ in the test-mass limit for small
eccentricity ${\cal O}(e_t^2)$.}

The evolution of orbital elements is 
\begin{subequations}\label{orbeltfinal}\begin{eqnarray}
\langle\frac{\ud n}{\ud t}\rangle &=&\frac{96 }{5 }\,\frac{c^6}{G^2 m^2}\,\nu\,x^{11/2}\,\left\{1+\left(-\frac{2423}{336} -\frac{11}{4} \nu
\right)x+4\pi\,x^{3/2}+\left(\frac{44\,201}{18\,144}+ \frac{74\,309}{2016} \nu + \frac{59}{18} \nu
^2\right)x^2+\right.\nonumber\\
&&+\pi\left(-{17\,599\over672}-{189\over8}\nu\right)x^{5/2}+\left({12\,720\,014\,063\over139\,708\,800}+\frac{16}{3} \,\pi^2-\frac{1712}{105}\,C\right.\nonumber\\
&&\left.+\left[
\frac{13\,392\,731}{217\,728} - \frac{205 }{96}\pi ^2 \right]\nu
-\frac{658\,843}{8064} \nu ^2 -\frac{5605}{2592} \nu ^3-\frac{856}{105} \ln
16x\right)x^3\nonumber\\
&&+e_t^2\left[\frac{157}{24}+\left(-\frac{2801}{112} -\frac{673}{16} \nu
\right)x+{2335\over48}\pi\,x^{3/2}\right.\nonumber\\
&&+\left(-\frac{1\,761\,523}{12096}+ \frac{1\,094\,473}{4032} \nu +
\frac{213\,539}{1728} \nu ^2\right)x^2+\pi \left(-{13\,445\over96}-{27\,645\over56}\nu\right)x^{5/2}\nonumber\\
&&+\left(\frac{82\,283\,570\,939}{46\,569\,600}+{18\,832\over45}\ln 2-{234\,009\over560}\ln 3+{62\over3}\left[\frac{16}{3} \,\pi^2 -\frac{1712}{105}\,C\right]\right.\nonumber\\
&&\left.\left.\left.+\left[ \frac{23\,736\,079}{24\,192} 
+ \frac{7103 }{576}\pi ^2\right] \nu -
\frac{391\,117}{336} \nu ^2 -\frac{6\,874\,115}{31\,104}\nu ^3-\frac{53\,072}{315}\ln 16x\right)x^3\right]\right\}\nonumber\\
&&+\mathcal{O}\left(e_t^4\right)\,,\\
%%%%%%%%%%%%%%%%%%%%%%%%%%%%%%%%%%%%%%%%%%%%%%%%%%%%%%%%
\langle \frac{\ud k}{\ud t}\rangle&=&
\frac{192}{5}\,\frac{c^3}{G m}\,\nu\,x^5\left\{ 1+x
\left(\frac{2281}{336}-\frac{89 }{12}\nu \right)+4\pi x^{3/2}+x^2
\left(\frac{897\,725}{18\,144}+\left[-\frac{342\,527 }{2016}+\frac{123 \pi
^2}{32}\right] \nu\right.\right.\nonumber\\ 
&&\left.+\frac{208 \nu ^2}{9}\right)+e_t^2\left[\frac{35}{8}+x \left(\frac{11\,595}{224}-\frac{851}{16} \nu
\right)+{257\over8}\pi\,x^{3/2}+x^2
\left(\frac{32\,245\,547}{72\,576}\right.\right.\nonumber\\
&&\left.\left.\left.+\left[-\frac{5\,515\,931}{4032}+\frac{9389 }{384}\pi ^2\right]\nu+\frac{162\,853}{576} \nu
^2\right)\right]\right\}\nonumber\\
&& +\mathcal{O}\left(e_t^4\right)\,,\\
%%%%%%%%%%%%%%%%%%%%%%%%%%%%%%%%%%%%%%%%%%%%%%%%%%%%%%%%%%%%%%%%
\langle\frac{\ud \omega}{\ud t}\rangle &=&\frac{96}{5}\,\frac{c^6}{G^2 m^2}\,\nu\,x^{11/2}\,
\left\{1+\left(-\frac{743}{336}-\frac{11  }{4}\nu\right)x+ 4\pi\,x^{3/2}+\left(\frac{34103}{18144}+\frac{13661}{2016}\nu+\frac{59 }{18}\nu ^2\right)x^2\right.\nonumber\\
&&+\left.\pi\left(-{4159\over672}-{189\over8}\nu\right)x^{5/2}+
\left(\frac{16447322263}{139708800}+\frac{16 }{3}\pi ^2+\left(-\frac{56198689}{217728}+\frac{451 }{48}\pi ^2\right) \nu \right.\right.\nonumber\\
&&\left.\left. +\frac{541 }{896}\nu
   ^2-\frac{5605 }{2592}\nu ^3 -\frac{1712}{105}\,C-\frac{856}{105}\ln \left(16\,x\right)\right)x^3\right.\nonumber\\
&&+e_t^2\left[\frac{157}{24}+\left(\frac{713}{112}-\frac{673 }{16}\nu \right)x+{2335\over48}\pi\,x^{3/2}\right.\nonumber\\
&&+\left(-\frac{519523}{12096}-\frac{143771 }{4032}\nu +\frac{213539 }{1728}\nu ^2\right)x^2+\pi \left({7885\over96}-{27\,645\over56}\nu\right)x^{5/2}\nonumber\\
&&+\left(\frac{276032869667}{139708800}-\frac{106144 }{315}C
+\frac{992 }{9}\pi ^2+\left(-\frac{59984699}{24192}
+\frac{227795 }{2304}\pi ^2\right) \nu\right. \nonumber\\
&&+\left.\left.\left.\frac{966019 }{4032}\nu ^2-\frac{6874115 }{31104}\nu ^3+\frac{18832 }{45}\ln 2-\frac{234009
  }{560} \ln 3-\frac{53072}{315}\ln\left(16\,x\right)\right)x^3\right]\right\}\nonumber\\
&& +\mathcal{O}\left(e_t^4\right)\,,\\
%%%%%%%%%%%%%%%%%%%%%%%%%%%%%%%%%%%%%%%%%%%%%%%%%%%%%%%%%%%%%%%%%%%%%%%%%%%
\langle\frac{\ud a_r}{\ud t}\rangle &=&
-\frac{64}{5} c x^3 \nu\left\{1+\left({-\frac{743}{336} -\frac{11}{4} \nu
}\right)x+4\pi\,x^{3/2}+\left(\frac{38\,639}{18\,144}+ \frac{11\,393}{2016} \nu + \frac{19}{6} \nu^2\right)x^2\right.\nonumber\\
&&+\pi\left(-{4159\over672}-{189\over8}\,\nu\right)x^{5/2}+\left({16\,439\,941\,813\over139\,708\,800}+\frac{16}{3} \,\pi^2 -\frac{1712}{105}\,C\right.\nonumber\\
&&\left.+\left(
-\frac{3\,635\,195}{15\,552} + \frac{245 }{32}\pi ^2\right) \nu+ \frac{167\,971}{24\,192} \nu ^2 -\frac{61}{32} \nu ^3-\frac{856}{105} \ln(16x)\right)x^3\nonumber\\
&&+e_t^2\left[{\frac{157}{24}}+\left(\frac{4267}{336} -\frac{673}{16} \nu
\right)x +\frac{2335  }{48}\pi\,x^{3/2}\right.\nonumber\\
&&+\left(-\frac{19\,423}{1728} -\frac{93\,607}{1008}
\nu +\frac{23\,587}{192}\nu ^2\right)x^2+\pi\,\left(\frac{11825}{96}
-\frac{27645  }{56}\nu\right)\,x^{5/2}\nonumber\\ &&+\left({144\,392\,349\,571\over69\,854\,400}+{18\,832\over45}\ln 2-{234\,009\over560}\ln 3+{62\over3}\left[\frac{16}{3} \,\pi^2 -\frac{1712}{105}\,C\right]\right.\nonumber\\
&&\left.\left.\left.+\left[-\frac{67\,044\,655}{24\,192} + 
\frac{11\,971 }{128}\pi^2 \right]\nu+\frac{3\,943\,769}{8064} \nu^2-\frac{83\,195}{384} \nu ^3-{53\,072\over315}\ln 16x\right)x^3\right]\right\}\nonumber\\
&&+ \mathcal{O}\left(e_t^4\right)\,,\\
%%%%%%%%%%%%%%%%%%%%%%%%%%%%%%%%%%%%%%%%%%%%%%%%%%%%%%%%%%%%%%%%%%%%%%%%%%%%%%%%
 \langle\frac{\ud e_t}{\ud t}\rangle
&=&-\frac{304}{15}\,\frac{c^3}{G m}\,e_t\, x^4 \,\nu \,\left\{1 +
\left(-\frac{2817}{2128} - \frac{1021}{228}\nu\right)x +{985\over152}\pi\,x^{3/2}\right.\nonumber\\
&&\left.+
\left(-\frac{120\,293}{38\,304} + \frac{33\,559}{4256}\nu + \frac{141}{19}\nu ^2
\right)x^2+\pi\left(-{55\,691\over4256}-{19\,067\over399}\nu\right)x^{5/2}\right. \nonumber\\
&&+
\left({245\,840\,579\,209\over884\,822\,400}+{4601\over105}\ln 2-{234\,009\over5320}\ln 3\right.\nonumber\\
&&+{769\over304}\left[\frac{16}{3} \,\pi^2 -\frac{1712}{105}\,C\right]+\left[
\frac{31\,417 }{3648}\pi ^2 - \frac{125\,449\,717 }{459\,648}\right]\nu+
\frac{42\,311}{51\,072} \nu ^2 - \frac{305\,005}{49\,248} \nu ^3 \nonumber\\
&& \left. - \frac{82\,283}{3990}\ln 16x\right)x^3+e_t^2\left[\frac{881}{304}+\left(\frac{40\,115}{4256} - \frac{5\,1847 
}{1824}\nu\right)x+{21\,729\over608}\pi\,x^{3/2}\right.\nonumber\\
&&+\left(-\frac{1\,538\,725}{51\,072} - \frac{1\,252\,109 }{17024}\nu +
\frac{274\,515 }{2432}\nu \ ^2\right)x^2+\pi\left({286\,789\over3584}-{7\,810\,371\over17\,024}\nu\right)x^{5/2}\nonumber\\
&&+\left(\frac{1\,306\,972\,981\,451}{589\,881\,600}-{3\,813\,587\over3990}\ln2 +{6\,318\,243\over21\,280}\ln3 \right.\nonumber\\
&& + \left[\frac{1\,571\,323  }{29\,184}\pi ^2 -
\frac{202\,888\,331 }{131328}\right]\nu +\frac{14\,915\,155 }{38\,304}\nu^2-\frac{100\,330\,729 }{393\,984}\nu ^3\nonumber\\
&&\left.\left.\left.+{14023\over608}\left[\frac{16}{3} \,\pi^2 -\frac{1712}{105}\,C\right]-\frac{1\,500\,461
}{7980}\ln(16x)\right)x^3\right]\right\}\nonumber\\
&& + \mathcal{O}\left(e_t^4\right)\,.\label{detdt}
\end{eqnarray}\end{subequations}
We find that in the limit when $e_t\rightarrow 0$, the evolution of the orbital frequency, i.e. $\langle\ud\omega/\ud t\rangle_\odot$, reproduces the known result for circular orbits at 3PN order as given in~\cite{BFIJ02,BDEI05}.

\acknowledgments 
K.G.A. was a VESF fellow of European Gravitational Observatory (EGO) during the intermediate stages of this work. K.G.A. also acknowledges support by the National Science Foundation, Grant No.\ PHY 06--52448, the National Aeronautics and Space Administration, Grant No.\ NNG-06GI60G, and the Centre National de la Recherche Scientifique, Programme International de Coop\'eration Scientifique (CNRS-PICS), Grant No. 4396 in the final stages of the project. B.R.I. acknowledges the hospitality of IHES and IAP during different stages of this work. K.G.A. and B.R.I. thank Moh'd S.S. Qusailah for discussions during the initial stages of this project. S.S. acknowledges S.N. Bose Centre for hospitality during the final stages of the work. All algebraic calculations were done with the software Mathematica.

\appendix

\section{The non-linear memory DC term}
\label{appA}
Our aim is to compute the zero-frequency part (or DC part) of the non-linear memory integral found in Sec.~\ref{memory}, namely
\begin{equation}
{\cal G}_i^{\rm memory}(t_0) = \frac{4}{35}\,\frac{G^2}{c^{10}}~
\varepsilon_{ijk}\,I_{ja}^{{(3)}}(t_0) \int_{t_1}^{t_0} \ud t
\,I_{kb}^{{(3)}}(t)~ I_{ab}^{{(3)}}(t)\,.
\label{mem1}
\end{equation}
For convenience in this Appendix we denote the current time of the observer by $t_0$ (formerly denoted $T_R$) and the earlier time over which the memory is integrated by $t$ (formerly $V$). In Eq.~\eqref{mem1} we shall suppose that the system was formed at some initial instant $t_1$ on some very eccentric orbit with initial eccentricity $e_1$ close to 1, but strictly less than 1. Then the system will evolve by radiation reaction until reaching the current eccentricity $e_0$ such that $e_0\ll e_1 <1$.

At this order of approximation we can replace in \eqref{mem1} the quadrupole moment and its time derivatives by Newtonian values. We express the result in terms of the current orbital separation $r_0\equiv r(t_0)$, radial velocity $\dot{r}_0$ and instantaneous orbital frequency $\dot{\varphi}_0$. The result will be an integral extending on the corresponding variables $r$, $\dot{r}$ and $\dot{\varphi}$ for the orbit at any instant in the past [recall our notation after Eq.~\eqref{Li}; here e.g. $r\equiv r(t)$]. For convenience we set the origin of the orbital phase $\varphi$ (or true anomaly) at the current binary's separation, i.e. we pose $\varphi_0=0$. A straightforward calculation gives the norm of the memory integral as
\begin{equation}
{\cal G}^{\rm memory} = \frac{64}{105}\frac{G^5m^6\nu^3}{c^{10}r_0}\int_{t_1}^{t_0} \ud t\,\left[
\left(\dot{\varphi}_0\frac{\dot{r}^2}{r^4}
-\frac{\dot{r}_0}{r_0}\frac{\dot{\varphi}\,\dot{r}}{r^3}\right)\cos2\varphi+\left(-4\dot{\varphi}_0\frac{\dot{\varphi}\,\dot{r}}{r^3}
-\frac{1}{4}\frac{\dot{r}_0}{r_0}\frac{\dot{r}^2}{r^4}\right)\sin2\varphi\right]\,.
\label{mem2}
\end{equation}
We then replace $r$, $\dot{r}$ and $\dot{\varphi}$ in this integral by the explicit solution for Keplerian motion,
\begin{subequations}
\begin{align}
r&=\frac{a(1-e^2)}{1+e\cos\varphi}\,,\\
\dot{r}&=\sqrt{\frac{G m}{a(1-e^2)}}\,e \sin\varphi\,,\\
\dot{\varphi}&=\sqrt{\frac{G m}{a^3(1-e^2)^{3}}}\,\left(1+e\cos\varphi\right)^2\,.
\end{align}\end{subequations}
The resulting integral is composed of many terms corresponding to different harmonics of the Keplerian motion. Thus, we find that the structure of the memory term is of the type
\begin{equation}
{\cal G}^{\rm memory} = \sum_n\int_{t_1}^{t_0}\ud t\,f_n(a,e)\,e^{\ii\,n\,\varphi}\,,
\label{mem2harm}
\end{equation}
where the $f_n$'s depend on the semi-major axis $a(t)$ and eccentricity $e(t)$ of the orbit in the past (and depend also on the current values $r_0$, $\dot{r}_0$, $\dot{\varphi}_0$). Here $\varphi(t)$ is the orbital phase which is oscillating at the orbital period. In contrast, $a(t)$ and $e(t)$ remain approximately constant during one period, but slowly evolve by radiation reaction during all the binary's past evolution. Now we expect that the oscillations of the orbital phase, due to the sequence of many orbital cycles in the entire life of the system, will more or less cancel each other in the memory integral, so that the true post-Newtonian order of the oscillating terms will be simply given by the power of $1/c$ they carry. The oscillating terms, having $n\not= 0$, are thus expected to have their normal 2.5PN order,\footnote{This expectation could be justified in more details using calculations similar to the ones in Sec. 4 of~\cite{ABIQ04}.} and we have shown in Sec.~\ref{memory} that these terms do not contribute to the \textit{averaged} angular momentum flux. 

With this, now there remains  the purely zero-frequency or DC component of the memory, corresponding to $n=0$, which is given by
\begin{equation}
{\cal G}^{\rm DC} = \int_{t_1}^{t_0}\ud t\,f_0(a,e)\,.
\label{DC0}
\end{equation}
This DC integral is \textit{cumulative} because it extends over some steadily increasing kernel and \textit{a priori} exhibits a strong dependence on the past (i.e. when $t_1\rightarrow -\infty$). The explicit calculation gives
\begin{equation}
{\cal G}^{\rm DC} = -\frac{16}{5}\frac{G^6m^7\nu^3}{c^{10}}\frac{\dot{\varphi}_0}{r_0}\int_{t_1}^{t_0} \ud t\,\frac{e^2}{a^5(1-e^2)^5}\left[1+\frac{20}{21}e^2+\frac{19}{336}e^4\right]\,.
\label{DC1}
\end{equation}
We now evaluate the DC term by inserting the secular evolution of orbital elements to Newtonian order consistently. We first convert the time integration into an integration over eccentricity by using the Peters~\cite{Pe64} formula
\begin{equation}
\dot{e} = - \frac{304}{15}\frac{G^3m^3\nu\,e}{c^5a^4(1-e^2)^{5/2}}\left(1+\frac{121}{304}e^2\right)\,.
\label{edotN}
\end{equation}
This is the dominant term in Eqs.~\eqref{evolecc}--\eqref{evoleccPN}; for simplicity we ignore the averaging procedure. Calling $e_1$ the initial eccentricity at the instant of formation of the binary system ($e_1>e_0$), we obtain from Eq.~\eqref{DC1} 
\begin{equation}
{\cal G}^{\rm DC} = -\frac{3}{19}\frac{G^3m^4\nu^2}{c^{5}}\frac{\dot{\varphi}_0}{r_0}\int_{e_0}^{e_1} \ud e\,\frac{e}{a(1-e^2)^{5/2}}\frac{1+\frac{20}{21}e^2+\frac{19}{336}e^4}{1+\frac{121}{304}e^2}\,.
\label{DC2}
\end{equation}
We discover here the memory effect: indeed, the real post-Newtonian order of the DC term \eqref{DC2} is found to be $1/c^5$ instead of the formal order $1/c^{10}$ exhibited in Eqs.~\eqref{mem1}--\eqref{mem2}, which means that it occurs at Newtonian level in the angular momentum flux instead of the 2.5PN order. This increase by a factor $\mathcal{O}(c^5)$, which corresponds to the inverse of the dominant order of radiation reaction, is clearly due to the cumulative integration over the past of the zero-frequency mode. 

We can simplify the result by using the relation for $a(e)$ deduced from the evolution equations for $e(t)$ and $a(t)$. Although the exact expression is known, we rather employ for simplicity an approximate relation given in Ref.~\cite{Pe64}, namely
\begin{equation}
\frac{a}{a_0}=\frac{1-e_0^2}{1-e^2}\left(\frac{e}{e_0}\right)^{12/19}\,.
\label{aeapprox}
\end{equation}
This finally yields
\begin{equation}
{\cal G}^{\rm DC} = -\frac{3}{19}\frac{G^3m^4\nu^2}{c^{5}}\frac{\dot{\varphi}_0}{r_0}\frac{e_0^{12/19}}{a_0(1-e_0^2)}\int_{e_0}^{e_1} \ud e\,\frac{e^{7/19}}{(1-e^2)^{3/2}}\frac{1+\frac{20}{21}e^2+\frac{19}{336}e^4}{1+\frac{121}{304}e^2}\,.
\label{DC3}
\end{equation}
We can see the importance of the dependence over the past by the fact that the integral \textit{diverges} when the initial eccentricity $e_1$ of the orbit approaches one [this would also be true had we used a more exact relation for $a(e)$]. For a binary system born on a very elongated elliptic orbit close to a parabolic one, the DC term is dominated by the initial evolution when $e_1\rightarrow 1$ and reads approximately
\begin{equation}
{\cal G}^{\rm DC}_{e_1\rightarrow 1} \sim -\frac{27}{119}\frac{G^3m^4\nu^2}{c^{5}}\frac{\dot{\varphi}_0}{r_0}\frac{e_0^{12/19}}{a_0(1-e_0^2)}\frac{1}{\sqrt{1-e_1^2}}\,.
\label{DC4}
\end{equation}
Finally we are interested in this paper in the averaged angular momentum flux. The formula \eqref{DC3}, averaged over the current orbit, gives immediately
\begin{equation}
\langle{\cal G}^{\rm DC}\rangle = -\frac{3}{19}m\,\nu^2\,c^{2}\,x_0^{7/2}\frac{e_0^{12/19}}{(1-e_0^2)^2}\int_{e_0}^{e_1} \ud e\,\frac{e^{7/19}}{(1-e^2)^{3/2}}\frac{1+\frac{20}{21}e^2+\frac{19}{336}e^4}{1+\frac{121}{304}e^2}\,.
\label{avDC}
\end{equation}
This indeed appears, comparing with \eqref{eq:AvAMFx}--\eqref{AMFxPN}, as a Newtonian-like term. This term will modify the enhancement factor at Newtonian order, namely $(1+\frac{7}{8}e_0^2)/(1-e_0^2)^2$, by an extra contribution coming from the binary's earlier  evolution. However we expect that this could be important only for very long-lived binary systems having started on a nearly parabolic orbit. For such systems the memory-induced modification of the Newtonian enhancement factor could be used to obtain an improved model of evolution of the orbital elements.

\section{Tables of numerical results}
\label{appB}
To facilitate the quantitative comparison of the PN results with high precision numerical computations of the inspiral and merger of eccentric binaries~\cite{HinderPNeccentric08}, we provide the numerical tables of all relevant enhancement functions in this Appendix.
\begin{table}[h]
\vspace{0.5cm}
\centering
\begin{tabular}{|c|c|c|c|c|}
\hline\hline
e & $\varphi(e)$ & $\psi(e)$ & $\zeta(e)$ & $\kappa(e)$\\\hline
0.00 & 1 & 1 & 1 & 1 \\
 0.05 & 1.031 & 0.9926 &
   1.052 & 1.066 \\
 0.10 & 1.127 & 0.9646 &
   1.221 & 1.282 \\
 0.15 & 1.304 & 0.8971 &
   1.540 & 1.703 \\
 0.20 & 1.588 & 0.7492 &
   2.082 & 2.446 \\
 0.25 & 2.027 & 0.4401 &
   2.981 & 3.738 \\
 0.30 & 2.702 & -0.1907 &
   4.480 & 6.020 \\
 0.35 & 3.757 & -1.468 &
   7.045 & 10.18 \\
 0.40 & 5.447 & -4.073 &
   11.589 & 18.07 \\
 0.45 & 8.254 & -9.499 &
   20.00 & 33.82 \\
 0.50 & 13.13 & -21.20 &
   36.44 & 67.20 \\
 0.55 & 22.07 & -47.70 &
   70.69 & 143.2 \\
 0.60 & 39.63 & -112.0 &
   148.0 & 332.4 \\
 0.65 & 77.23 & -282.5 &
   341.2 & 858.8 \\
 0.70 & 167.3 & -794.0 &
   890.4 & 2550 \\
 0.75 & 417.9 & -2611 &
   2753 & 9159 \\
 0.80 & 1282 & -10867 &
   10879 & 43276 \\
 0.85 & 5440 & -66117 &
   63317 & 315331 \\
 0.90 & 41628 & -810188 &
   746378 & 5.058$\times 10^{6}$\\
\hline
\end{tabular}
\vspace{0.5cm}
\caption{Tables for the numerical enhancement functions  appearing in the expression of the 3PN hereditary energy flux in Eq.~\eqref{Ftailfinal}. 
\label{tab1}}
\end{table}
\begin {table}[h]
\vspace{0.5cm}
\centering
\begin{tabular}{|c|c|c|c|c|}
\hline
\hline
 $e$  & $\tilde{\varphi}(e)$ & $\tilde{\psi}(e)$& $\tilde{\zeta}(e)$ &$\tilde{\kappa}(e)$\\
\hline
0.00 & 1     & 1      & 1     & 1     \\
0.05 & 1.016 & 0.9945 & 1.032 & 1.040 \\
0.10 & 1.067 & 0.9759 & 1.131 & 1.166 \\
0.15 & 1.157 & 0.9373 & 1.312 & 1.399 \\
0.20 & 1.294 & 0.8646 & 1.599 & 1.783 \\
0.25 & 1.493 & 0.7328 & 2.040 & 2.393 \\
0.30 & 1.775 & 0.4971 & 2.710 & 3.363 \\
0.35 & 2.176 & 0.0770 & 3.737 & 4.933 \\
0.40 & 2.752 &-0.6770 & 5.351 & 7.552 \\
0.45 & 3.597 & -2.051 & 7.964 & 12.09 \\
0.50 & 4.877 & -4.622 & 12.37 & 20.32\\
0.55 & 6.887 & -9.610 & 20.18 & 36.18\\
0.60 & 10.21 & -19.79 & 34.92 & 68.94\\
0.65 & 16.05 & -42.09 & 64.96 & 142.9\\
0.70 & 27.20 & -95.67 & 132.8 & 330.9\\
0.75 & 51.01 & -242.7 & 308.4 & 888.8\\
0.80 & 110.6 & -733.1 & 861.4 & 2957\\
0.85 & 301.0 & -2961  & 3218  & 13790\\
0.90 & 1239  & -21327 & 20453 & 1.19$\times10^5$\\
\hline
\end{tabular}
\vspace{0.5cm}
\caption{Tables for the numerical enhancement functions  appearing in the expression of the 3PN hereditary angular momentum flux in Eq.~\eqref{Gtailfinal}.
\label{tab2}}
\end{table}
\begin {table}[h]
\vspace{0.5cm}
\centering
\begin{tabular}{|c|c|c|c|c|c|}
\hline
\hline
 $e$ & $\psi_{n}(e)$& $\zeta_{n}(e)$ & $\varphi_{k}(e)$ & $\psi_{\w}(e)$ 
\\\hline
0.00 & 1     & 1      & 1     & 1     \\
0.05 & 1.013 & 1.053  & 1.020 &0.9657 \\
0.10 & 1.053 & 1.223  & 1.083 &0.8488 \\
0.15 & 1.119 & 1.546  & 1.197 &0.6018 \\
0.20 & 1.207 & 2.096  & 1.375 &0.1246 \\
0.25 & 1.308 & 3.007  & 1.644 &-0.7787\\
0.30 & 1.396 & 4.530  & 2.044 &-2.490 \\
0.35 & 1.405 & 7.137  & 2.646 &-5.776 \\
0.40 & 1.174 & 11.76  & 3.574 &-12.22 \\
0.45 & 0.311 & 20.33  & 5.051 &-25.27 \\
0.50 & -2.179& 37.09  & 7.508 &-52.86 \\
0.55 & -8.942& 72.06  & 11.82 &-114.4 \\
0.60 & -27.51& 151.1  & 19.93 &-262.2 \\
0.65 & -81.56& 348.6  & 36.56 &-651.7 \\
0.70 & -255.5& 910.8  & 74.68 & -1813 \\
0.75 & -909.7& 2819   & 176.2 &-5929  \\
0.80 & -4023 & 11149  & 512.0 &-24595 \\
0.85 & -25701& 64950  & 2059  &-1.49$\times10^5$  \\
0.90 & -3.28$\times10^5$ & 7.66$\times10^5$ & 14961& -1.83$\times10^6$ \\
\hline
\end{tabular}
\vspace{0.5cm}
\caption{Tables for the numerical enhancement functions  appearing in the 3PN hereditary part of the evolution of orbital elements $n$, $k$ and $\w$ in Eqs.~\eqref{ar-n-evo}. Recall that $\zeta_{\w}(e)=\zeta_{n}(e)$.
\label{tab3}} 
\end{table}
\vskip 10cm
\begin {table}[h]
\vspace{0.5cm}
\centering
\begin{tabular}{|c|c|c|c|c|c|}
\hline
\hline
 $e$ & $\psi_{a}(e)$& $\varphi_{e}(e)$ & $\psi_{e}(e)$& $\zeta_{e}(e)$ &$\kappa_{e}(e)$\\
\hline
0    & 1     & 1     & 1     &1    &1    \\
0.05 & 0.9488& 1.013 &0.9845 &1.024&1.027\\
0.10 &0.7773 & 1.056 &0.9338 &1.099&1.114\\
0.15 &0.4250 & 1.132 &0.8352 &1.237&1.273\\
0.20 &-0.2348& 1.248 &0.6629 &1.456&1.532\\
0.25 &-1.447 & 1.417 &0.3708 &1.790&1.938\\
0.30 &-3.686 & 1.658 &-0.1233&2.297&2.573\\
0.35 &-7.889 & 2.002 &-0.9688&3.076&3.590\\
0.40 &-15.98 & 2.502 &-2.444 &4.304&5.266\\
0.45 &-32.12 & 3.243 &-5.092 &6.307&8.152\\
0.50 & -65.77& 4.382 &-10.01 &9.726&13.38\\
0.55 & -140.0& 6.210 &-19.63 &15.89&23.50\\
0.60 & -316.5& 9.310 &-39.58 &27.81&44.67\\
0.65 & -777.3& 14.95 &-84.39 &52.93&93.54\\
0.70 & -2141& 26.20 &-196.2 &112.1&221.6\\
0.75 & -6936& 51.66 &-519.4 &274.7&620.2\\
0.80 & -28538& 120.3 &-1670  &827.6&2201\\
0.85 &-1.72$\times10^5$& 364.4 &-7363  &3451 &11332\\
0.90 & -2.09$\times10^6$& 1773\, &-58039&25971&    1.14$\times10^5$\\
\hline
\end{tabular}
\vspace{0.5cm}
\caption{Tables for the numerical enhancement functions  appearing in the 3PN hereditary part of the evolution of orbital elements $a_r$ and $e_t$ in Eqs.~\eqref{ar-n-evo}--\eqref{et-evo}. Recall that $\zeta_{a}(e)=\zeta_{n}(e)$.
\label{tab4}}
\end{table}
%############################################################
\section{Angular momentum flux and evolution of orbital elements in modified harmonic coordinates.}\label{appC}
%############################################################
In this paper we have first obtained the angular momentum flux
in the standard harmonic coordinates [Eq. (\ref{fluxSH})] 
- in terms of $r, {\dot r}, {\dot \varphi}$ - 
and transformed it to expressions in the ADM coordinates [Eq.(\ref{fluxADM})]. 
All subsequent formulas of the averaged flux and evolution of the 
orbital elements under gravitational radiation reaction refer
only to ADM coordinates.
Recent studies of binaries moving in  elliptical orbits
also employ the modified
harmonic coordinates and for the convenience of such investigations
we provide in this appendix explicit forms of the important equations
in modified harmonic coordinates.

Following the prescription
outlined in Sec. VI A of Ref.~\cite{ABIQ07} (see Eqs. (6.2) and (6.3) there), 
one can compute the 
angular momentum flux in modified harmonic (MH) coordinates starting from
the corresponding expression in the  standard harmonic (SH) coordinates
in Eq.~(\ref{fluxSH}).  The difference 
between the standard harmonic and modified harmonic coordinate
is a  3PN term.
The final result in MH coordinates may be written as
%\begin{equation}
${\cal G}_{\rm MH}={\cal G}_{\rm SH}+\delta_{\xi}{\cal G}$,
%\end{equation}
where
\begin{eqnarray}
\delta_{\xi}{\cal G}&=&-\frac{704}{15}\,\frac{G^5 m^6 \nu ^3}{c^{11}\,r^4}{\dot \varphi}\left[\frac{\dot{r}^2}{2}+\left(-\frac{v^2}{2}-\frac{3 \dot{r}^2}{4}+\frac{G m}{ r}\right)\ln \left(\frac{r}{r_0'}\right)\right].
\end{eqnarray}
As expected, in the expression for the angular momentum flux in MH coordinates
 the $\ln \left(r/r_0'\right) $ terms will be gauged away.

We next provide the expressions for the angular momentum flux averaged over
an orbit and the  secular evolution of various orbital elements 
in the modified harmonic coordinates. 
Notice that, in the 3PN accurate expressions, only the instantaneous terms 
at 2PN and 3PN will be different from the corresponding ADM expressions 
given earlier since  the difference between the  ADM and MH coordinates 
are at  order 2PN  and higher. 
Thus all the hereditary terms (which start at 1.5PN) will
be the same in the ADM and MH coordinates up to 3PN.
%, because their corrections will be 3.5PN and higher. 
Starting from the ADM expressions, one can obtain the corresponding 
results  for  various quantities 
by the sole transformation of  the eccentricity parameter $e_t$ as given 
in Eq.~\eqref{ADMMH}.
% (Notice that $x^{\rm ADM}=x^{\rm MH}$). 
Listed below  are the corresponding expressions at 2PN and 3PN orders 
in which $e_t$  now stands for $e_t^{\rm MH}$. As is obvious,
in the equation for  evolution of $K$, only the 3PN term will be different.
\begin{eqnarray}
{\cal H}_{\rm 2PN}^{\rm MH} &=& \frac{1}{(1-e_t^2)^4}\left\{-\frac{135431}{1134}+\frac{11287}{63}\nu+\frac{260}{9}\nu^2+\left(-\frac{598435}{756}+\frac{9497}{84}\nu+\frac{1546}{3}\nu^2\right)e_t^2\right.\nonumber\\
&&+\left.\left(\frac{30271}{432}-\frac{106381}{168}\nu+569\nu^2\right)e_t^4+\left(\frac{30505}{2016}-\frac{2201}{56}\nu+\frac{1519}{36}\nu^2\right)e_t^6\right.\nonumber\\
&&+\left.\sqrt{1-e_t^2}\left[80-32 \nu +e_t^2(335-134 \nu)+e_t^4(35-14 \nu)\right]\right\}\,,\\
{\cal H}_{\rm 3PN}^{\rm MH} &=&\frac{1}{(1-e_t^2)^5}\left\{\frac{2017023341}{1247400}+\frac{4340155}{6804}\nu-\frac{167483}{378}\nu^2-\frac{1550}{81}\nu^3\right.\nonumber\\
&&+e_t^2\left(\frac{540428354}{155925}+\left[\frac{621344957}{68040} + \frac{41 \pi ^2}{2}\right]\nu-\frac{416621}{108}\nu^2-\frac{96973}{81}\nu^3\right)\nonumber\\
&&+e_t^4\left(-\frac{6350078491}{1663200}+\left[\frac{1034477929}{90720} - \frac{11521 \pi ^2}{256}\right]\nu+\frac{720619}{1008}\nu^2-\frac{438907}{108}\nu^3\right)\nonumber\\
&&+e_t^6\left(-\frac{272636461}{554400}+\left[\frac{4987541}{18144} - \frac{615 \pi ^2}{128}\right]\nu+\frac{1885945}{1008}\nu^2-\frac{283205}{162}\nu^3\right)\nonumber\\
&&+e_t^8\left(-\frac{10305073}{709632}+\frac{417923}{12096}\nu+\frac{95413}{8064}\nu^2-\frac{146671}{2592}\nu^3\right)\nonumber\\
&&+\sqrt{1-e_t^2}\left[-\frac{379223}{630}+\left[-\frac{48907}{63} + \frac{41 \pi ^2}{6}\right]\nu+\frac{580}{3}\nu^2\right.\nonumber\\
&&+\left.e_t^2\left(\frac{309083}{315}+\left[-\frac{456250}{63} + \frac{2747 \pi ^2}{96}\right]\nu+1902\nu^2\right)\right.\nonumber\\
&&+\left.e_t^4\left(\frac{13147661}{5040}+\left[-\frac{2267795}{504} + \frac{287 \pi ^2}{96}\right]\nu+\frac{2703}{2}\nu^2\right)\right.\nonumber\\
&&+\left.e_t^6\left(70-\frac{203}{3}\nu+\frac{77}{3}\nu^2\right)\right]\nonumber\\
&&+\left(\frac{13696}{105}+\frac{98012}{105}e_t^2+\frac{23326}{35}e_t^4
+\frac{2461}{70}e_t^6\right)\,\ln\left[\frac{x}{x_0}\frac{1
+\sqrt{1-e_t^2}}{2(1-e_t^2)}\right]\Biggr\}\,,
\end{eqnarray}

\begin{eqnarray}
{\cal N}_{\rm 2PN}^{\rm MH}&=& \frac{1}{(1-e_t^2)^{11/2}}\left\{-\frac{1159}{945}+\frac{15265}{21}\nu+\frac{944}{15}\nu^2+e_t^2\left(-\frac{4819994}{945}+3231\nu+\frac{182387}{90}\nu^2\right)\right.\nonumber\\
&&+\left.e_t^4\left(-\frac{203957}{90}-\frac{764357}{140}\nu+\frac{396443}{72}\nu^2\right)+e_t^6\left(\frac{4760347}{1680}-\frac{993011}{240}\nu+\frac{192943}{90}\nu^2\right)\right.\nonumber\\
&&+\left.e_t^8\left(\frac{391457}{3360}-\frac{6037}{56}\nu+\frac{2923}{45}\nu^2\right)\right.\nonumber\\
&&+\left.\sqrt{1-e_t^2}\left[48-\frac{96}{5}\nu+e_t^2\left(2134-\frac{4268}{5}\nu\right)+e_t^4\left(2193-\frac{4386}{5}\nu\right)\right.\right.\nonumber\\
&&+\left.\left.e_t^6\left(\frac{175}{2}-35\nu\right)\right]\right\}\,,\\
{\cal N}_{\rm 3PN}^{\rm MH} &=&\frac{1}{(1-e_t^2)^{13/2}}\left\{\frac{4915859933}{1039500}+\left[\frac{1463719}{2268} - \frac{369 \pi ^2}{10}\right]\nu-\frac{711931}{420}\nu^2-\frac{1121}{27}\nu^3\right.\nonumber\\
&&+e_t^2\left(\frac{10928916619}{297000}+\left[\frac{4697941919}{113400} - \frac{1599 \pi ^2}{80}\right]\nu-\frac{23667137}{840}\nu^2-\frac{1287385}{324}\nu^3\right)\nonumber\\
&&+e_t^4\left(-\frac{127363208627}{8316000}+\left[\frac{9286298159}{64800} - \frac{94423 \pi ^2}{160}\right]\nu-\frac{5331901}{224}\nu^2-\frac{33769597}{1296}\nu^3\right)\nonumber\\
&&+e_t^6\left(-\frac{82502370763}{5544000}+\left[\frac{2015302783}{100800} - \frac{12751 \pi ^2}{32}\right]\nu+\frac{18129215}{448}\nu^2-\frac{3200965}{108}\nu^3\right)\nonumber\\
&&+e_t^8\left(\frac{59641969601}{7392000}+\left[-\frac{2779943}{210} - \frac{12177 \pi ^2}{640}\right]\nu+\frac{5399701}{420}\nu^2-\frac{982645}{162}\nu^3\right)\nonumber\\
&&+e_t^{10}\left(\frac{33332681}{197120}-\frac{1874543}{10080}\nu+\frac{109733}{840}\nu^2-\frac{8288}{81}\nu^3\right)\nonumber\\
&&+\sqrt{1-e_t^2}\left[-\frac{2667319}{1125}+\left[\frac{56242}{105} - \frac{41 \pi ^2}{10}\right]\nu+\frac{632}{5}\nu^2\right.\nonumber\\
&&+e_t^2\left(-\frac{2673296}{375}+\left[-\frac{10074037}{315} + \frac{45961 \pi ^2}{240}\right]\nu+\frac{125278}{15}\nu^2\right)\nonumber\\
&&+e_t^4\left(\frac{700397951}{21000}+\left[-\frac{4767517}{60} + \frac{6191 \pi ^2}{32}\right]\nu+\frac{317273}{15}\nu^2\right)\nonumber\\
&&+e_t^6\left(\frac{708573457}{31500}+\left[-\frac{6849319}{252} + \frac{287 \pi ^2}{960}\right]\nu+\frac{232177}{30}\nu^2\right)\nonumber\\
&&+\left.e_t^8\left(\frac{56403}{112}-\frac{427733}{840}\nu+\frac{4739}{30}\nu^2\right)\right]\nonumber\\
&&+\left.\left(\frac{54784}{175} + \frac{465664 }{105}e_t^2 + \frac{4426376
  \ }{525}e_t^4 + \frac{1498856 }{525}e_t^6 + \frac{31779
  }{350}e_t^8\right)\right.\nonumber\\
&&\left.\times\ln\left[\frac{x}{x_0}\frac{1
  +\sqrt{1-e_t^2}}{2(1-e_t^2)}\right]\right\}\,,
\end{eqnarray}
\begin{eqnarray}
{\cal O}_{\rm 2PN}^{\rm MH} &=& \frac{1}{(1-e_t^2)^{11/2}}\left\{-\frac{11257}{945}+\frac{15677}{105}\nu+\frac{944}{15}\nu^2+e_t^2\left(-\frac{580291}{189}+\frac{2557}{5}\nu+\frac{182387}{90}\nu^2\right)\right.\nonumber\\
&&+e_t^4\left(\frac{32657}{1260}-\frac{959279}{140}\nu+\frac{396443}{72}\nu^2\right)+e_t^6\left(\frac{4634689}{1680}-\frac{977051}{240}\nu+\frac{192943}{90}\nu^2\right)\nonumber\\
&&+e_t^8\left(\frac{391457}{3360}-\frac{6037}{56}\nu+\frac{2923}{45}\nu^2\right)\nonumber\\
&&+\sqrt{1-e_t^2}\left[48-\frac{96}{5}\nu+e_t^2\left(2134-\frac{4268}{5}\nu\right)+e_t^4\left(2193-\frac{4386}{5}\nu\right)\right.\nonumber\\
&&+\left.\left.e_t^6\left(\frac{175}{2}-35\nu\right)\right]\right\}\,,\\
%%%%%%%%%%%%%%%%%%%%%%%
{\cal O}_{\rm 3PN}^{\rm MH} &=& \frac{1}{(1-e_t^2)^{13/2}}\left\{\frac{614389219}{148500}+\left[-\frac{57265081}{11340} + \frac{369 \pi^2}{2}\right]\nu-\frac{16073}{140}\nu^2-\frac{1121}{27}\nu^3\right.\nonumber\\
&&+e_t^2\left(\frac{19898670811}{693000}+\left[\frac{2678401319}{113400} + \frac{3239 \pi^2}{16}\right]\nu-\frac{9657701}{840}\nu^2-\frac{1287385}{324}\nu^3\right)\nonumber\\
&&+e_t^4\left(\frac{8036811073}{8316000}+\left[\frac{43741211273}{453600} - \frac{197087 \pi ^2}{320}\right]\nu+\frac{1306589}{672}\nu^2-\frac{33769597}{1296}\nu^3\right)\nonumber\\
&&+e_t^6\left(\frac{985878037}{5544000}+\left[\frac{54136669}{14400} - \frac{261211 \pi ^2}{640}\right]\nu+\frac{62368205}{1344}\nu^2-\frac{3200965}{108}\nu^3\right)\nonumber\\
&&+e_t^8\left(\frac{2814019181}{352000}+\left[-\frac{4342403}{336} - \frac{12177 \pi ^2}{640}\right]\nu+\frac{3542389}{280}\nu^2-\frac{982645}{162}\nu^3\right)
\nonumber\\
&&+e_t^{10}\left(\frac{33332681}{197120}-\frac{1874543}{10080}\nu+\frac{109733}{840}\nu^2-\frac{8288}{81}\nu^3\right)\nonumber\\
&&+\sqrt{1-e_t^2}\left[-\frac{1425319}{1125}+\left[\frac{9874}{105} - \frac{41 \pi ^2}{10}\right]\nu+\frac{632}{5}\nu^2\right.\nonumber\\
&&+e_t^2\left(\frac{933454}{375}+\left[-\frac{2257181}{63} + \frac{45961 \pi ^2}{240}\right]\nu+\frac{125278}{15}\nu^2\right)\nonumber\\
&&+e_t^4\left(\frac{840635951}{21000}+\left[-\frac{4927789}{60} + \frac{6191 \pi ^2}{32}\right]\nu+\frac{317273}{15}\nu^2\right)\nonumber\\
&&+e_t^6\left(\frac{702667207}{31500}+\left[-\frac{6830419}{252} + \frac{287 \pi ^2}{960}\right]\nu+\frac{232177}{30}\nu^2\right)\nonumber\\
&&+\left.e_t^8\left(\frac{56403}{112}-\frac{427733}{840}\nu+\frac{4739}{30}\nu^2\right)\right]\nonumber\\
&&+\left(\frac{54784}{175} + \frac{465664 }{105}e_t^2 + \frac{4426376 }{525}e_t^4 + \frac{1498856 }{525}e_t^6 + \frac{31779
  }{350}e_t^8\right)\nonumber\\
&&\times\ln\left[\frac{x}{x_0}\frac{1
  +\sqrt{1-e_t^2}}{2(1-e_t^2)}\right]\Biggr\}\,,
\end{eqnarray}
\begin{eqnarray}
{\cal A}_{\rm 2PN}^{\rm MH} &=& \frac{1}{(1-e_t^2)^{11/2}}\left\{\frac{194882}{2835}-\frac{34882}{315}\nu-\frac{608}{15}\nu^2+e_t^2\left(\frac{938912}{567}+\frac{14093}{45}\nu-\frac{20243}{15}\nu^2\right)\right.\nonumber\\
&&+e_t^4\left(-\frac{417341}{270}+\frac{7203533}{1260}\nu-\frac{73549}{20}\nu^2\right)+e_t^6\left(-\frac{5422741}{2520}+\frac{1069759}{360}\nu-\frac{64169}{45}\nu^2\right)\nonumber\\
&&+e_t^8\left(-\frac{366593}{5040}+\frac{9703}{126}\nu-\frac{1924}{45}\nu^2\right)\nonumber\\
&&+\sqrt{1-e_t^2}\left[-96+\frac{192}{5}\nu+e_t^2\left(-1452+\frac{2904}{5}\nu\right)+e_t^4\left(-1353+\frac{2706}{5}\nu\right)\right.\nonumber\\
&&+\left.\left.e_t^6\left(-74+\frac{148}{5}\nu\right)\right]\right\}\,,\\
%%%%%%%%%%%%%%%%%%%%%%%%%%%%%%%%%%%%%%%%%%
{\cal A}_{\rm 3PN}^{\rm MH} &=& \frac{1}{(1-e_t^2)^{13/2}}\left\{-\frac{4121173183}{1559250}+\left[\frac{3347773}{1701} - \frac{449 \pi ^2}{5}\right]\nu+\frac{285629}{1890}\nu^2+\frac{122}{5}\nu^3\right.\nonumber\\
&&+e_t^2\left(-\frac{17773786511}{1039500}+\left[-\frac{525272336}{42525} - \frac{1763 \pi ^2}{8}\right]\nu+\frac{20208193}{3780}\nu^2+\frac{78437}{30}\nu^3\right)\nonumber\\
&&+e_t^4\left(-\frac{77462249023}{12474000}+\left[-\frac{1804146458}{42525} + \frac{155681 \pi ^2}{480}\right]\nu-\frac{183216629}{15120}\nu^2+\frac{2089273}{120}\nu^3\right)\nonumber\\
&&+e_t^6\left(-\frac{14962838591}{1188000}+\left[\frac{88290809}{5400} + \frac{73269 \pi ^2}{320}\right]\nu-\frac{391717217}{10080}\nu^2+\frac{1781461}{90}\nu^3\right)\nonumber\\
&&+e_t^8\left(-\frac{10456078343}{1584000}+\left[\frac{373304}{35} + \frac{12271 \pi ^2}{960}\right]\nu-\frac{24022093}{2520}\nu^2+\frac{180428}{45}\nu^3\right)\nonumber\\
&&+e_t^{10}\left(-\frac{81086491}{887040}+\frac{109847}{864}\nu-\frac{66209}{630}\nu^2+\frac{592}{9}\nu^3\right)\nonumber\\
&&+\sqrt{1-e_t^2}\left[\frac{17249966}{23625}+\left[\frac{21500}{21} - \frac{41 \pi ^2}{5}\right]\nu-240\nu^2\right.\nonumber\\
&&+e_t^2\left(-\frac{39170981}{7875}+\left[\frac{2741416}{105} - \frac{4961 \pi ^2}{40}\right]\nu-\frac{29964}{5}\nu^2\right)\nonumber\\
&&+e_t^4\left(-\frac{975995201}{31500}+\left[\frac{5741363}{105} - \frac{18491 \pi ^2}{160}\right]\nu-\frac{68913}{5}\nu^2\right)\nonumber\\
&&+e_t^6\left(-\frac{2779019203}{189000}+\left[\frac{22404341}{1260} - \frac{1517 \pi ^2}{240}\right]\nu-\frac{24347}{5}\nu^2\right)\nonumber\\
&&+\left.e_t^8\left(-\frac{17933}{42}+\frac{47459}{105}\nu-\frac{666}{5}\nu^2\right)\right]\nonumber\\
&&-\left(\frac{109568}{525}+\frac{931328 }{315}e_t^2+\frac{8852752}{1575}
e_t^4+\ \frac{2997712 }{1575}e_t^6+\frac{10593 }{175}e_t^8\right)\nonumber\\
&&\times
\ln\left[\frac{x}{x_0}\frac{1+\sqrt{1-e_t^2}}{2(1-e_t^2)}\right]
\Biggr\}\,,
\end{eqnarray}
\begin{eqnarray}
{\cal E}_{\rm 2PN}^{\rm MH} &=& \frac{1}{(1-e_t^2)^{9/2}}\left\{-\frac{952397}{1890}+\frac{5937}{14}\nu+\frac{752}{5}\nu^2\right.\nonumber\\
&&+e_t^2\left(-\frac{3113989}{2520}-\frac{388419}{280}\nu+\frac{64433}{40}\nu^2\right)+e_t^4\left(\frac{4656611}{3024}-\frac{13057267}{5040}\nu+\frac{127411}{90}\nu^2\right)\nonumber\\
&&+e_t^6\left(\frac{420727}{3360}-\frac{362071}{2520}\nu+\frac{821}{9}\nu^2\right)\nonumber\\
&&+\left.\sqrt{1-e_t^2}\left[\frac{1336}{3}-\frac{2672}{15}\nu+e_t^2\left(\frac{2321}{2}-\frac{2321}{5}\nu\right)+e_t^4\left(\frac{565}{6}-\frac{113}{3}\nu\right)\right]\right\}\,\\
%%%%%%%%%%%%%%%%%%%%%%%%%%%%%%%%%%%%%%%%%%
{\cal E}_{\rm 3PN}^{\rm MH} &=& \frac{1}{(1-e_t^2)^{11/2}}\left\{\frac{7742634967}{891000}+\left[\frac{43386337}{113400} + \frac{1017 \pi ^2}{10}\right]\nu-\frac{4148897}{2520}\nu^2-\frac{61001}{486}\nu^3\right.\nonumber\\
&&+e_t^2\left(\frac{6556829759}{891000}+\left[\frac{770214901}{25200} - \frac{15727 \pi ^2}{192}\right]\nu-\frac{80915371}{15120}\nu^2-\frac{86910509}{19440}\nu^3\right)\nonumber\\
&&+e_t^4\left(-\frac{17072216761}{2376000}+\left[\frac{8799500893}{907200} - \frac{295559 \pi ^2}{1920}\right]\nu+\frac{351962207}{20160}\nu^2-\frac{2223241}{180}\nu^3\right)\nonumber\\
&&+e_t^6\left(\frac{17657772379}{3696000}+\left[-\frac{91818931}{10080} - \frac{6519 \pi ^2}{640}\right]\nu+\frac{2495471}{252}\nu^2-\frac{11792069}{2430}\nu^3\right)\nonumber\\
&&+e_t^8\left(\frac{302322169}{1774080}-\frac{1921387}{10080}\nu+\frac{41179}{216}\nu^2-\frac{193396}{1215}\nu^3\right)\nonumber\\
&&+\sqrt{1-e_t^2}\left[-\frac{22713049}{15750}+\left[-\frac{5526991}{945} + \frac{8323 \pi ^2}{180}\right]\nu+\frac{54332}{45}\nu^2\right.\nonumber\\
&&+e_t^2\left(\frac{89395687}{7875}+\left[-\frac{38295557}{1260} 
+ \frac{94177 \pi ^2}{960}\right]\nu+\frac{681989}{90}\nu^2\right)\nonumber\\
&&+e_t^4\left(\frac{5321445613}{378000}+\left[-\frac{26478311}{1512} + \frac{2501 \pi ^2}{2880}\right]\nu+\frac{225106}{45}\nu^2\right)\nonumber\\
&&+\left.e_t^6\left(\frac{186961}{336}-\frac{289691}{504}\nu+\frac{3197}{18}\nu^2\right)\right]+\frac{730168
}{23625}\,\frac{1}{1+\sqrt{1-e_t^2}}\nonumber\\
&&+{304\over 15}
\left(\frac{82283}{1995}
+ \frac{297674 }{1995}e_t^2 + \frac{1147147 \ }{15960}e_t^4 + \frac{61311
}{21280}e_t^6\right)\,\ln\left[\frac{x}{x_0}\frac{1
+\sqrt{1-e_t^2}}{2(1-e_t^2)}\right]\Biggr\}\,,
\end{eqnarray}
\begin{eqnarray}
%%%%%%%%%%%%%%%%%%%
{\cal K}_{\rm 3PN}^{\rm MH} &=& \frac{1}{(1-e_t^2)^{11/2}}\left\{\frac{232082}{189}+\left[-\frac{131366}{21} + \frac{738 \pi ^2}{5}\right]\nu+\frac{13312}{15}\nu^2\right.\nonumber\\
&&+e_t^2\left(\frac{2895119}{630}+\left[-\frac{1509994}{105} + \frac{1271 \pi ^2}{10}\right]\nu+\frac{29879}{5}\nu^2\right)\nonumber\\
&&+e_t^4\left(\frac{1651297}{252}+\left[-\frac{1229554}{105} + \frac{5371 \pi ^2}{320}\right]\nu+\frac{54133}{10}\nu^2\right)\nonumber\\
&&+e_t^6\left(\frac{1850407}{1680}-\frac{388799}{280}\nu+\frac{19573}{30}\nu^2\right)\nonumber\\
&&+\left.\sqrt{1-e_t^2}\left[672-\frac{1344}{5}\nu+e_t^2\left(2436-\frac{4872}{5}\nu\right)+e_t^4\left(672-\frac{1344}{5}\nu\right)\right]\right\}\,.
\end{eqnarray}

We conclude this appendix  with two useful relations, the  analogues of Eqs. (\ref{ADMMH}) and (\ref{xtozetaADM}) but
expressed in  MH  coordinates. We have,
\begin{eqnarray}
e_t^{\rm ADM}&=&e_t^{\rm MH}\left\{1+\frac{\zeta^{4/3}}{1-e_t^2}
\left(\frac{1}{4}+\frac{17}{4}\nu\right)+\frac{\zeta^2}{(1-e_t^2)^2}\left[\frac{3}{2}+\left(\frac{45299}{1680}-\frac{21 }{16}\pi ^2\right) \nu 
-\frac{83 }{24}\nu ^2\right.\right.\nonumber\\
&&\qquad+\left.\left.e_t^2\left(\frac{1}{2}+\frac{249  }{16}\nu
-\frac{241 }{24}\nu ^2\right)\right]\right\}\,,
\end{eqnarray}
\begin{eqnarray}
x &=& \zeta^{2/3}\left\{1+\frac{2}{1-e_t^2}\zeta^{2/3}+
\frac{1}{(1-e_t^2)^2}\left[12-\frac{14}{3}\nu+e_t^2\left(\frac{17}{2}-\frac{13}{3}\nu\right)\right]\zeta^{4/3}\right.\nonumber\\
&&+\left.\frac{1}{(1-e_t^2)^3}\left[\frac{250}{3}+\left(-\frac{255}{2}+\frac{41 }{16}\pi ^2\right) \nu +\frac{14 }{3}\nu ^2+e_t^2\left(138+\left(-\frac{422}{3}+\frac{41 }{64}\pi ^2\right) \nu +\frac{80 }{3}\nu ^2\right)\right.\right.\nonumber\\
&&+\left.\left.e_t^4\left(13-\frac{55  }{6}\nu+\frac{65 }{12}\nu ^2\right)+\sqrt{1-e_t^2}\left(10-4 \nu +e_t^2(20-8 \nu ) \right)\right]\zeta^{2}\right\}.
\end{eqnarray}

\bibliography{/home/arun/tphome/arun/ref-list}
\end{document}